\documentclass[aps,twocolumn]{revtex4}

\usepackage{amssymb}
\usepackage{amsmath}
\usepackage{bm}
\usepackage[dvips]{epsfig}
\usepackage{graphicx}
\usepackage{textcomp}

\begin{document}

\title{The second boundaries of Stability Zones and the angular
diagrams of conductivity for metals having complicated Fermi 
surfaces.}

\author{A.Ya. Maltsev.}

\affiliation{
\centerline{\it L.D. Landau Institute for Theoretical Physics}
\centerline{\it 142432 Chernogolovka, pr. Ak. Semenova 1A,
maltsev@itp.ac.ru}}

\begin{abstract}
We consider some general aspects of dependence of 
magneto-conductivity on a magnetic field in metals having 
complicated Fermi surfaces. As it is well known, a nontrivial 
behavior of conductivity in metals in strong magnetic fields is 
connected usually with appearance of non-closed quasiclassical 
electron trajectories on the Fermi surface in a magnetic field.
The structure of the electron trajectories depends strongly
on the direction of the magnetic field  and usually the 
greatest interest is caused by open trajectories that are 
stable to small rotations of the direction of $\, {\bf B} $.
The geometry of the corresponding Stability Zones on the angular 
diagram in the space of directions of $\, {\bf B} \, $ represents 
a very important characteristic of the electron spectrum in 
a metal linking the parameters of the spectrum to the 
experimental data. Here we will consider some very general 
features inherent in the angular diagrams of metals with 
Fermi surfaces of the most arbitrary form. In particular,
we will show here that any Stability Zone actually has a second 
boundary, restricting a larger region with a certain behavior of 
conductivity. Besides that, we  shall discuss here general 
questions of complexity of the angular diagrams for the 
conductivity and propose a theoretical scheme for dividing 
the angular diagrams into ``simple'' and ``complex'' diagrams. 
The proposed scheme will in fact also be closely related to 
behavior of the Hall conductivity in a metal in strong magnetic 
fields. In conclusion, we will also discuss the relationship 
of the questions under consideration to the general features 
of an (abstract) angular diagram describing the behavior of 
quasiclassical electron trajectories at all energy levels 
in the conduction band.
\end{abstract}

\maketitle

\section{Introduction.}

 We will consider here phenomena connected with some special
properties of the quasiclassical electron motion along the
Fermi surface in normal metals. According to standard approach,
we will describe the electron states in a metal with the aid
of the number of the energy band and the value of the
quasimomentum $\, {\bf p} $. As it is well known, any value
of $\, {\bf p} \, $ should in fact be considered as an infinite 
set of ``equivalent'' values $\, {\bf p} \in \mathbb{R}^{3} \, $,
differing by vectors of the reciprocal lattice $\, {L}^{*} $.
The basis of the reciprocal lattice can be given by the vectors

$${\bf a}_{1} \,\,\, = \,\,\, 2 \pi \hbar \,\,
{{\bf l}_{2} \, \times \, {\bf l}_{3} \over
({\bf l}_{1}, \, {\bf l}_{2}, \, {\bf l}_{3} )} \,\,\, , \quad \quad
{\bf a}_{2} \,\,\, = \,\,\, 2 \pi \hbar \,\,
{{\bf l}_{3} \, \times \, {\bf l}_{1} \over
({\bf l}_{1}, \, {\bf l}_{2}, \, {\bf l}_{3} )} \,\,\, ,  $$
$${\bf a}_{3} \,\,\, = \,\,\, 2 \pi \hbar \,\,
{{\bf l}_{1} \, \times \, {\bf l}_{2} \over
({\bf l}_{1}, \, {\bf l}_{2}, \, {\bf l}_{3} )} \,\,\, , $$
where $\, ({\bf l}_{1}, \, {\bf l}_{2}, \, {\bf l}_{3} ) \, $
represent a basis of the direct lattice of a metal. The energy
of an electron state in the band $\, s \, $ is given by a 
three-periodic function $\, \epsilon_{s} ({\bf p}) \, $
in the $\, {\bf p}$ - space, which can also be considered as a
function on the three-dimensional torus $\, \mathbb{T}^{3} \, $,
given by the factorization of the $\, {\bf p}$ - space with
respect to the reciprocal lattice vectors
$$\mathbb{T}^{3} \,\,\, = \,\,\, \mathbb{R}^{3} / L^{*} $$

\vspace{1mm}

 As it is also well known, the electron states in normal metals
are occupied, if their energy is less than the Fermi energy
$\, \epsilon_{s} ({\bf p}) \, < \, \epsilon_{F} \, $, and are
empty if $\, \epsilon_{s} ({\bf p}) \, > \, \epsilon_{F} \, $.
So, we have in any normal metal a finite number of completely
filled energy bands, a finite number of partially filled 
energy bands (conduction bands) and an infinite number of
empty energy bands. The Fermi surface $\, S_{F} \, $ of a
normal metal is defined as the union of the surfaces
$\, \epsilon_{s} ({\bf p}) \, = \, \epsilon_{F} \, $ in the
$\, {\bf p}$ - space, where the index $\, s \, $ numerates
all the conduction bands. In all our considerations here
we will assume that the Fermi surface represents a smooth
3-periodic non-selfintersecting surface in the 
$\, {\bf p}$ - space, which can contain different connected
components.

\vspace{1mm}

 More precisely, a collective electron state in a metal can be
described in the kinetic approximation by a one-particle 
distribution function $\, f ({\bf p}) $, which can be different
from $0$ or $1$ just in a narrow region near the Fermi surface.
So, all the kinetic properties of a normal metal are defined
in fact by electron states near the surface $\, S_{F} \, $.

 The quasiclassical evolution of electron states in the
presence of external electric and magnetic fields can be
described with the aid of the adiabatic system
$${\dot {\bf p}} \,\,\,\, = \,\,\,\, {e \over c} \,\,
\left[ {\bf v}_{\rm gr} ({\bf p}) \, \times \, {\bf B} \right]
\,\,\, + \,\,\, e \, {\bf E} \,\,\, , $$
which can be considered as a dynamical system on the torus
$\, \mathbb{T}^{3} $.

 The function $\, f ({\bf p}) \, $ satisfies
the one-particle kinetic equation
\begin{multline*}
f_{t} \,\,\, + \,\,\, {e \over c} \, \sum_{l=1}^{3} \,
\left[ \nabla \epsilon ({\bf p}) \times {\bf B} \right]^{l} \,\,
{\partial f \over \partial p^{l}} \,\,\, + \,\,\, 
e \, \sum_{l=1}^{3} \, E^{l} \, {\partial f \over \partial p^{l}} 
\,\,\,\, =  \\ 
 = \,\,\,\, I [f] ({\bf p}, \, t) \quad ,  
\end{multline*}
where $\, I [f] \, $ represents the collision integral.

\vspace{1mm}

 In our considerations here we will consider the electric field
$\, {\bf E} \, $ as infinitely small. At the same time, the
value $\, B \, $ of the magnetic field is supposed to be large
enough. More precisely, we will assume here the condition
$\, \omega_{B} \tau \, \gg \, 1 \, $, where $\, \tau \, $
plays the role of the mean free electron motion time and
$\, \omega_{B} \, = \, e B / m^{*} c \, $ has the meaning of the
cyclotron frequency of the electron motion in crystal. As a
corollary, the phenomena considered here will be actually 
connected with the geometry of trajectories of the system

\begin{equation}
\label{MFSyst}
{\dot {\bf p}} \,\,\,\, = \,\,\,\, {e \over c} \,\,
\left[ {\bf v}_{\rm gr} ({\bf p}) \, \times \, {\bf B} \right]
\,\,\,\, = \,\,\,\, {e \over c} \,\, \left[ \nabla \epsilon ({\bf p})
\, \times \, {\bf B} \right]
\end{equation}
 
\vspace{1mm}
 
\begin{figure}[t]
\begin{center}
\includegraphics[width=0.9\linewidth]{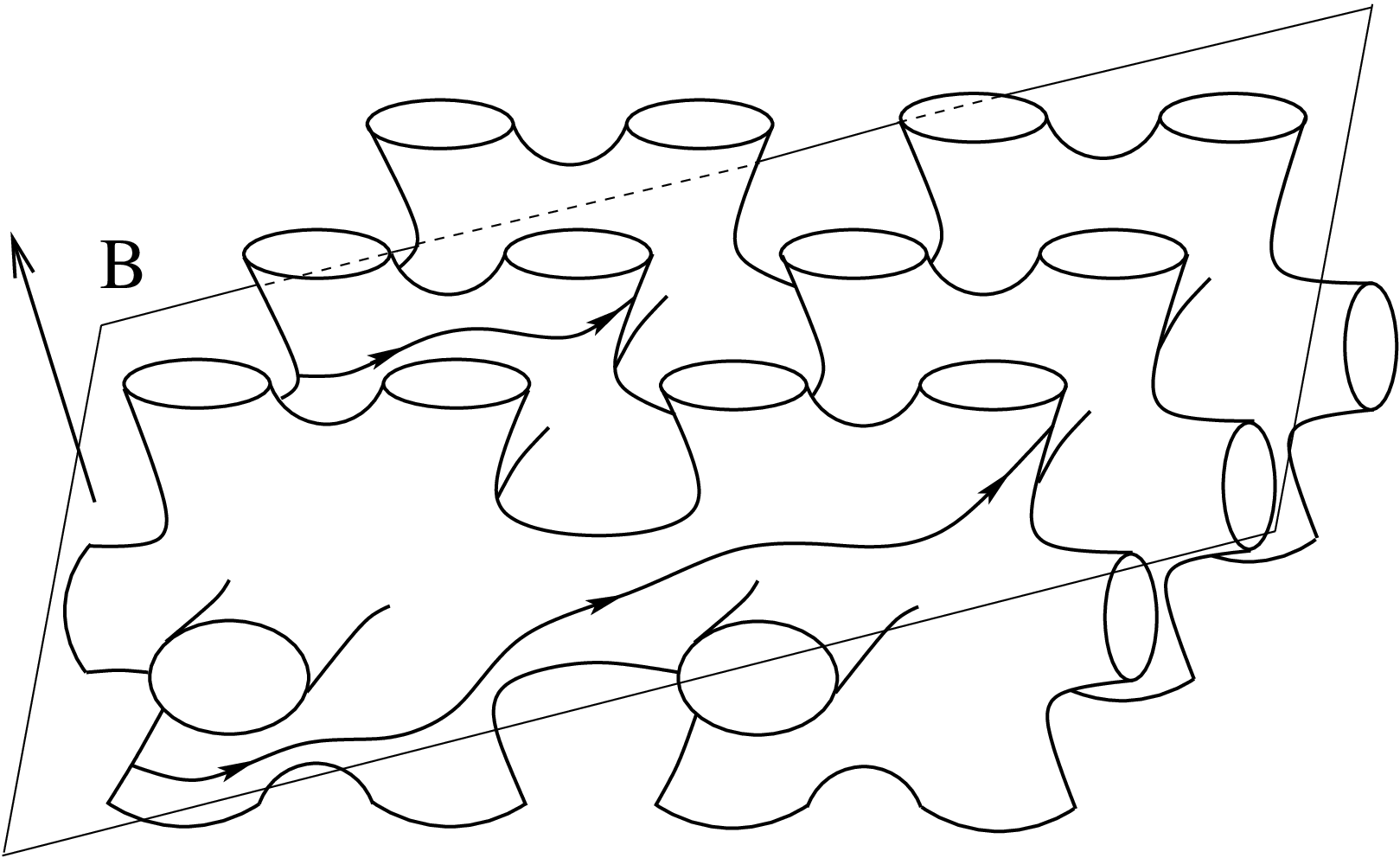}
\end{center}
\caption{Quasiclassical trajectories on a constant energy surface
of a complicated form.}
\label{ComplSurf}
\end{figure}

 System (\ref{MFSyst}) conserves the energy of an electron state
and also the projection of the quasimomentum $\, {\bf p} \, $ on
the directions of $\, {\bf B} $. So, the trajectories of system
(\ref{MFSyst}) in the $\, {\bf p}$ - space can be obtained as the
intersections of the constant energy levels 
$\, \epsilon_{s} ({\bf p}) \, = \, const \, $ with the planes
orthogonal to $\, {\bf B} $. Let us say, however, that in spite 
of this analytical integrability of system (\ref{MFSyst})
the geometry of its trajectories in $\, {\bf p}$ - space can
in fact be rather complicated for the energy levels having
complicated form (Fig. \ref{ComplSurf}). It is easy to see also
that in kinetic phenomena in normal metals only the trajectories
lying on the Fermi surface can play an essential role.

\vspace{1mm}

 An important role of the geometry of trajectories of system
(\ref{MFSyst}) for the magneto-transport phenomena in metals
was first revealed by the school of I.M. Lifshitz 
(I.M. Lifshitz, M.Ya. Azbel, M.I. Kaganov, V.G. Peschansky)
in 1950's. In particular, a crucial difference between the
closed and open trajectories of system (\ref{MFSyst}), which
is manifested in the behavior of the magneto-conductivity
in strong magnetic fields, was first described in detail in the
papers \cite{lifazkag,lifpes1,lifpes2}. Thus, it was first
pointed out in the paper \cite{lifazkag} that the closed and
the open periodic trajectories of system (\ref{MFSyst}) give
quite different contributions to the magneto-conductivity
in the plane, orthogonal to $\, {\bf B} $, in the limit
$\, \omega_{B} \tau \, \gg \, 1 \, $. More precisely, the
contribution of the closed trajectories in the plane,
orthogonal to $\, {\bf B} $, is almost isotropic in the limit
$\, \omega_{B} \tau \, \gg \, 1 \, $ and decreases rapidly
when $\, \omega_{B} \tau \, \rightarrow \infty \, $. The 
general conductivity tensor $\, \sigma^{kl} (B) \, $ in the 
presence of only closed trajectories on the Fermi level 
can be expressed by the following asymptotic formula

\begin{multline}
\label{Closed}
\sigma^{kl} \,\,\,\, \simeq \,\,\,\,
{n e^{2} \tau \over m^{*}} \, \left(
\begin{array}{ccc}
( \omega_{B} \tau )^{-2}  &  ( \omega_{B} \tau )^{-1}  &
( \omega_{B} \tau )^{-1}  \cr
( \omega_{B} \tau )^{-1}  &  ( \omega_{B} \tau )^{-2}  &
( \omega_{B} \tau )^{-1}  \cr
( \omega_{B} \tau )^{-1}  &  ( \omega_{B} \tau )^{-1}  &  *
\end{array}  \right) \,\, ,   \\
\omega_{B} \tau \,\, \rightarrow \,\, \infty  \,\,\, ,
\end{multline}
where $\, {\hat z} \, = \, {\bf B} / B \, $, and is analogous
to the same formula for the free electron gas.

 At the same time, the contribution of the open periodic
trajectories to the conductivity is strongly anisotropic
in the plane orthogonal to $\, {\bf B} \, $ and decreases just
in one direction (coinciding with the mean direction of the 
open trajectories in $\, {\bf p}$ - space) in the limit
$\, \omega_{B} \tau \, \rightarrow \infty \, $. The general 
formula for the asymptotic behavior of the conductivity 
tensor can be written in this case in the form 

\begin{multline}
\label{Periodic}
\sigma^{kl} \,\,\,\, \simeq \,\,\,\,
{n e^{2} \tau \over m^{*}} \, \left(
\begin{array}{ccc}
( \omega_{B} \tau )^{-2}  &  ( \omega_{B} \tau )^{-1}  &
( \omega_{B} \tau )^{-1}  \cr
( \omega_{B} \tau )^{-1}  &  *  &  *  \cr
( \omega_{B} \tau )^{-1}  &  *  &  *
\end{array}  \right) \,\, ,   \\
\omega_{B} \tau \,\, \rightarrow \,\, \infty  \,\,\, ,
\end{multline}
where the direction of the $\, x$ - axis coincides with the mean 
direction of the open trajectories in the $\, {\bf p}$ - space 
(here and everywhere the notation $\, * \, $ means just some
constant of the order of $1$).

 The open trajectories of system (\ref{MFSyst}), considered in
\cite{lifpes1,lifpes2}, have more general form and are not 
periodic in the $\, {\bf p}$ - space. However, all the 
(quasiperiodic) trajectories, considered in \cite{lifpes1,lifpes2}, 
have a mean direction in the $\, {\bf p}$ - space and also give 
anisotropic contribution to the conductivity in strong magnetic 
fields. Generally speaking, the asymptotic behavior of conductivity
tensor can not be described here by a simple formula
(\ref{Periodic}), however, it also shows vanishing of the 
conductivity just along one direction in the plane, orthogonal
to $\, {\bf B}$, and a finite conductivity
in the orthogonal direction in the limit
$\, \omega_{B} \tau \, \rightarrow \infty \, $.
Many aspects, connected with the geometry of trajectories
of system (\ref{MFSyst}), were described in the papers
\cite{lifkag1,lifkag2,lifkag3,etm,KaganovPeschansky}.
Certainly, the references given here do not give a complete 
list of papers, which could be mentioned in connection with
this rather wide area.

\vspace{1mm}

 The general problem of classification of open trajectories
of system (\ref{MFSyst}) was set by S.P. Novikov 
(\cite{MultValAnMorseTheory}) and was intensively studied
in his topological school (S.P. Novikov, A.V. Zorich, 
S.P. Tsarev, I.A. Dynnikov). Due to rather deep mathematical
results (see \cite{zorich1,dynn1992,Tsarev,dynn1,dynn2}), 
obtained in the study of this problem from the topological 
point of view, the problem of description of possible types 
of open trajectories of system (\ref{MFSyst}) is now solved, 
so we can use in fact the full classification of trajectories 
of system (\ref{MFSyst}) in our study of transport 
phenomena in normal metals. 

\begin{figure}[t]
\begin{center}
\includegraphics[width=0.9\linewidth]{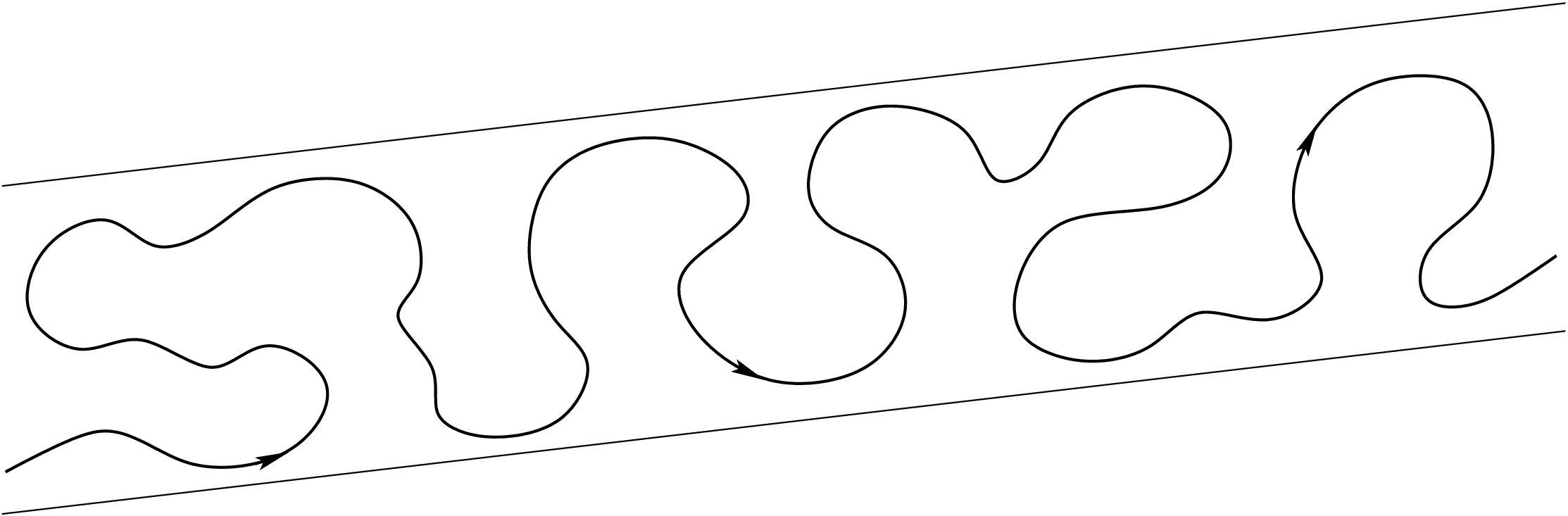}
\end{center}
\caption{The form of the stable open trajectories of system
(\ref{MFSyst}) in the planes orthogonal to $\, {\bf B} \, $
in the $\, {\bf p}$ - space.}
\label{StableTr}
\end{figure}

 The most important role in the classification of open 
trajectories  of system (\ref{MFSyst}) is played in fact by 
a description of the stable open trajectories of 
(\ref{MFSyst}), i.e. the open trajectories which do not change
essentially their global geometry under small rotations of
$\, {\bf B} \, $ or variations of the energy level 
$\, \epsilon ({\bf p}) \, = \, \epsilon_{0} \, $.
As follows from the results of \cite{zorich1,dynn1992,dynn1},
the stable open trajectories of system (\ref{MFSyst})
possess the following remarkable properties:

\vspace{2mm}

1) Every stable open trajectory of system (\ref{MFSyst})
in $\, {\bf p}$ - space lies in a straight line of a finite
width in the plane orthogonal to $\, {\bf B} \, $ and
passes through it from $\, - \infty \, $ to
$\, + \infty \, $ (Fig. \ref{StableTr});

\vspace{2mm}

2) The mean direction of all stable open trajectories in
$\, {\bf p}$ - space is given by the intersection of the 
plane orthogonal to $\, {\bf B} \, $ with some locally
stable integral plane $\, \Gamma \, $ in the 
$\, {\bf p}$ - space.

\vspace{2mm}

 Let us note here also that the property (1) was in fact  
first formulated by S.P. Novikov in \cite{MultValAnMorseTheory}
in the form of a conjecture and was later proved in the papers
\cite{zorich1,dynn1992,dynn1} for the open trajectories
having stability properties in the sense pointed above.

\vspace{1mm}

 The remarkable properties of the stable open trajectories
of system (\ref{MFSyst}) can be observed in direct measurements
of magneto-conductivity of normal metals in strong magnetic
fields. Thus, due to the geometric properties of the stable
open trajectories, their contribution to the conductivity 
in the plane orthogonal to $\, {\bf B} \, $ is strongly 
anisotropic in the limit $\, \omega_{B} \tau \, \gg \, 1 \, $, 
such that the direction of the lowest conductivity 
(in $\, {\bf x}$ - space) coincides with the mean direction 
of the trajectories in $\, {\bf p}$ - space. Due to the stability
of these trajectories with respect to small rotations of
$\, {\bf B} \, $, this situation takes place also for all
close directions of $\, {\bf B} \, $, which allows 
to observe also the integral plane $\, \Gamma \, $, defined
by the property (2), in the case of the presence of stable open 
trajectories on the Fermi surface. Let us note here that a plane 
is called integral in the $\, {\bf p}$ - space, if it is generated 
by two reciprocal lattice vectors 
$\, {\bf q}_{1}$, $\, {\bf q}_{2} \, $:

$$\begin{array}{c}
\Gamma \,\,\, = \,\,\, \{ \lambda \, {\bf q}_{1} \,\, + \,\,
\mu \, {\bf q}_{2} \, , \quad \lambda, \mu \, \in \, \mathbb{R} \}
\,\,\, ,   \\  \\
{\bf q}_{1} \,\, = \,\, n_{1} \, {\bf a}_{1} \, + \, 
n_{2} \, {\bf a}_{2} \, + \, n_{3} \, {\bf a}_{3} \,\,\, ,  \\  \\
{\bf q}_{2} \,\, = \,\, m_{1} \, {\bf a}_{1} \, + \, 
m_{2} \, {\bf a}_{2} \, + \, m_{3} \, {\bf a}_{3} \,\,\, ,   \\  \\
(n_{1}, n_{2}, n_{3}, m_{1}, m_{2}, m_{3} \, \in \, \mathbb{Z})
\end{array}  $$

 In the $\, {\bf x}$ - space the same plane
can be represented by an indivisible triple
of integer numbers $\, (M_{1}, M_{2}, M_{3}) \, $ 
with the aid of the equation
$$M_{1} \, \left( {\bf x} , {\bf l}_{1} \right) \,\, + \,\,
M_{2}   \, \left( {\bf x} , {\bf l}_{2} \right) \,\, + \,\,
M_{3}   \, \left( {\bf x} , {\bf l}_{3} \right) 
\,\,\, = \,\,\, 0 
\,\,\, , $$
where $\, ({\bf l}_{1}, {\bf l}_{2}, {\bf l}_{3}) \, $ represent
the basis of the direct lattice.

\vspace{1mm}

 The numbers $\, (M_{1}, M_{2}, M_{3}) \, $ were introduced
in \cite{PismaZhETF} as topological characteristics of electron
spectra in metals, which can be observed in the transport
phenomena in strong magnetic fields. According to
\cite{PismaZhETF}, these numbers can be called topological
quantum numbers observable in the conductivity of normal
metals. In accordance to the picture represented above,
we can in general distinguish special ``Stability Zones''
in the space of directions of $\, {\bf B} \, $
(the unit sphere $\, \mathbb{S}^{2}$), corresponding to
the presence of stable open trajectories on the Fermi surface,
which can be observed in the study of magneto-conductivity
in strong magnetic fields.

\vspace{1mm}

\begin{figure}[t]
\begin{center}
\includegraphics[width=0.9\linewidth]{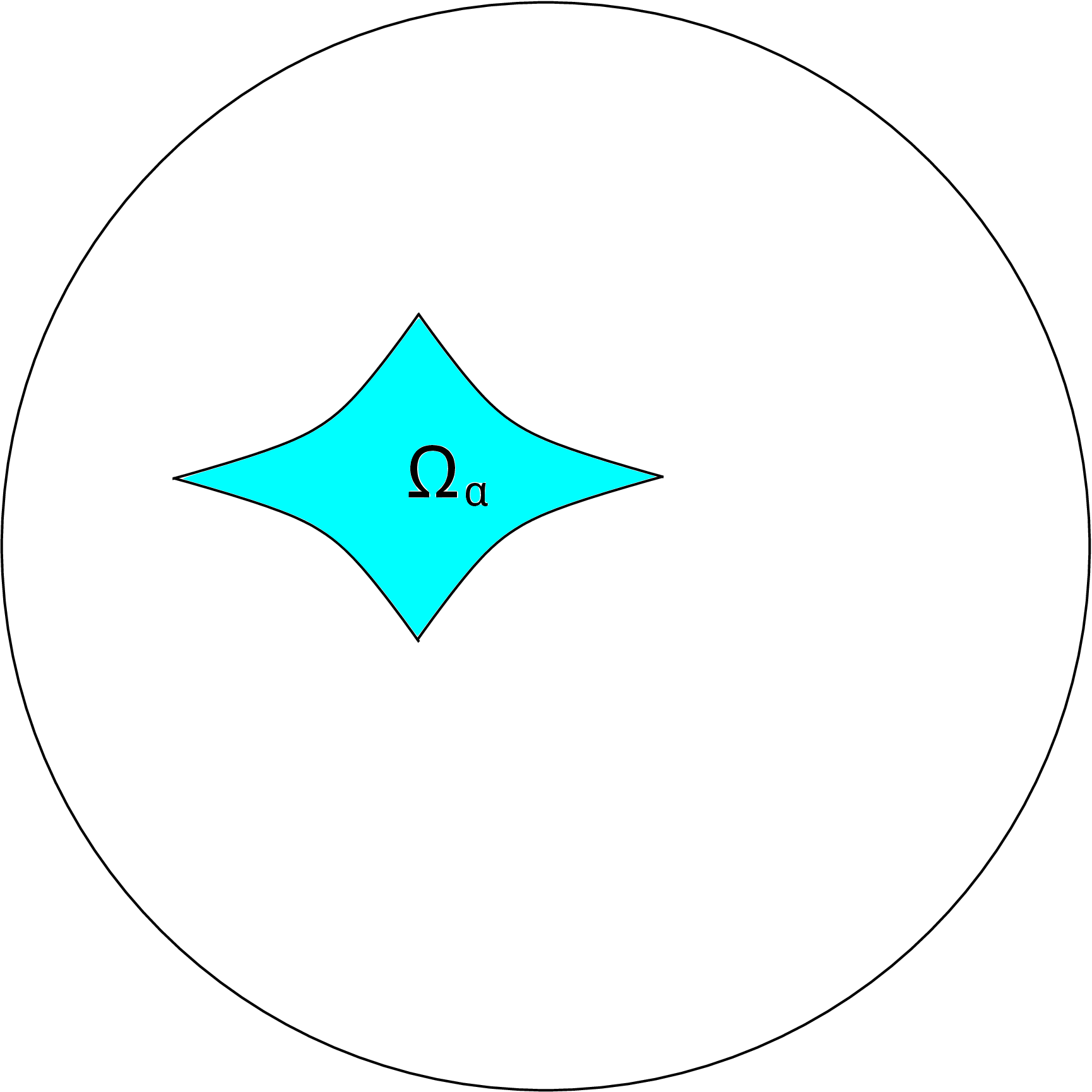}
\end{center}
\caption{An exact mathematical Stability Zone on the unit
sphere $\, \mathbb{S}^{2}$.}
\label{ExactMathZone}
\end{figure}

 Just by definition, we will call the Stability Zone 
$\, \Omega_{\alpha} \, \in \, \mathbb{S}^{2} \, $ a complete 
region in the space of directions of $\, {\bf B} \, $,
where we have stable open trajectories of system 
(\ref{MFSyst}) on the Fermi surface,
corresponding to the same topological quantum numbers 
$\, (M_{1}^{\alpha}, M_{2}^{\alpha}, M_{3}^{\alpha}) \, $.
The Stability Zones, defined in this way, can be also called
exact mathematical Stability Zones and represent some regions
with piecewise smooth boundaries on the unit sphere
(Fig. \ref{ExactMathZone}). It is not difficult to see 
that any Stability Zone is invariant under the transformation
$\, {\bf B} \, \rightarrow - {\bf B} \, $ and in most 
cases consists of two opposite regions on the unit sphere.

\vspace{1mm}

 Let us say here, that the geometry of the Stability Zones 
and the values of
$\, (M_{1}^{\alpha}, M_{2}^{\alpha}, M_{3}^{\alpha}) \, $
represent rather nontrivial characteristics of the 
electron spectra in metals, which are not defined 
just by the topology of the Fermi surface and depend
also on its geometrical parameters in rather nontrivial way.
Let us note also that the full sets of 
exact mathematical Stability Zones $\, \Omega_{\alpha} \, $ 
and the topological numbers 
$\, (M_{1}^{\alpha}, M_{2}^{\alpha}, M_{3}^{\alpha}) \, $
can be rather nontrivial for a metal having complicated
Fermi surface. We would like to give here a reference to the 
paper \cite{dynn3}, containing a description of many 
mathematical results related to the behavior of trajectories 
of system (\ref{MFSyst}) with an arbitrary dispersion relation 
and the geometry of the corresponding Stability Zones 
in the space of directions of $\, {\bf B} $. 
We would also like to specially mention here the paper 
\cite{DeLeoPhysB} that presents a rather convenient method 
for investigating the geometry of the Stability Zones for 
specific dispersion relations and also containing applications 
of this method to important analytical examples.
Let us also say that a large number of physical consequences 
arising from the results of mathematical study of the system 
(\ref{MFSyst}) were presented in the papers 
\cite{UFN,BullBrazMathSoc,JournStatPhys,ZhETF1997}.

\vspace{1mm}

 The numbers 
$\, (M_{1}^{\alpha}, M_{2}^{\alpha}, M_{3}^{\alpha}) \, $
describe the ``geometric properties'' of the tensor
$\, \sigma^{kl} ({\bf B}) \, $ for 
$\, {\bf B}/B \, \in \,  \Omega_{\alpha} \, $ 
and represent well observed experimental quantities in the 
limit $\, \omega_{B} \tau \, \gg \, 1 $. At the same time,
the analytic dependence of the tensor 
$\, \sigma^{kl} ({\bf B}) \, $ both on the value and 
the direction of $\, {\bf B} \, $ for 
$\, {\bf B}/B \, \in \, \Omega_{\alpha} \, $ can be in
fact rather nontrivial even under the implementation of
the condition $\, \omega_{B} \tau \, \gg \, 1 $
(\cite{JETP2017}). The reason for such behavior of the 
conductivity is actually the appearance of periodic 
open trajectories of system (\ref{MFSyst}) for special 
directions $\, {\bf B}/B \, \in \,  \Omega_{\alpha} \, $,
which introduce irregularity in the limiting values of
$\, \sigma^{kl} ({\bf B}) \, $ when
$\, B \rightarrow \infty \, $. As a result, the angular 
dependence of the tensor $\, \sigma^{kl} ({\bf B}) \, $
is actually rather irregular for 
$\, {\bf B}/B \, \in \,  \Omega_{\alpha} \, $ even for big 
enough values of the parameter $\, \omega_{B} \tau \, $.
Let us say, that the appearance of periodic open
trajectories of system (\ref{MFSyst}) for irrational
directions of $\, {\bf B} \, $ represents in general
a nontrivial fact, closely connected with arising
of a special integral plane $\, \Gamma_{\alpha} \, $
for every Stability Zone. Thus, the periodic open 
trajectories of system (\ref{MFSyst}) arise on the Fermi 
surface at $\, {\bf B}/B \, \in \,  \Omega_{\alpha} \, $
whenever the intersection of the plane orthogonal
to $\, {\bf B} \, $ and the plane $\, \Gamma_{\alpha} \, $
has a rational direction. Its not difficult to see, that
the corresponding directions of $\, {\bf B} \, $ form
in general an everywhere dense set in the Zone 
$\, \Omega_{\alpha} \, $.

\begin{figure}[t]
\begin{center}
\includegraphics[width=0.9\linewidth]{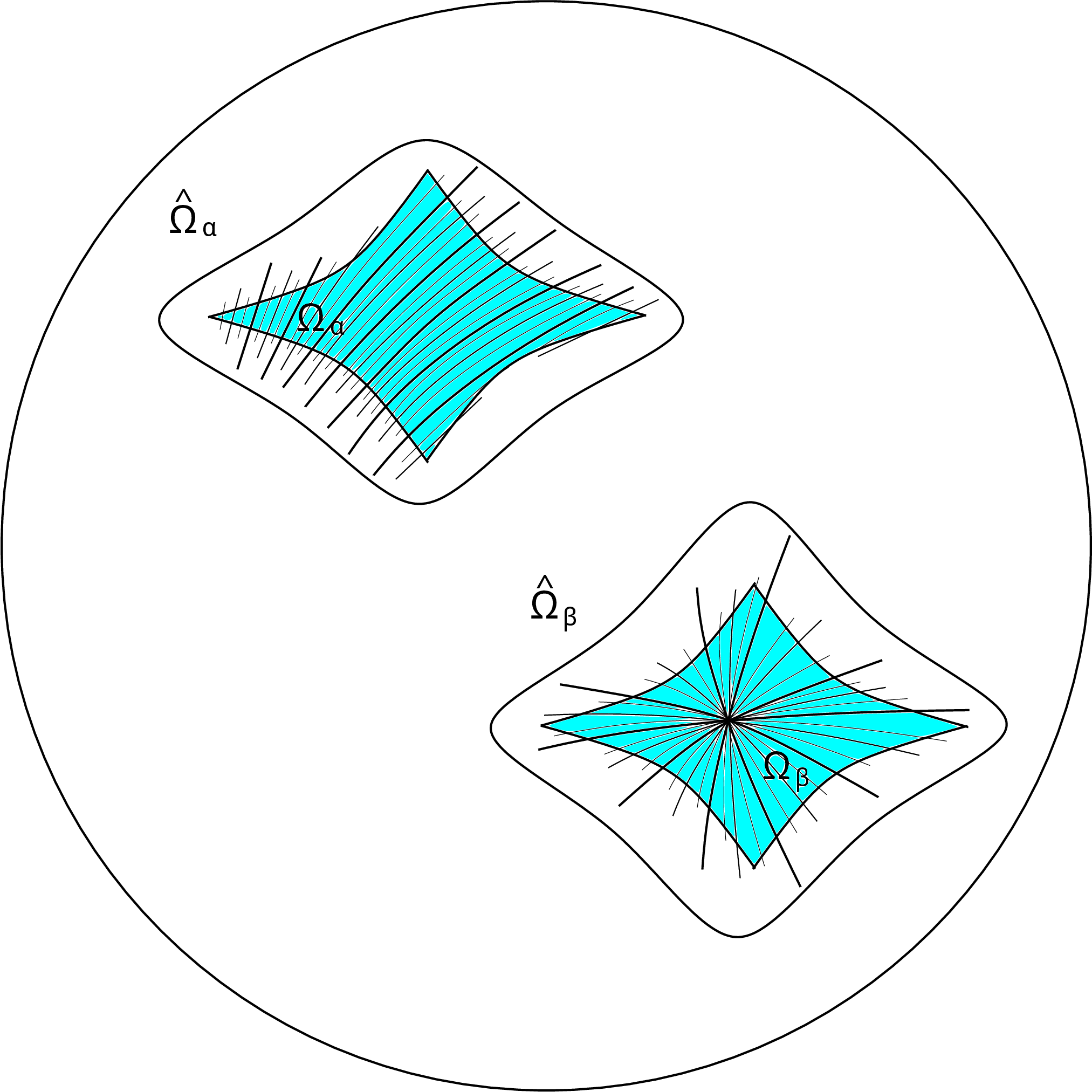}
\end{center}
\caption{Mathematical Stability Zones, special directions of 
$\, {\bf B} \, $, corresponding to appearance of periodic
open trajectories of system (\ref{MFSyst}) on the Fermi
surface, and ``experimentally observable'' Stability Zones
on the angular diagram.}
\label{ExpZone}
\end{figure}

 Another important feature of the periodic open 
trajectories of system (\ref{MFSyst}), following from
the topological structure of the system (\ref{MFSyst})
for $\, {\bf B}/B \, \in \,  \Omega_{\alpha} \, $,
is that they actually exist also on some special sets 
outside the Stability Zone near its boundary (let us note 
also here that this phenomenon can appear also for some 
special examples of simple Fermi surfaces, see e.g. 
\cite{GurzhyKop}). The corresponding directions of 
$\, {\bf B} \, $ are not included in the Stability Zones 
since the periodic trajectories are no longer stable 
outside the Zones $\, \Omega_{\alpha} \, $. In general, 
we can have either periodic open trajectories or very 
long closed trajectories on the Fermi surface for directions 
of $\, {\bf B} \, $, close to the boundary of a Stability
Zone $\, \Omega_{\alpha} \, $. As a result, the exact boundary
of a Stability Zone $\, \Omega_{\alpha} \, $ is usually
unobservable in direct conductivity measurements even
under the condition $\, \omega_{B} \tau \, \gg \, 1 \, $.
As was pointed out in \cite{JETP2017}, we have to introduce 
actually extended ``experimentally observable'' Stability
Zones $\, \hat{\Omega}_{\alpha} \, $ on the angular diagram,
which arise as observable Stability Zones in the direct
measurements of conductivity (Fig. \ref{ExpZone}).

 At the same time, the exact boundary of a Stability Zone
represents an important characteristic of the electron
spectrum in metal being closely connected with the geometric
parameters of the Fermi surface. As was pointed out in 
\cite{OscPhen}, the boundary of an exact mathematical
Stability Zone $\, \Omega_{\alpha} \, $ can be actually
determined in the study of oscillation phenomena in metals
like the cyclotron resonance phenomenon or different
types of quantum oscillations. As was shown in \cite{OscPhen},
the picture of (classical or quantum) oscillations undergoes
sharp changes at the boundary of exact mathematical
Stability Zone, which can be clearly registered in
the dependence of the oscillations on the value of $\, B \, $.
More precisely, the change in the dependence of the 
oscillating values on the value of $\, 1/B \, $ is given
by abrupt disappearance of one of the main oscillating
terms (one oscillating harmonic) after crossing the 
boundary of $\, \Omega_{\alpha} \, $ on the angular diagram.

\vspace{1mm}

 In the present paper we will consider some general aspects
of the angular diagram of conductivity and the structure
of the Stability Zones on $\, \mathbb{S}^{2} \, $ for 
metals having (arbitrary) complicated Fermi surfaces. 
In particular, we are going to show here that the 
Stability Zones have in fact ``the second boundaries'' on the 
angular diagram in the space of directions of $\, {\bf B} \, $,
which can be also detected in the same study of oscillation
phenomena in strong magnetic fields. The region 
$\, \Sigma_{\alpha} \, $, restricted by the second boundary
of a Stability Zone $\, \Omega_{\alpha} \, $, always contains
the Zone $\, \Omega_{\alpha} \, $ and can have different 
sizes depending on the geometry of the Fermi surface.
Thus, we can have both the situations when the second 
boundary is close to the (first) boundary of a Stability Zone 
or it is located rather far from the boundary of 
$\, \Omega_{\alpha} \, $. In particular, the region
$\, \Sigma_{\alpha} \, $ can contain either the whole
``experimentally observable Stability Zone''
$\, \hat{\Omega}_{\alpha} \, $ in a given experimental
study of conductivity or just a part of the Zone
$\, \hat{\Omega}_{\alpha} \, $.

 One of the essential properties of the domain between the 
first and the second boundary of a Stability Zone is that 
it can not contain directions of $\, {\bf B} \, $ 
corresponding to the presence of any stable open 
trajectories on the Fermi surface (at the same time, 
the unstable periodic trajectories can arise here for special 
directions of $\, {\bf B} \, $ mentioned above). 
In the present paper we consider different
phenomena arising in the domain 
$\, \Sigma_{\alpha} \backslash \Omega_{\alpha} \, $
at different directions of magnetic field. As will be shown,
the picture of (classical and quantum) oscillation phenomena
has in this domain rather specific features which can in fact
be used for the investigation of the geometry of the Fermi
surface. At the same time, the direct conductivity
measurements (at $\, \Omega = 0$) also have here special
features which will be discussed in the next sections.
We will also consider here the most general features of 
the location of the Stability Zones and the corresponding 
additional areas 
$\, \Sigma_{\alpha} \backslash \Omega_{\alpha} \, $
on the angular diagram.

\vspace{1mm}

 Another aspect which we will touch in the present
paper is the distribution of the Stability Zones on a general
angular diagram. As we will show, it is natural to divide
actually all the angular diagrams into ``simpler'' (Type A)
and ``more complicated'' (Type B) diagrams in the general case.
Both Types (A and B) represent completely general classes
that cover the whole set of angular diagrams and are equally
possible from theoretical point of view. At the same time,
it seems that for most of the real materials we should 
actually expect the appearance of Type A diagrams with a 
much higher probability than diagrams of the Type B.
In most cases Type A angular diagrams are characterized by 
the presence of only a finite number of Stability Zones
on $\, \mathbb{S}^{2} \, $ while the angular diagrams of 
Type B generically contain an infinite number of different
Stability Zones on the unit sphere. Another important
feature of the angular diagrams of Type B is that we should
also expect here special directions of $\, {\bf B} \, $
corresponding to appearance of more complicated (chaotic)
open trajectories on the Fermi surface, which leads 
to the appearance of very special asymptotic regimes of 
conductivity behavior in strong magnetic fields. As a 
consequence, the appearance of a Type B angular diagram 
gives a much greater variety of different types of 
conductivity behavior, observed experimentally. As we will
also show here, the angular diagrams of different types 
can be conveniently distinguished by a study of the Hall 
conductivity outside the Stability Zones at
$\, \omega_{B} \tau \, \rightarrow \, \infty \, $.

\vspace{1mm}

 In sections II and III we will represent a detailed 
consideration of the structure of system (\ref{MFSyst}) for 
$\, {\bf B}/B \, \in \, 
\Sigma_{\alpha} \backslash \Omega_{\alpha} \, $ 
and consider phenomena connected with special
features of its trajectories in this domain. Our considerations
here will be based on topological results, obtained in recent
topological investigations of the S.P. Novikov problem
(\cite{zorich1,dynn1992,dynn1,dynn2,dynn3}) and describing 
the topology of ``carriers of open trajectories'' in the 
case of the presence of stable open trajectories on the
Fermi surface. As we will see, the topological description
of the structure of system (\ref{MFSyst}) in the Stability
Zones gives in fact also a good possibility to describe
the properties of its trajectories in the domain
$\, \Sigma_{\alpha} \backslash \Omega_{\alpha} \, $.
As we have said already, the main considerations will be
connected with the oscillation phenomena in this domain,
having classical or quantum nature.

 In section IV we discuss different Types of the angular
diagrams and their connection with types of the boundaries
of Stability Zones and the Hall conductivity behavior
outside the Stability Zones. 

 In section V we discuss a connection between the angular
diagrams for conductivity of metals and the angular 
diagrams for a full dispersion relation 
$\, \epsilon ({\bf p}) \, $, which describe the behavior
of trajectories of system (\ref{MFSyst}) at all 
energy levels $\, \epsilon \, $. Let us note here that 
although the introduction of an angular diagram for the 
full dispersion relation has a somewhat abstract character 
in the theory of normal metals, such a comparison is, 
as it seems to us, quite useful from a methodological 
point of view.

\vspace{1cm}

\section{The structure of the Fermi surface and the 
geometry of trajectories of system (\ref{MFSyst})
for $\, {\bf B}/B \, \in \, 
\Sigma_{\alpha} \backslash \Omega_{\alpha} \, $.}
\setcounter{equation}{0}

\begin{figure}[t]
\begin{center}
\vspace{5mm}
\includegraphics[width=\linewidth]{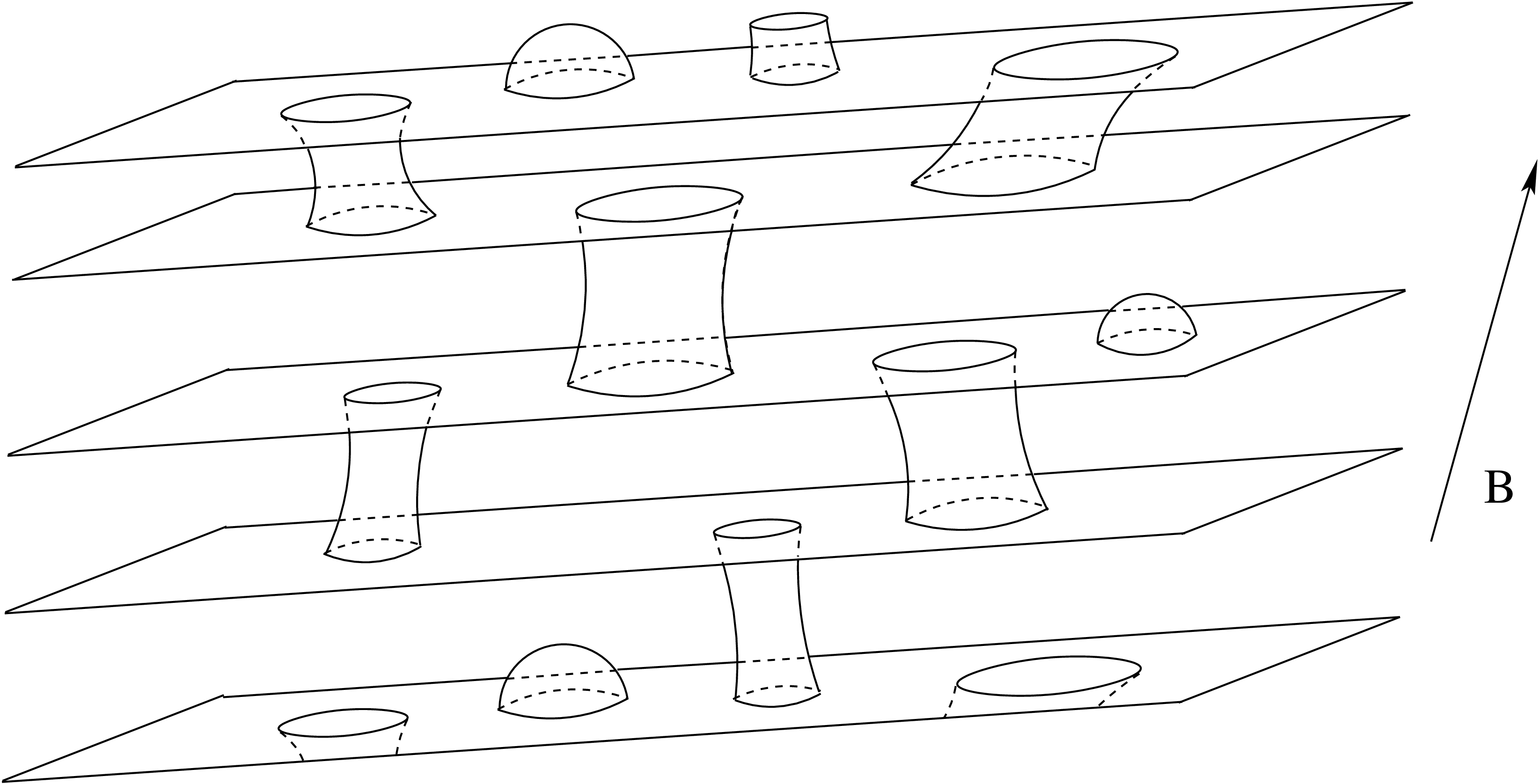}
\end{center}
\caption{The structure of a connected component of the Fermi 
surface, carrying stable open trajectories of system
(\ref{MFSyst}).}
\label{ComplFermiSurf}
\end{figure}

 Let us start with the description of the structure of the
Fermi surface for 
$\, {\bf B}/B \, \in \,  \Omega_{\alpha} \, $,
which will also give a description of its structure in the
domain $\, \Sigma_{\alpha} \backslash \Omega_{\alpha} \, $.
As follows from rather deep topological theorems proved in
\cite{zorich1,dynn1992,dynn1,dynn2,dynn3}, in the presence
of stable open trajectories on the level 
$\, \epsilon ({\bf p}) \, = \, \epsilon_{F} \, $
the structure of the corresponding connected component of 
the Fermi surface can be described in the following way:

\vspace{2mm}

 Let us imagine a set of parallel integral planes in the 
$\, {\bf p}$ - space, connected by cylinders of finite
heights (Fig. \ref{ComplFermiSurf}).

\begin{figure}[t]
\begin{center}
\vspace{5mm}
\includegraphics[width=\linewidth]{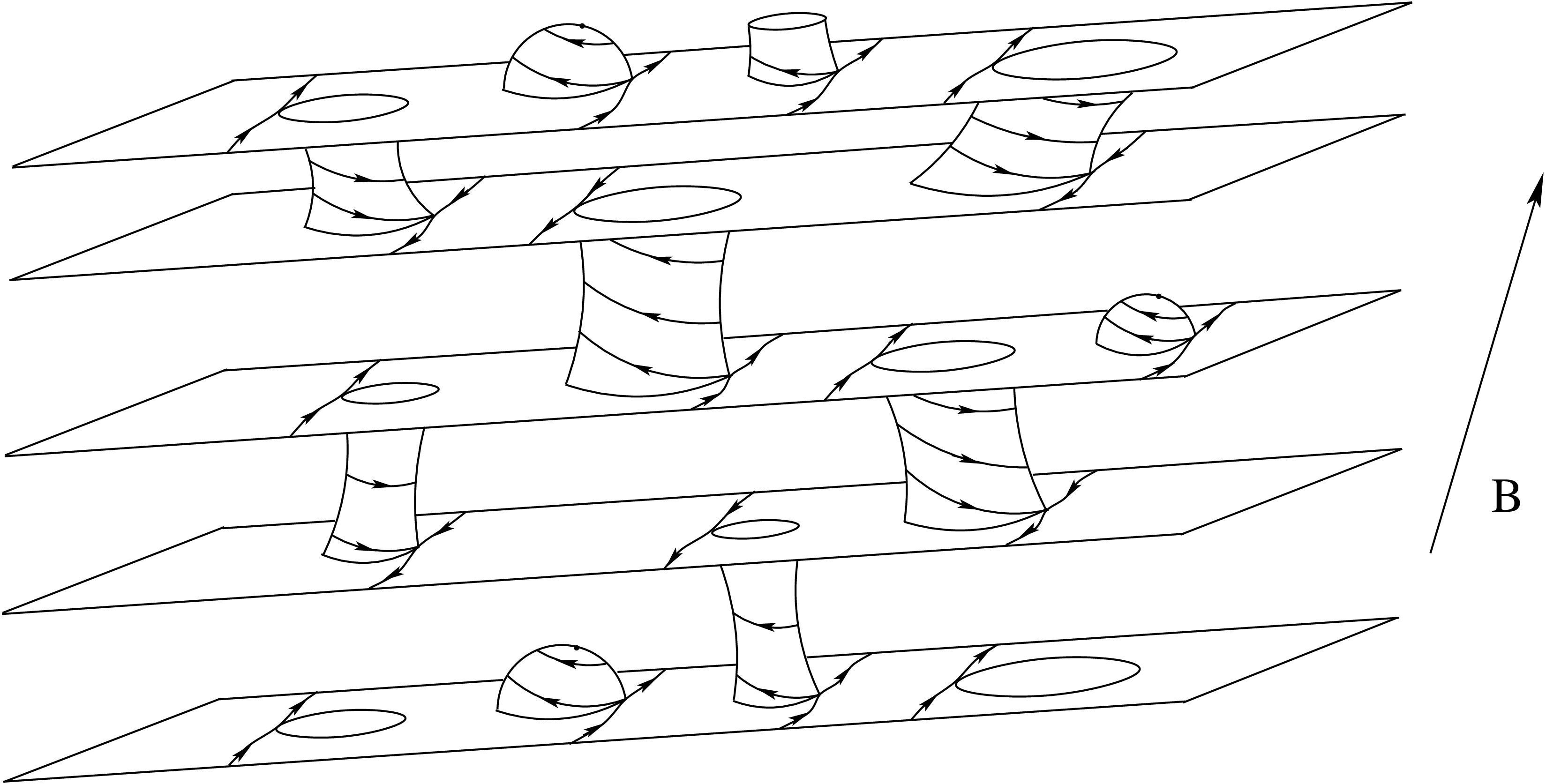}
\end{center}
\caption{Open and closed trajectories of system (\ref{MFSyst})
on the Fermi surface for 
$\, {\bf B}/B \, \in \,  \Omega_{\alpha} \, $.}
\label{TrOnComplFermiSurf}
\end{figure}

It is not difficult to see, that if we choose the direction 
of $\, {\bf B} \, $ close to the directions of the axes of
the cylinders we will have a cylinder of closed trajectories
on every cylinder, shown at Fig. \ref{ComplFermiSurf}, 
restricted by singular closed trajectories on the Fermi surface
(Fig. \ref{TrOnComplFermiSurf}). Let us note, that we
admit here also the appearance of cylinders of closed 
trajectories having a point as a base. The cylinders of the 
last type can be called trivial from the topological point
of view. Theoretically, it is also possible to admit a compound 
structure of cylinders connecting planes. This, however, does not 
affect further conclusions and is extremely unlikely for real 
Fermi surfaces, so we will not consider this complication here.

\begin{figure}[t]
\begin{center}
\vspace{5mm}
\includegraphics[width=0.9\linewidth]{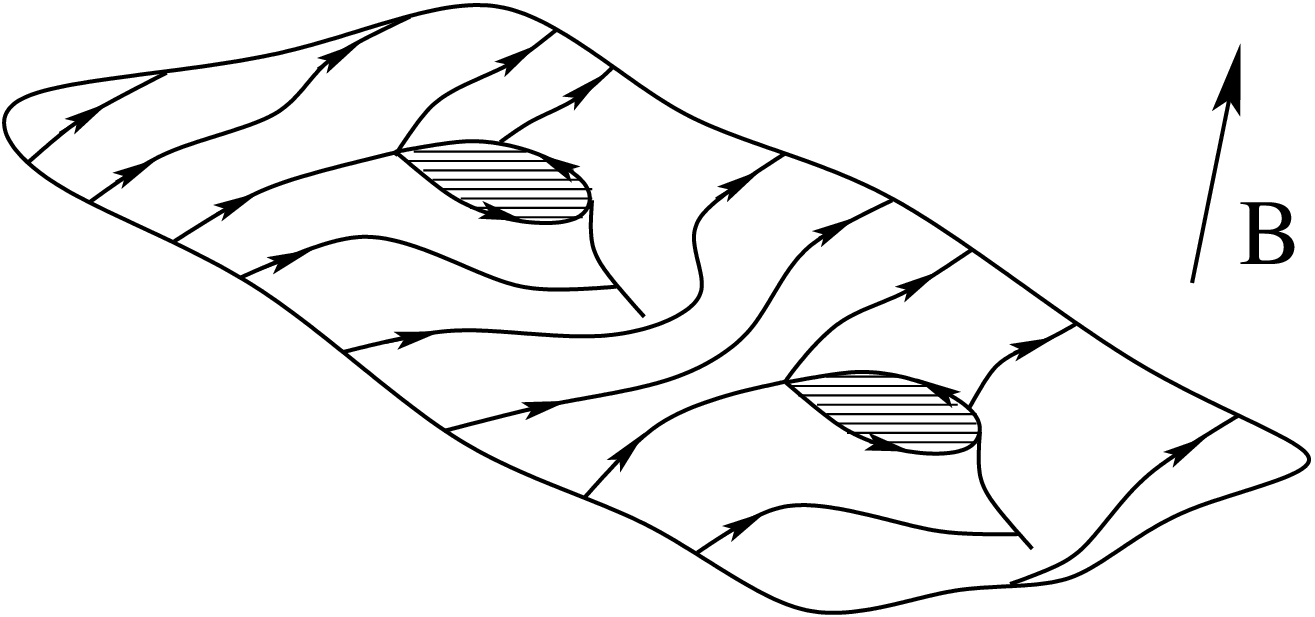}
\end{center}
\caption{A carrier of the stable open trajectories on
the Fermi surface for
$\, {\bf B}/B \, \in \,  \Omega_{\alpha} \, $.}
\label{DeformedPlane}
\end{figure}

 We can see then, that the integral planes at 
Fig. \ref{TrOnComplFermiSurf} are actually separated from
each other by the cylinders of closed trajectories and
represent ``carriers of open trajectories'' of
system (\ref{MFSyst}) on the Fermi surface. Thus, every
carrier of open trajectories represents a periodically
deformed integral plane (with holes) in the 
$\, {\bf p}$ - space, which is locally stable under small 
rotations of $\, {\bf B} \, $ (Fig. \ref{DeformedPlane}). 
It is easy to see here, that every open trajectory at 
Fig. \ref{TrOnComplFermiSurf} has a mean direction in
the $\, {\bf p}$ - space, given by the intersection of the 
plane, orthogonal to $\, {\bf B} \, $, and the integral 
plane, corresponding to the carriers of open trajectories.

 Let us say here, that Fig. \ref{ComplFermiSurf}
and \ref{TrOnComplFermiSurf} represent just a topological 
structure of the Fermi surface carrying stable open
trajectories and can be visually much more complicated.
The representation of the Fermi surface shown at 
Fig. \ref{ComplFermiSurf}, \ref{TrOnComplFermiSurf}
is not uniquely defined, thus, we get different 
representations (of the same Fermi surface) in the form
Fig. \ref{ComplFermiSurf}, \ref{TrOnComplFermiSurf}
for directions of $\, {\bf B} \, $ lying in two different 
Stability Zones
$\, \Omega_{\alpha} \, $, $\, \Omega_{\beta} \, $.
At the same time, the existence of the structure, shown
at Fig. \ref{ComplFermiSurf}, \ref{TrOnComplFermiSurf},
on any Fermi surface carrying stable open trajectories 
of system (\ref{MFSyst}) represents a rigorous topological
theorem, valid in the most general case 
(see \cite{zorich1,dynn1992,dynn1,dynn2,dynn3}).

 It is easy to see that the integral planes at 
Fig. \ref{TrOnComplFermiSurf} can be actually divided 
into two types according to the direction of the electron
motion (back or forward) on the corresponding open
trajectories. In the same way, all the cylinders of
closed trajectories can be divided into two different
types according to the type of the closed trajectories
(electron type or hole type) on them. We will assume
here that we have just two nonequivalent integral planes
(having different types) in the representation of the 
Fermi surface according to 
Fig. \ref{ComplFermiSurf}, \ref{TrOnComplFermiSurf}. 
As can be shown, the Fermi surfaces, having bigger
number of nonequivalent integral planes in the 
$\, {\bf p}$ - space in their representation 
above, should have in fact very high genus, so this
situation should be actually quite rare for real metals.
At the same time, we will assume here that we have
at least two or more nonequivalent cylinders of closed
trajectories in the representation of the Fermi surface
according to 
Fig. \ref{ComplFermiSurf}, \ref{TrOnComplFermiSurf}
(where the number of topologically nontrivial cylinders 
is actually equal to $\, g - 1 \, $).

\begin{figure}[t]
\begin{center}
\vspace{5mm}
\includegraphics[width=0.9\linewidth]{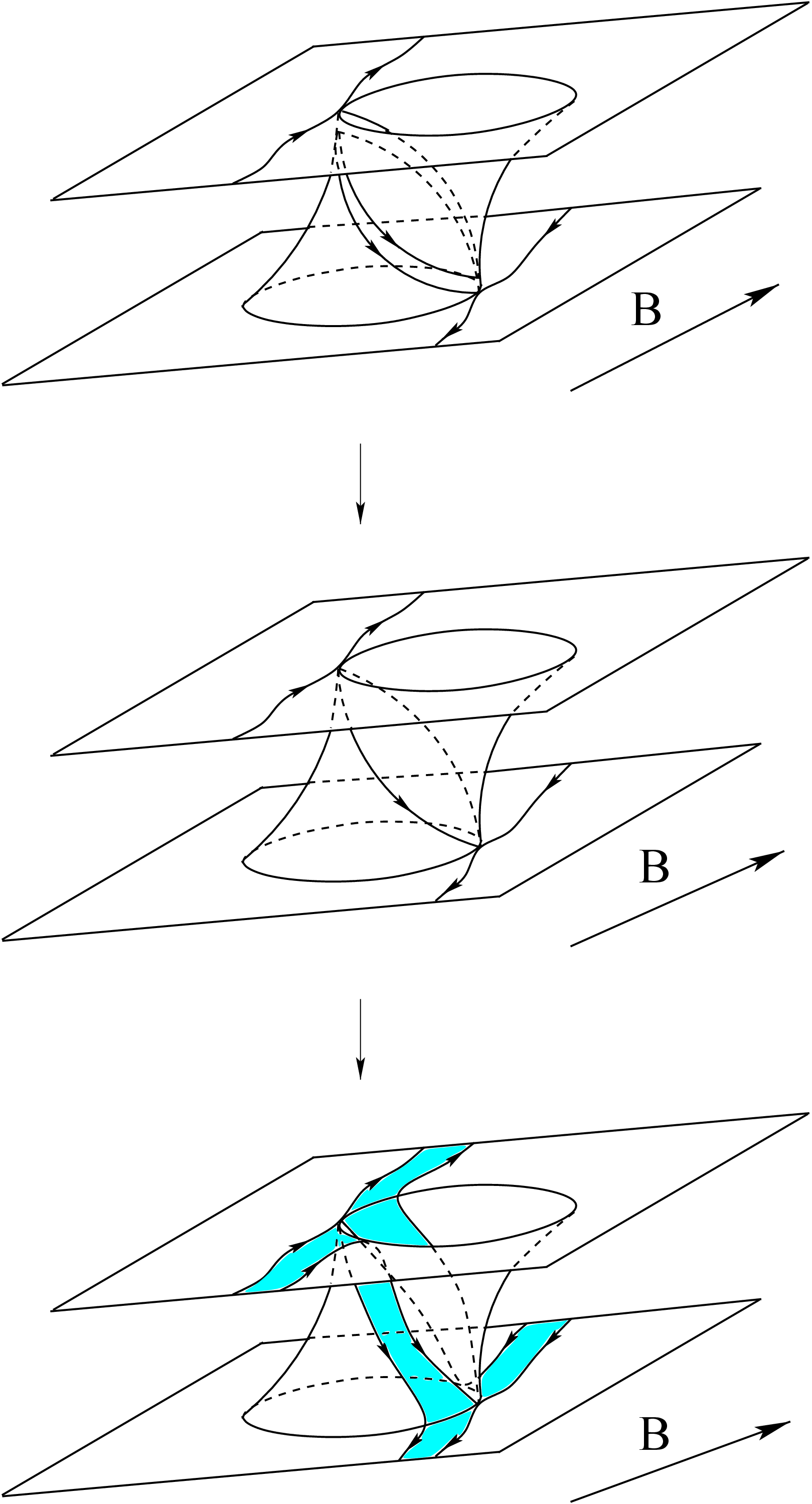}
\end{center}
\caption{The disappearance of a cylinder of closed 
trajectories on the Fermi surface after crossing the
boundary of a Stability Zone on the angular diagram.}
\label{Cylind}
\end{figure}

 The exact boundary of a Stability Zone 
$\, \Omega_{\alpha} \, $ is defined by disappearance 
of one of the cylinders from the set of nonequivalent 
nontrivial cylinders of closed trajectories on the Fermi 
surface. Thus, for directions of $\, {\bf B} \, $,
close to the boundary of a Stability Zone, the height of 
the corresponding cylinder of closed trajectories become 
very small and vanishes at the boundary of 
$\, \Omega_{\alpha} \, $. After crossing the boundary of 
$\, \Omega_{\alpha} \, $ the corresponding cylinder does 
not separate a pair of integral planes anymore, so 
the trajectories can now jump from one plane to another
(Fig. \ref{Cylind}).

\begin{figure}[t]
\vspace{5mm}
\begin{tabular}{lc}
\includegraphics[width=0.6\linewidth]{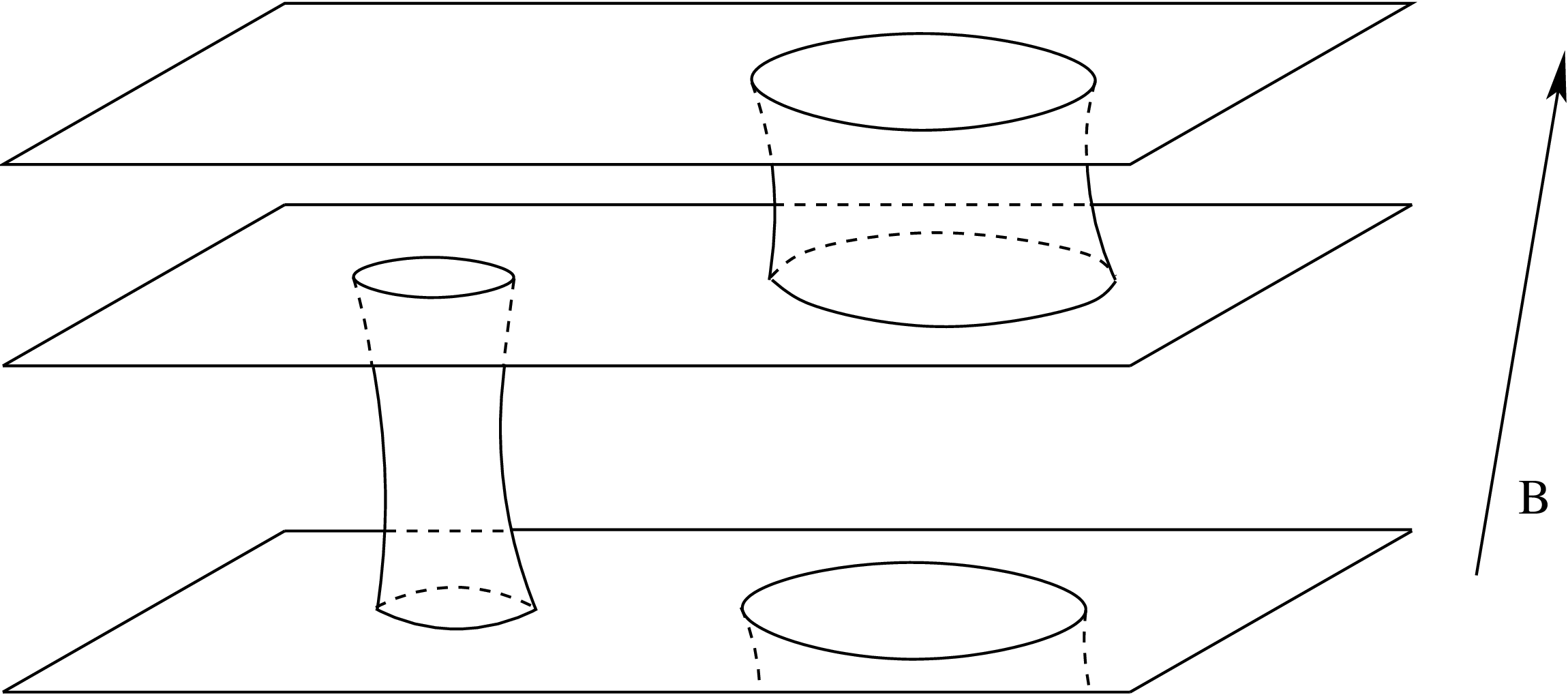}  &
\hspace{5mm}
\includegraphics[width=0.3\linewidth]{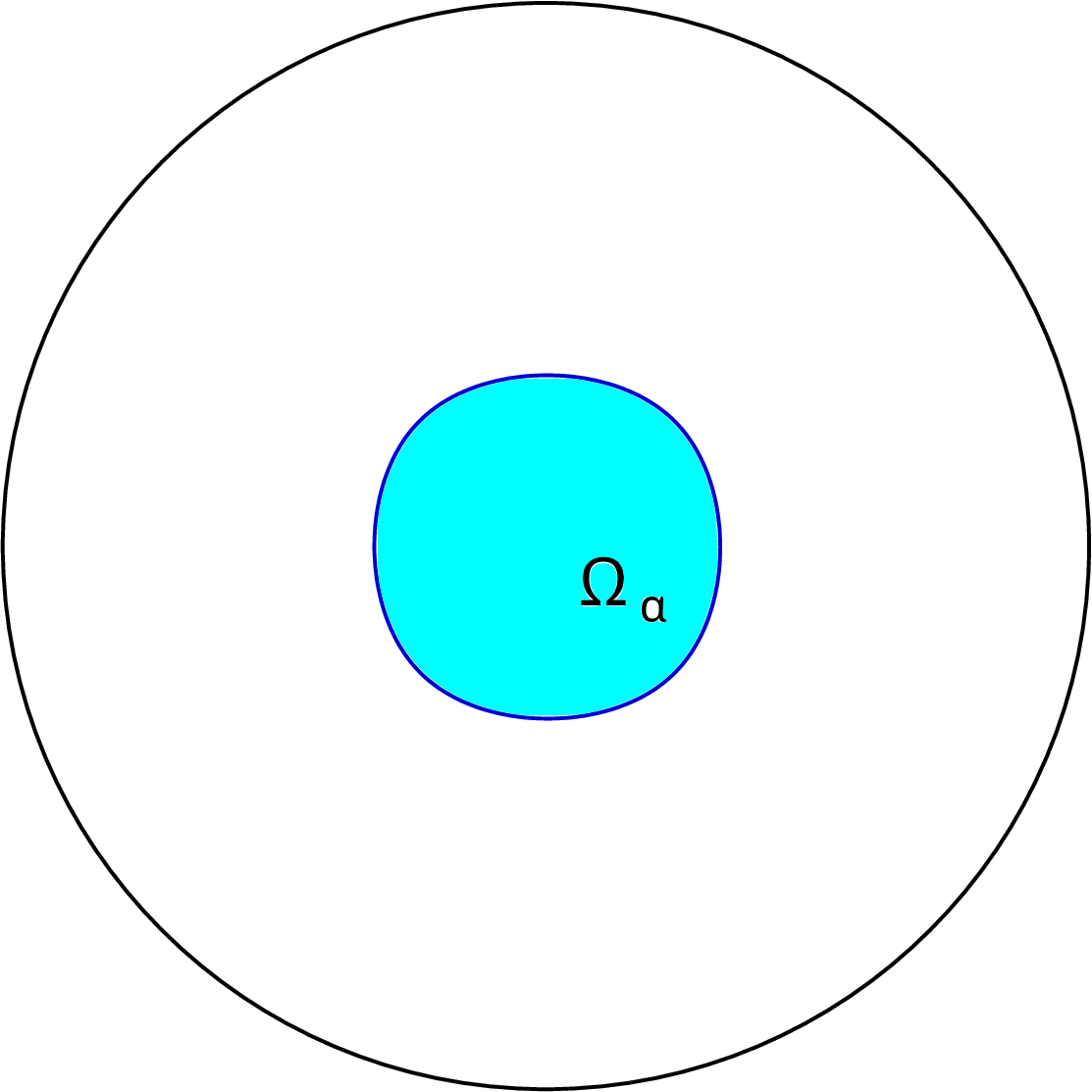}
\end{tabular}
\caption{An example of the Fermi surface having a Stability 
Zone $\, \Omega_{\alpha} \, $ with a simple boundary.}
\label{SimpleFirstBound}
\end{figure}

\begin{figure}[t]
\vspace{5mm}
\begin{tabular}{lc}
\includegraphics[width=0.6\linewidth]{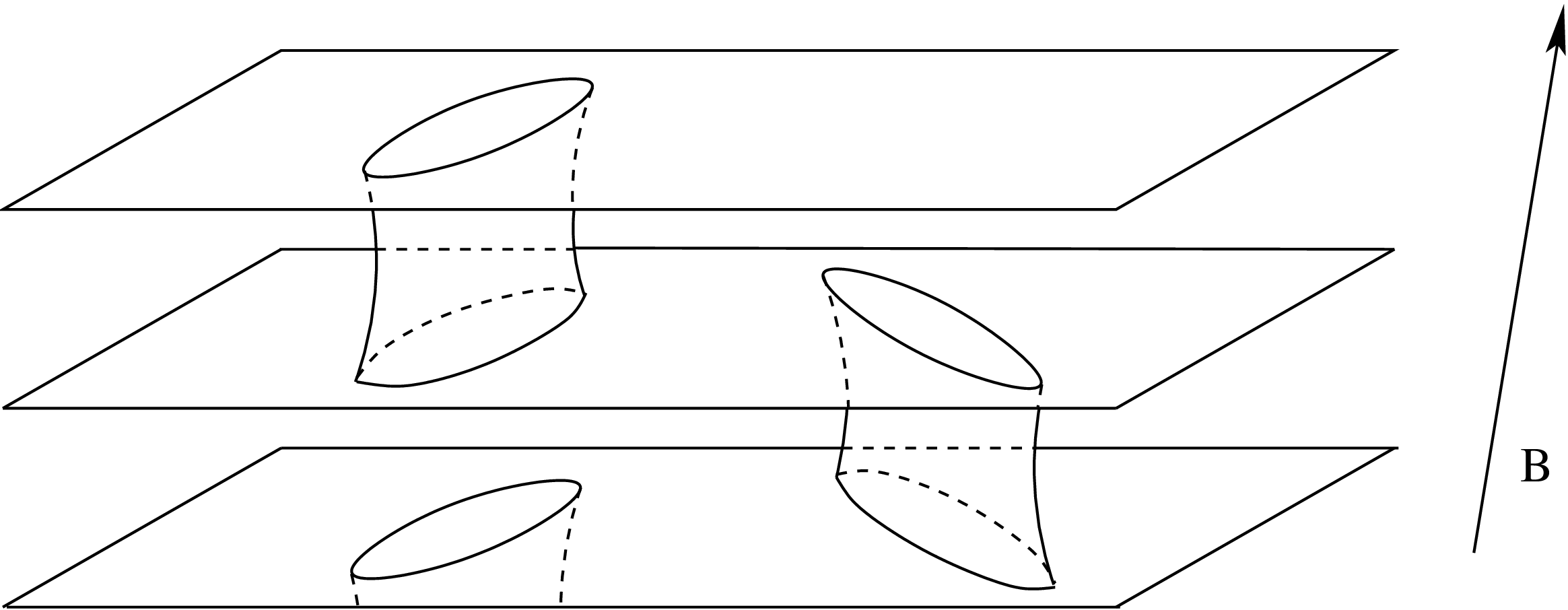}  &
\hspace{5mm}
\includegraphics[width=0.3\linewidth]{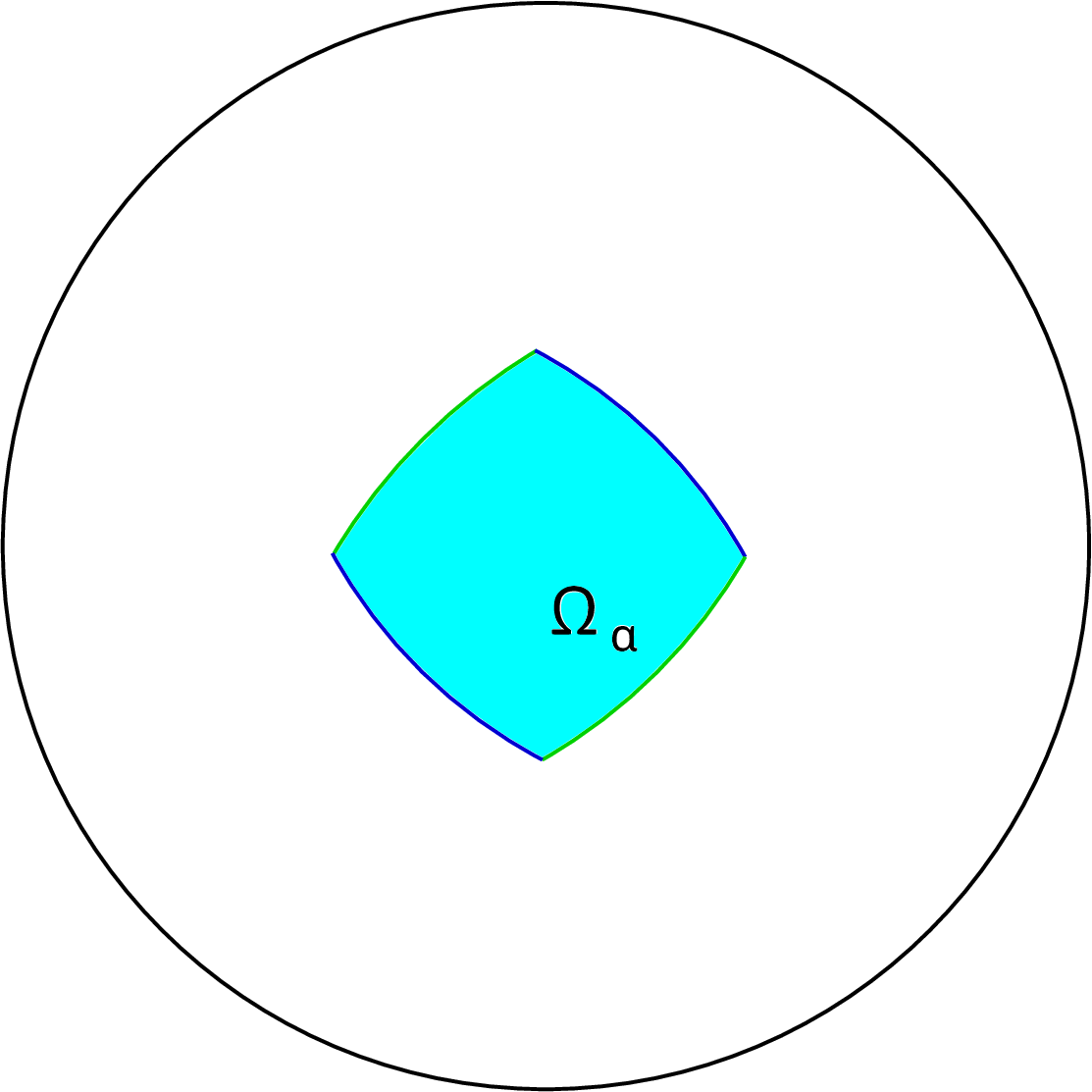}
\end{tabular}
\caption{An example of the Fermi surface having a Stability 
Zone $\, \Omega_{\alpha} \, $ with a compound boundary.}
\label{CompoundFirstBound}
\end{figure}

 Let us note here, that in general we can have both the 
situation when the whole boundary of a Stability Zone
is defined by disappearance of the same cylinder 
of closed trajectories and the situation when different
parts of the boundary of $\, \Omega_{\alpha} \, $
correspond to disappearance of different cylinders
from the set. Let us say here that in the first case
the Stability Zone has a simple boundary 
(Fig. \ref{SimpleFirstBound}) and in the 
second case the Stability Zone has a compound boundary
(Fig. \ref{CompoundFirstBound}). In general, our 
considerations here will be applicable to both of the 
above cases.

\begin{figure}[t]
\vspace{5mm}
\begin{tabular}{lc}
\includegraphics[width=0.45\linewidth]{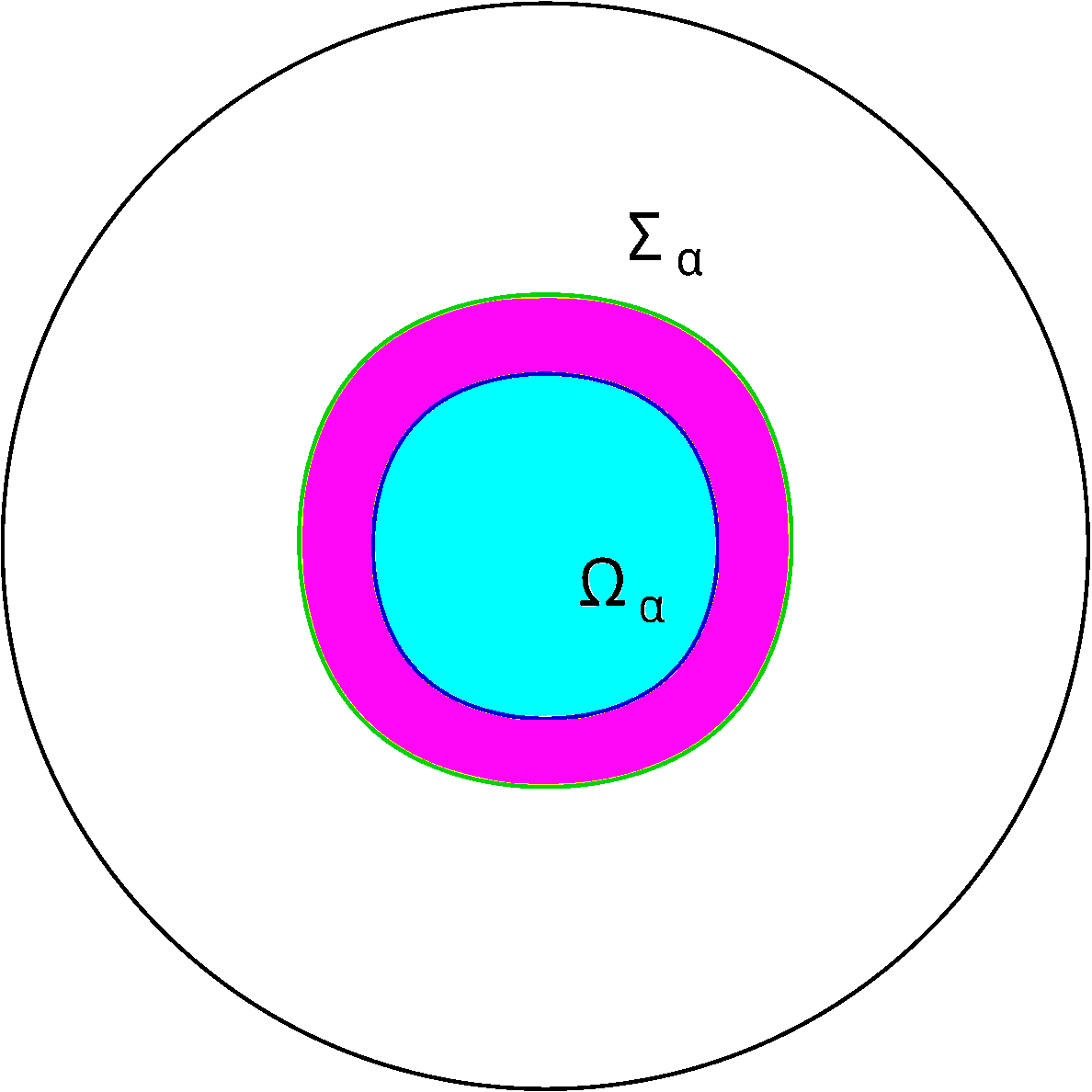}  &
\hspace{5mm}
\includegraphics[width=0.45\linewidth]{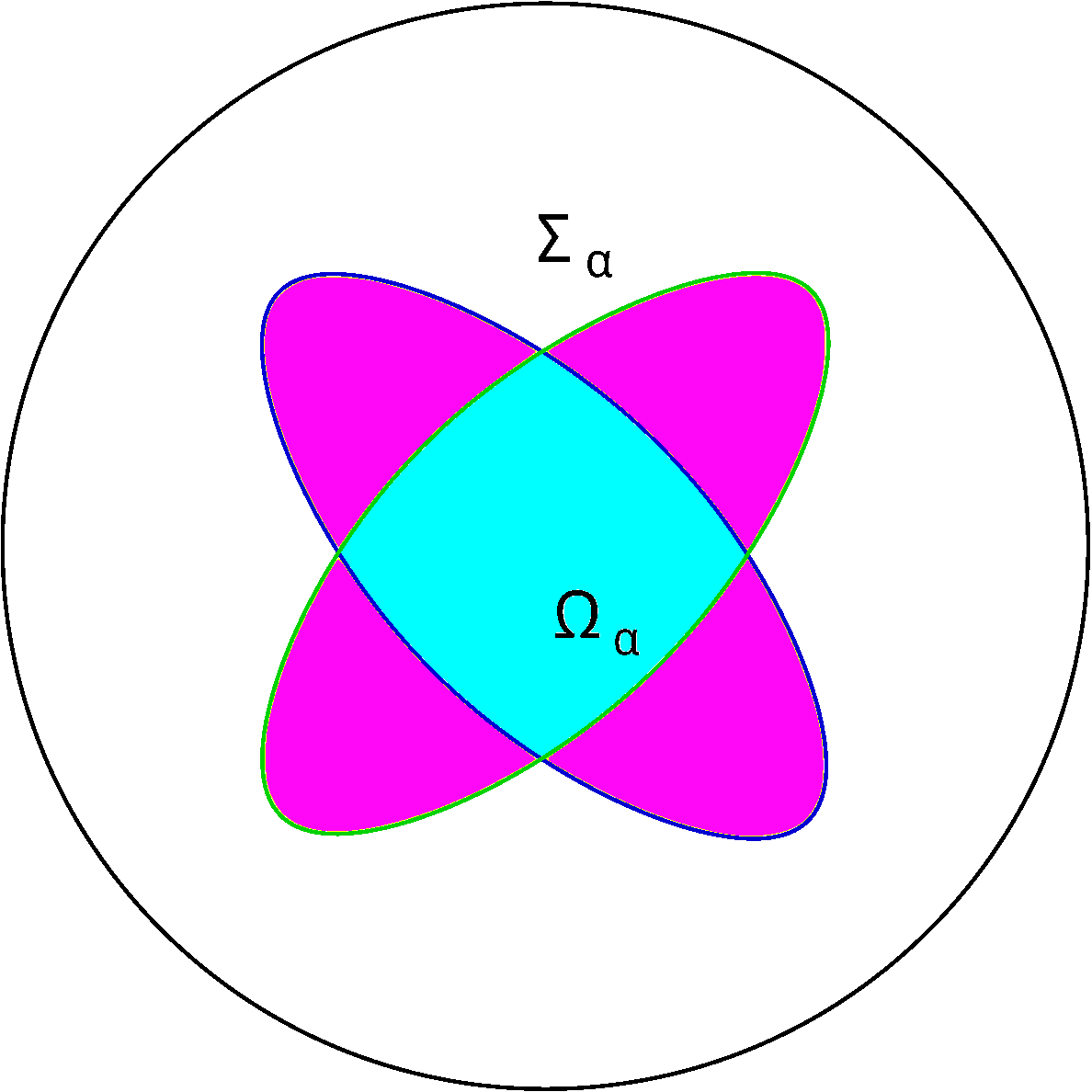}
\end{tabular}
\caption{Stability Zones having simple and compound second 
boundary on the angular diagram.}
\label{Boundaries}
\end{figure}

\begin{figure*}
\begin{center}
\includegraphics[width=175mm]{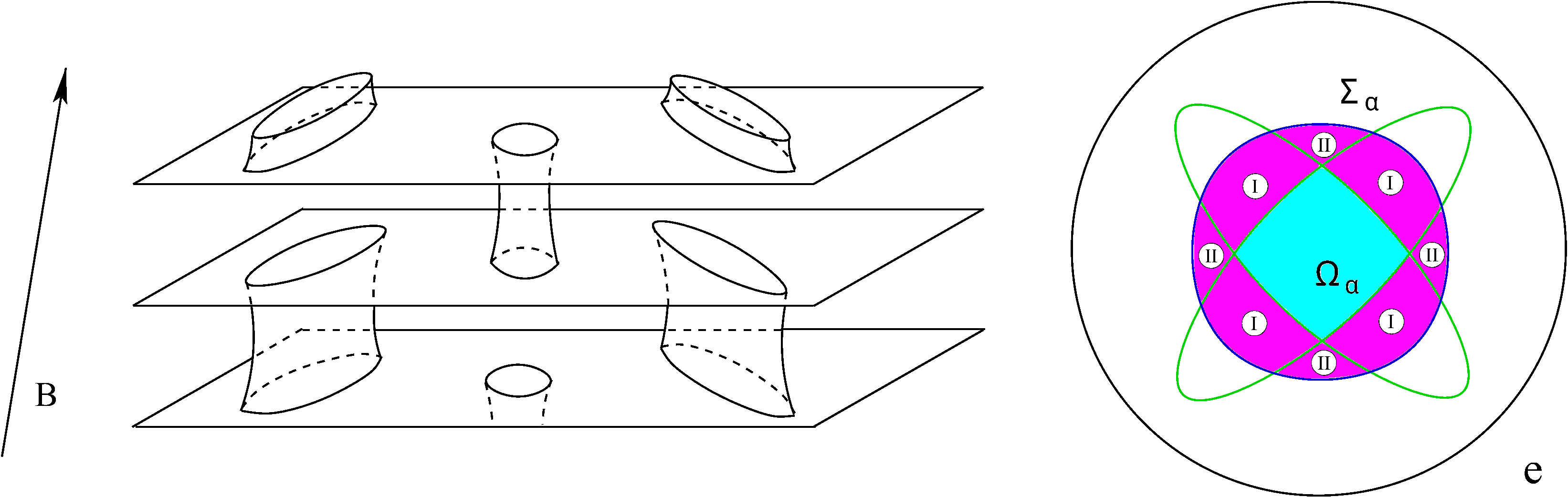}
\end{center}
\begin{center}
\includegraphics[width=175mm]{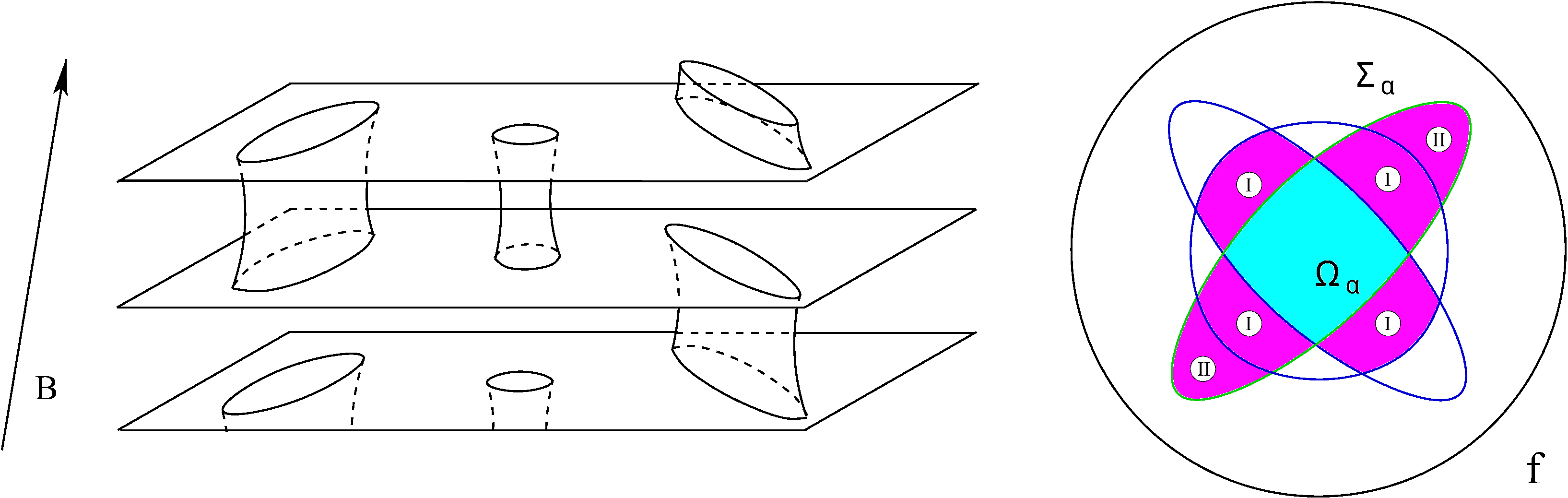}
\end{center}
\begin{center}
\includegraphics[width=175mm]{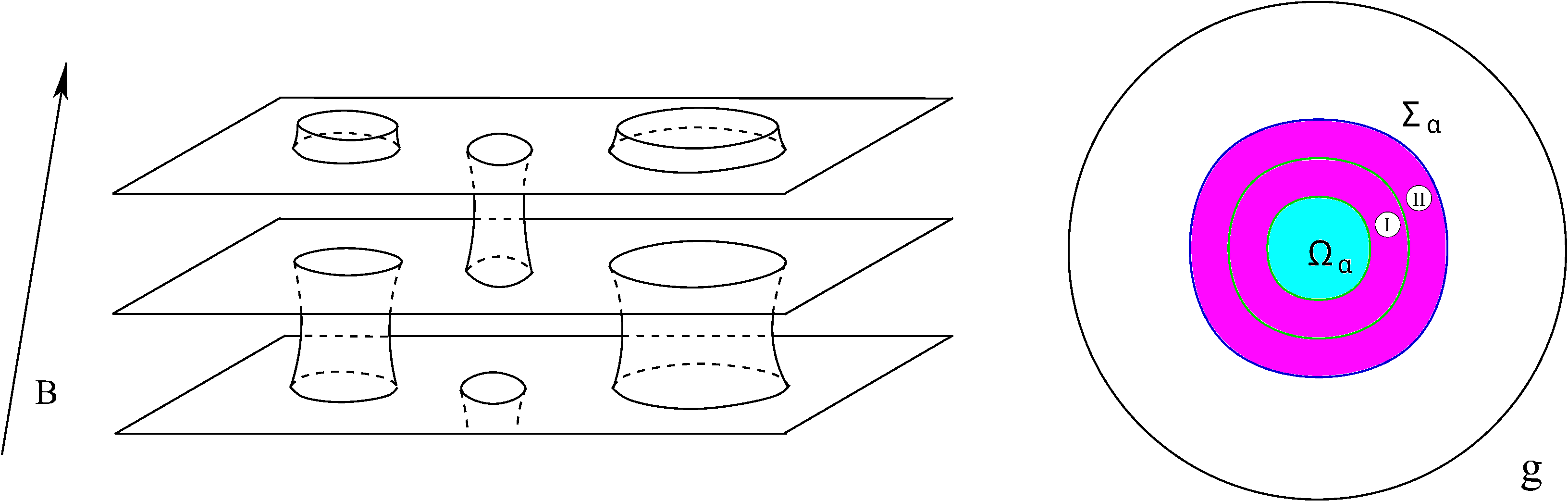}
\end{center}
\caption{Examples of the Fermi surfaces having
different types of the Zones $\, \Omega_{\alpha} \, $
(central parts) and $\, \Sigma_{\alpha} \, $
(all shaded areas).}
\label{Examples}
\end{figure*}

 As we said already, after the disappearance of one of the 
cylinders of closed trajectories at Fig. \ref{TrOnComplFermiSurf}
the integral planes are not separated from each other anymore
and the trajectories of system (\ref{MFSyst}) can now jump
from one plane to another. At the same time, the geometry
of trajectories of (\ref{MFSyst}) still can be easily described
in the vicinity of the boundary of the Zone 
$\, \Omega_{\alpha} \, $. Indeed, in the immediate vicinity 
of the boundary of $\, \Omega_{\alpha} \, $ the Fermi surface
shown at Fig. \ref{TrOnComplFermiSurf} is still divided
into pairs of connected integral planes, which are separated
from each other by the remaining cylinders of closed trajectories
of system (\ref{MFSyst}). This situation takes place until
the second cylinder of closed trajectories on the Fermi surface 
disappears, so that  jumps of the trajectories of system 
(\ref{MFSyst}) between connected pairs of former carriers of 
open trajectories become possible. Let us call the corresponding
curve in the space of directions of $\, {\bf B} \, $ the second
boundary of a Stability Zone $\, \Omega_{\alpha} \, $ and denote
by $\, \Sigma_{\alpha} \, $ the region, restricted by this 
curve on the angular diagram. Let us note here, that the second
boundary of a Stability Zone $\, \Omega_{\alpha} \, $ can
be also simple or compound (Fig. \ref{Boundaries}) depending 
on the geometry of the Fermi surface. In general, the form of 
the second boundary of a Stability Zone is not closely connected 
with the form of its first boundary, as can be seen from the 
examples shown at Fig. \ref{Examples}.

\begin{figure*}
\begin{center}
\includegraphics[width=175mm]{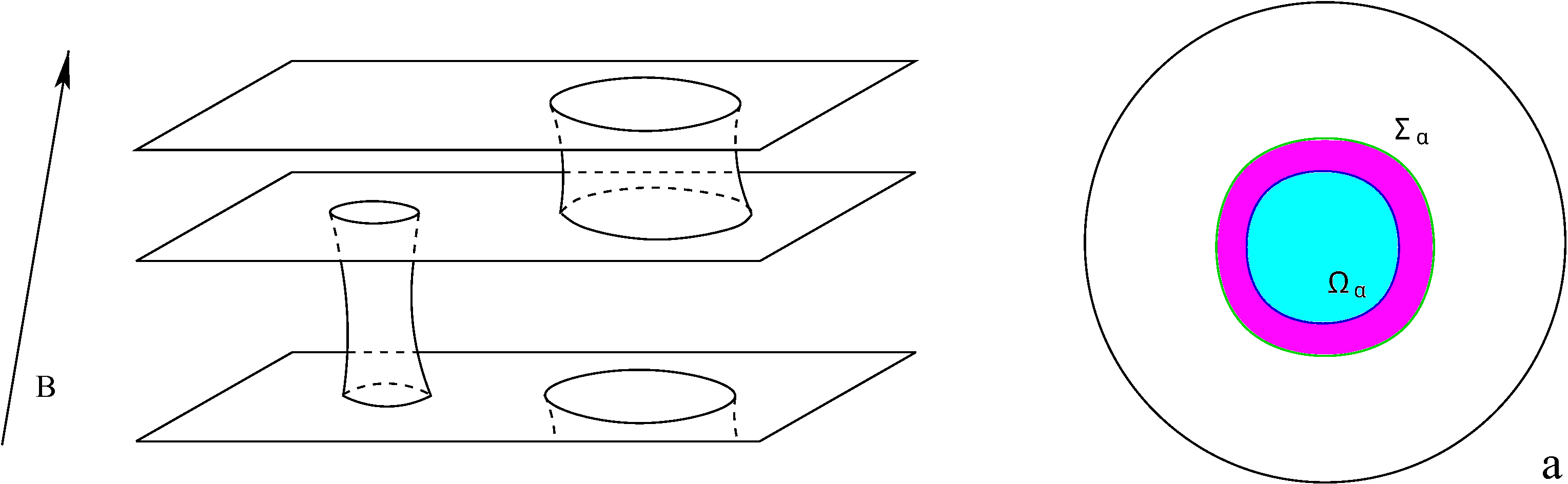}
\end{center}
\begin{center}
\includegraphics[width=175mm]{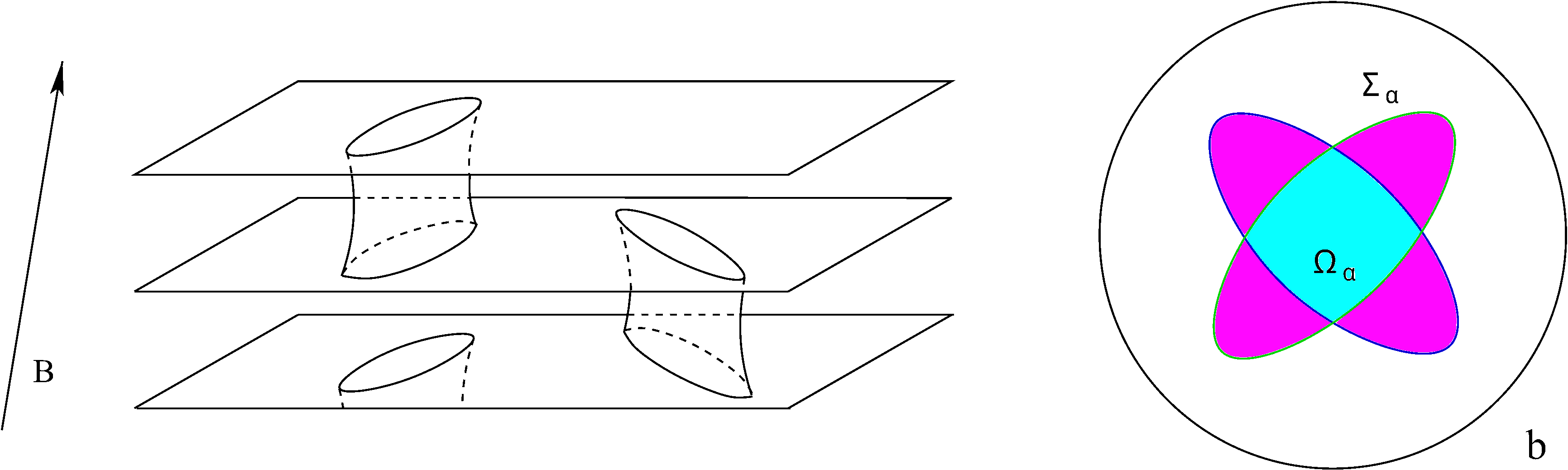}
\end{center}
\begin{center}
\includegraphics[width=175mm]{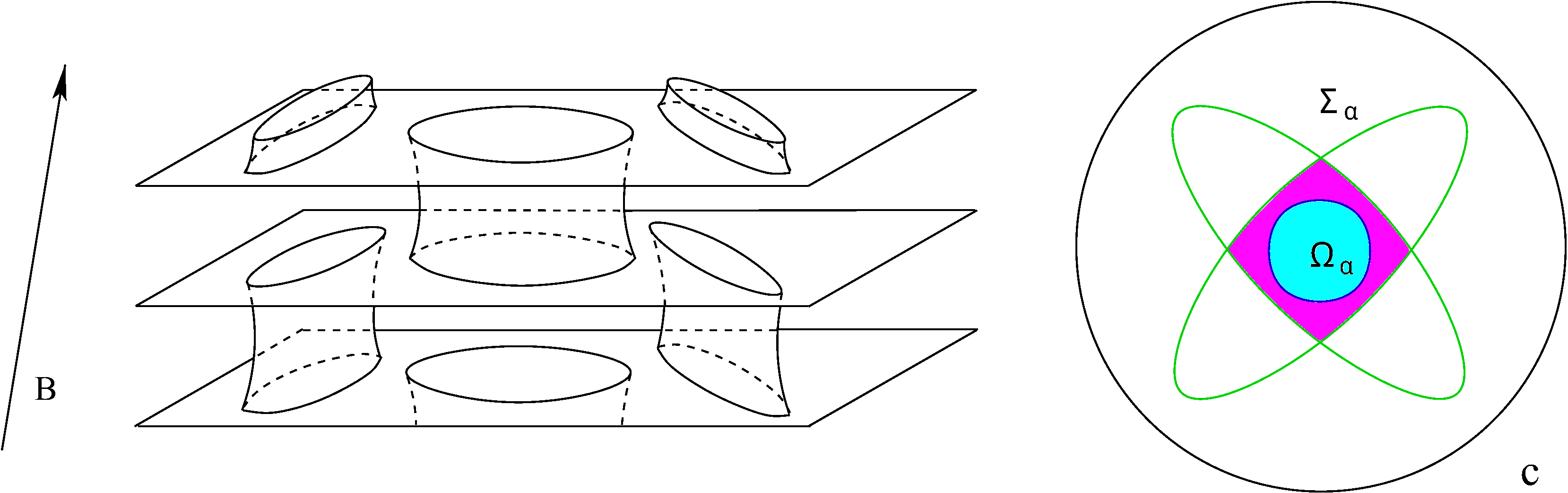}
\end{center}
\begin{center}
\includegraphics[width=175mm]{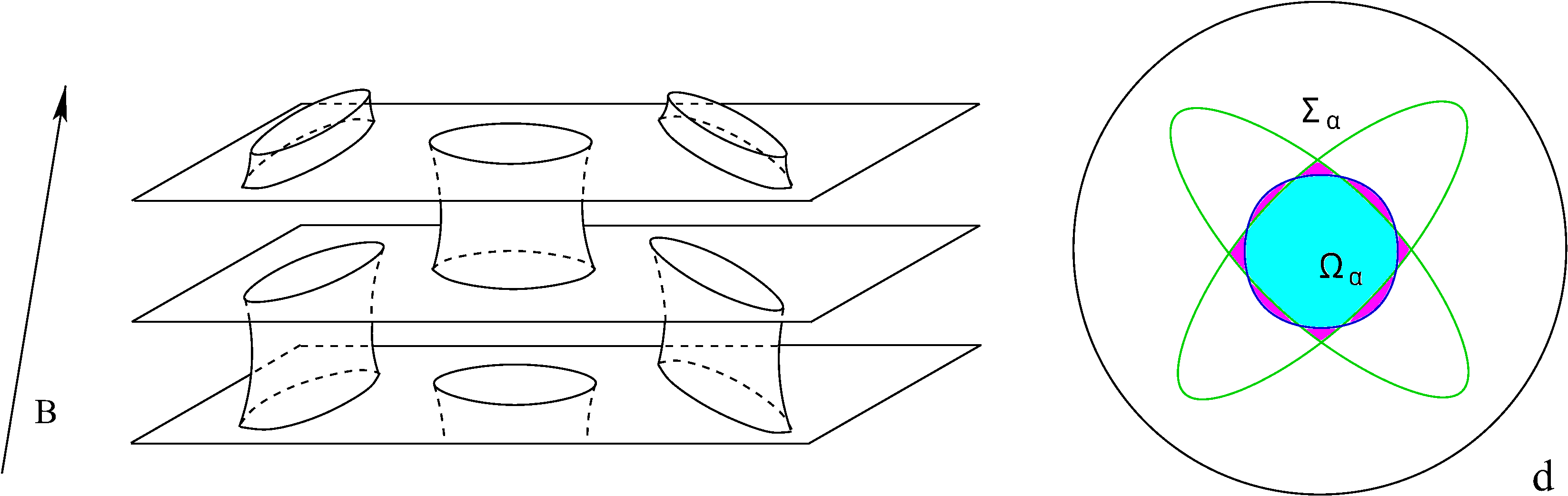}
\end{center}
\end{figure*}

\vspace{5mm}

 Just by definition, the domain 
$\, \Sigma_{\alpha} \backslash \Omega_{\alpha} \, $
is defined as the domain, adjacent to $\, \Omega_{\alpha} \, $,
where the Fermi surface can be represented as a set of 
pairs of connected integral planes with Topological Numbers
$\, (M^{\alpha}_{1}, M^{\alpha}_{2}, M^{\alpha}_{3})$, 
which are separated by cylinders of closed trajectories of system 
(\ref{MFSyst}). As can be seen, the domain 
$\, \Sigma_{\alpha} \backslash \Omega_{\alpha} \, $
can be connected (Fig. \ref{Examples}, a, c, e, g) 
or can consist of several connected components
(Fig. \ref{Examples}, b, d, f).
The boundary of every connected component of the domain
$\, \Sigma_{\alpha} \backslash \Omega_{\alpha} \, $
can be divided into two parts, i.e. the internal boundary
(adjacent to $\, \Omega_{\alpha} \, $) and the external
boundary. What is important here is that for every connected
component of $\, \Sigma_{\alpha} \backslash \Omega_{\alpha} \, $
its internal and external boundaries correspond to disappearance
of cylinders of closed trajectories of opposite types 
(electron type or hole type) (Fig. \ref{Examples}). 
It can be seen also that we can have both the situations
when the connected pairs of former carriers of open
trajectories are defined uniquely for the whole
domain $\, \Sigma_{\alpha} \backslash \Omega_{\alpha} \, $
(Fig. \ref{Examples}, a, c, e, g), or represented
in different ways in different parts of
$\, \Sigma_{\alpha} \backslash \Omega_{\alpha} \, $
(Fig. \ref{Examples}, b, d, f). One more feature
which can be noted here is that even for a given
set of connected pairs of the former carriers of
open trajectories we can have parts of 
$\, \Sigma_{\alpha} \backslash \Omega_{\alpha} \, $
where they are connected through one former
cylinder of closed trajectories or where they 
are connected through two (or more) former cylinders
of closed trajectories (parts I and II at
Fig. \ref{Examples}, e, f, g). The corresponding 
regions are also separated by lines where we have
disappearance or appearance of a cylinder of
(short) closed trajectories, which can be also
observed in the study of oscillation phenomena
in the corresponding frequency range.

 The domain 
$$\Omega^{\prime}_{\alpha} \,\,\, = \,\,\, 
\Sigma_{\alpha} \backslash \Omega_{\alpha} $$
can be called the ``derivative'' of the Stability Zone 
$\, \Omega_{\alpha} \, $ since the Fermi surface keeps here
a part of its special representation in the Zone 
$\, \Omega_{\alpha} \, $, which still gives an effective 
description of trajectories of system (\ref{MFSyst}) 
in this domain. 

\begin{figure}[t]
\begin{center}
\vspace{5mm}
\includegraphics[width=\linewidth]{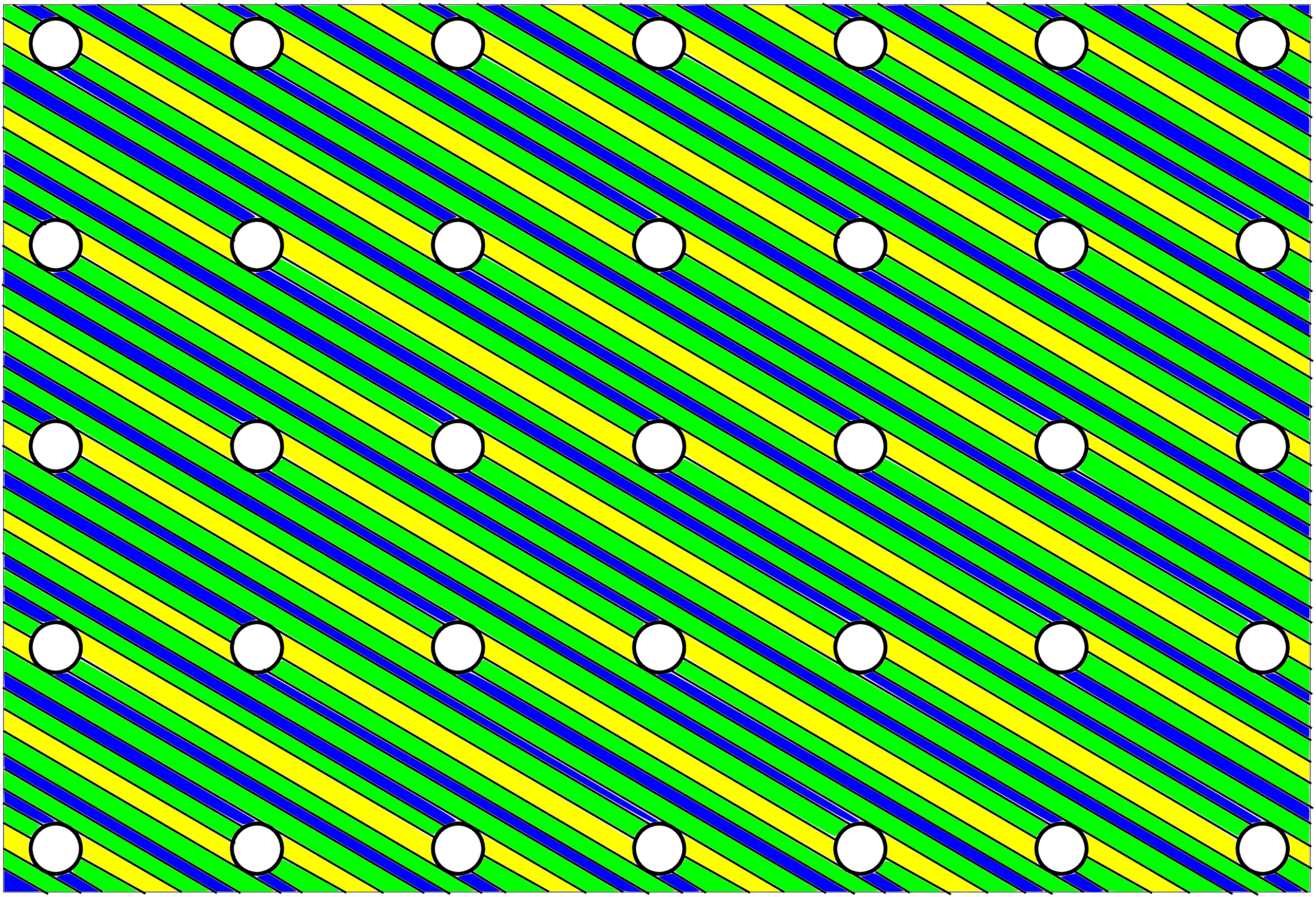}
\end{center}
\caption{Schematic division of a pair of former
carriers of open trajectories into layers of long
closed trajectories, separated by singular closed
trajectories.}
\label{PerNet}
\end{figure}

 To describe the trajectories of system (\ref{MFSyst}) on the 
connected pairs of the former carriers of open trajectories
we have to note first that the trajectories that have jumps
on a fixed cylinder connecting a pair of integral planes 
(Fig. \ref{Cylind}) form narrow strips restricted by
two singular trajectories with close values of $\, p_{z} \, $.
The value of $\, p_{z} \, $ is constant at every trajectory
of system (\ref{MFSyst}), so all these strips are represented
just by straight narrow strips after the projection on a
plane, containing the vector $\, {\bf B}/B \, $. Using the 
projection on the plane, containing the vector $\, {\bf B}/B \, $, 
we can mark the corresponding cylinders just by circles 
(or segments) of the corresponding size ($\Delta p_{z}$), which 
form a periodic structure in the plane (Fig. \ref{PerNet}). 
Every trajectory of system (\ref{MFSyst}) will be represented
now by a straight line $\, p_{z} \, = \, const \, $, 
which can have endpoints at the circles, introduced above.
The singular trajectories of system (\ref{MFSyst}) are 
represented by lines, tangential to the circles, and separate
trajectories of substantially different geometry on the Fermi 
surface. It is not difficult to see that for generic directions 
of $\, {\bf B} \, $, such that the intersection of the plane
orthogonal to $\, {\bf B} \, $ and the plane 
$\, \Gamma_{\alpha} \, $ has an irrational direction in
$\, {\bf p}$ - space, every trajectory of system 
(\ref{MFSyst}) will be actually represented by a line of a 
finite size, restricted by two circles at both ends. 
As a result, the pair of the former carriers of open 
trajectories will be divided itself into cylinders of
closed trajectories of system (\ref{MFSyst}), which will
have rather big sizes if $\, {\bf B}/B \, $ is close 
to the boundary of the Zone $\, \Omega_{\alpha} \, $.

\begin{figure}[t]
\begin{center}
\vspace{5mm}
\includegraphics[width=0.9\linewidth]{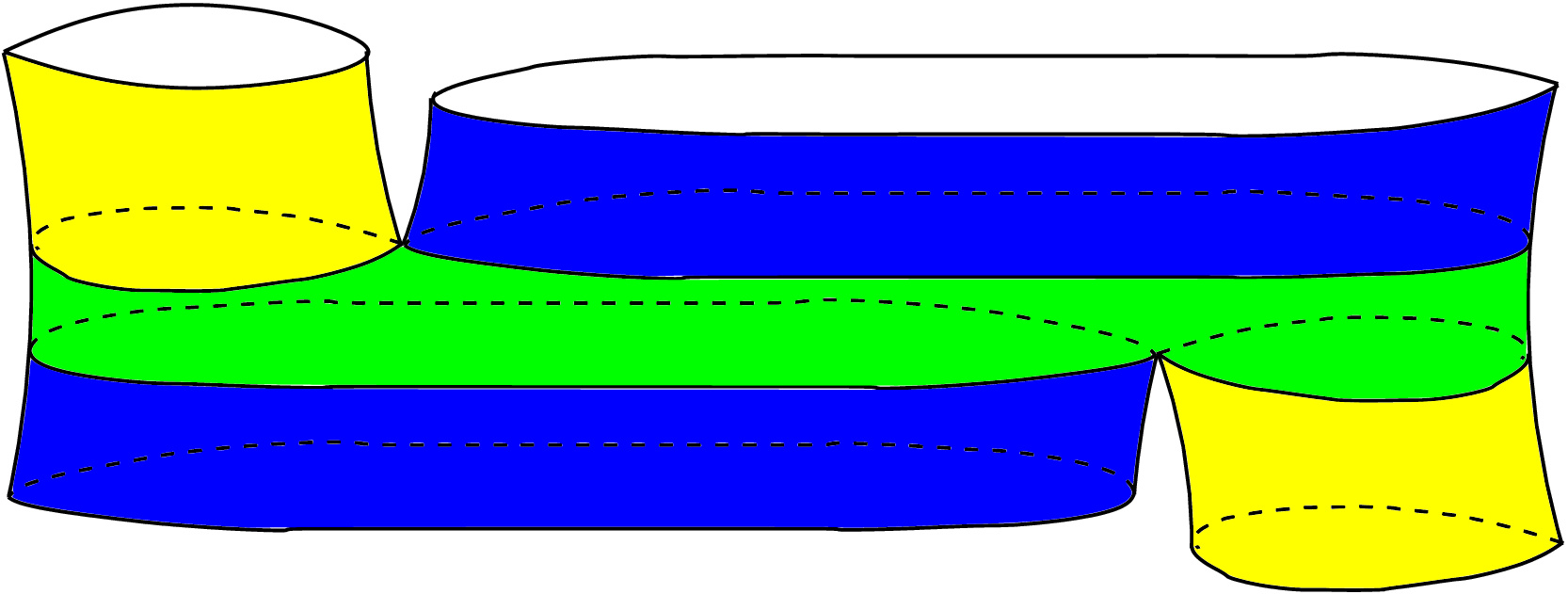}
\end{center}
\caption{Schematic representation of layers of long
closed trajectories on a connected pair of former carriers 
of open trajectories in the $\, {\bf p}$ - space.}
\label{LongCyl}
\end{figure}

 Let us consider here the simplest situation when the jumps 
of the trajectories between two former carriers of open 
trajectories occur just on one cylinder from the set of 
the former cylinders of closed trajectories, connecting 
these carriers. Fig. \ref{PerNet} represents schematic division 
of a connected pair of the former carriers of open trajectories
into layers of long closed trajectories after crossing
the boundary of the Zone $\, \Omega_{\alpha} $. Every 
colored strip at Fig. \ref{PerNet} represents the
projection of trajectories of approximately the same form 
on a plane ($\Pi$) containing the vector $\, {\bf B}/B \, $. 
As can be easily seen, we will have here exactly three 
nonequivalent layers of significantly different long closed 
trajectories, separated by singular closed trajectories, 
on every connected pair of former carriers of open 
trajectories in the $\, {\bf p}$ - space (Fig. \ref{LongCyl}).

\begin{figure}[t]
\begin{center}
\vspace{5mm}
\includegraphics[width=\linewidth]{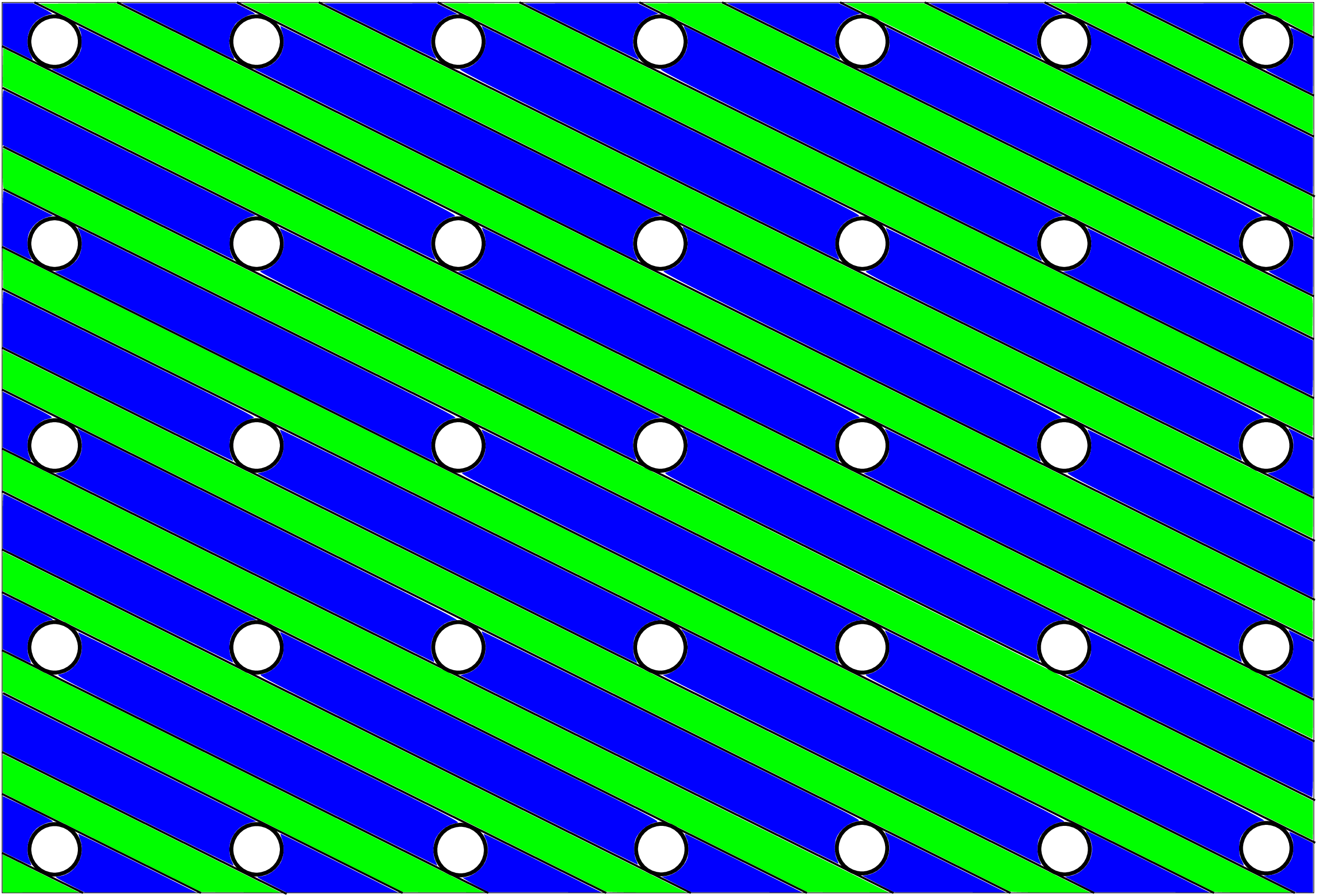}
\end{center}
\caption{Layers of periodic open trajectories and 
of long closed trajectories on a pair of 
connected former carriers of open trajectories.}
\label{PerNet2}
\end{figure}

 At the same time, if the intersection of the plane, 
orthogonal to $\, {\bf B} \, $, and the plane 
$\, \Gamma_{\alpha} \, $ has a rational direction in the 
$\, {\bf p}$ - space and the value of $\, \Delta p_{z} \, $
is small enough we can still have periodic open trajectories
on the former carriers of open trajectories after crossing
the boundary of the Zone $\, \Omega_{\alpha} \, $ 
(Fig. \ref{PerNet2}). As can be easily seen, we will have 
in this case just one layer of long closed trajectories
and two (opposite) layers of periodic open trajectories 
on a connected pair of former carriers of open trajectories.
It is easy to see also, that in the picture described above
we can not have open trajectories of system (\ref{MFSyst})
which could be stable with respect to all small rotations
of the direction of $\, {\bf B} \, $.

\begin{figure}[t]
\begin{center}
\vspace{5mm}
\includegraphics[width=0.9\linewidth]{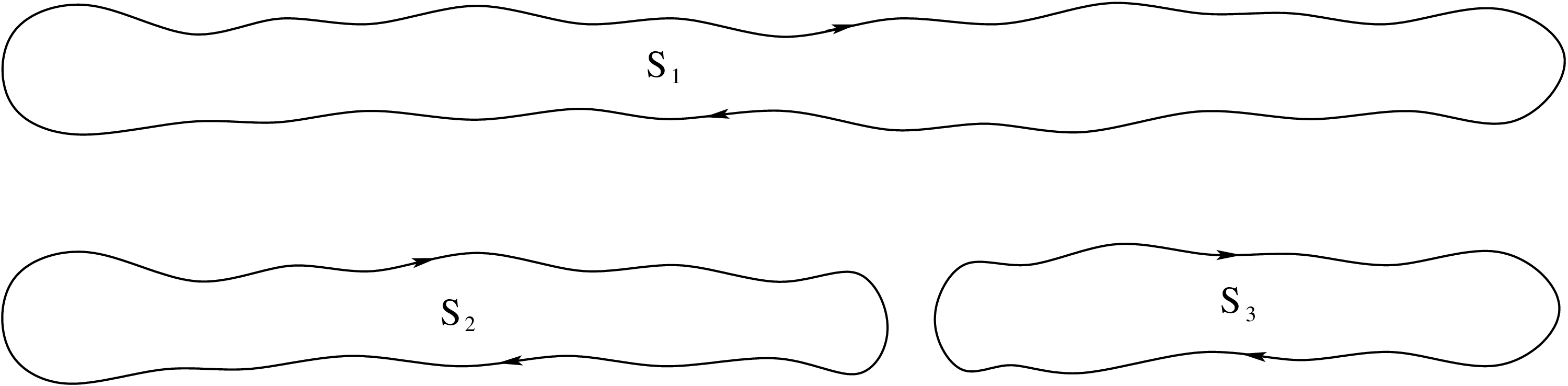}
\end{center}
\caption{Three different long closed trajectories arising
on the Fermi surface near the boundary of the Zone
$\, \Omega_{\alpha} \, $ in the domain 
$\, \Omega^{\prime}_{\alpha} \,\,\, = \,\,\, 
\Sigma_{\alpha} \backslash \Omega_{\alpha} \, $.}
\label{ThreeTraject}
\end{figure}

 Easy to see that the long closed trajectories arising
near the boundary of the Stability Zone 
$\, \Omega_{\alpha} \, $ have a specific form shown at
Fig. \ref{ThreeTraject}. We can see, that we should have
here three different ``sets'' of the long closed trajectories
according to the number of nonequivalent layers of
significantly different closed trajectories on the Fermi 
surface. We can see also, that we can write the approximate 
equality
$$S_{1} \,\,\, \simeq \,\,\, S_{2} \, + \, S_{3} $$
for the areas, restricted by the trajectories, shown
at Fig. \ref{ThreeTraject}, in the plane, orthogonal to
$\, {\bf B} \, $ in the $\, {\bf p}$ - space.  

 In the same way, we can write also
$$T_{1} \,\,\, \simeq \,\,\, T_{2} \, + \, T_{3} $$
for the periods of motion along the trajectories shown
at Fig. \ref{ThreeTraject}, or
$$1/\omega_{1} \,\,\, \simeq \,\,\, 1/\omega_{2} \, + \, 
1/\omega_{3} $$
for the corresponding cyclotron frequencies.

\vspace{1mm}

\begin{figure}[t]
\begin{center}
\vspace{5mm}
\includegraphics[width=0.9\linewidth]{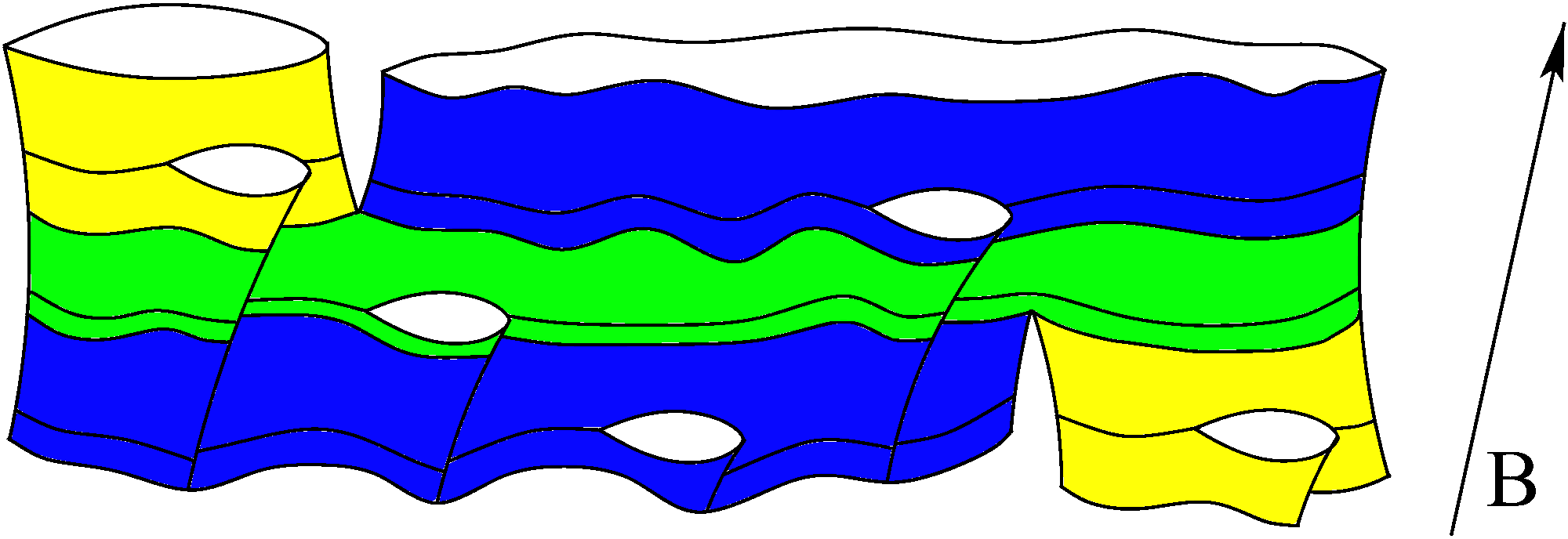}
\end{center}
\caption{The geometry of cylinders of long closed trajectories 
on a connected pair of former carriers of open trajectories
in the $\, {\bf p}$ - space (schematically).}
\label{FullLongCyl}
\end{figure}

 More precisely, the layers of long closed trajectories,
shown at Fig. \ref{LongCyl}, should contain also additional
singularities, arising at the bases of cylinders of short
closed trajectories, separating the pairs of the former
carriers of open trajectories (Fig. \ref{FullLongCyl}).
These singularities always appear at the boundaries
of singular two-dimensional discs, orthogonal to 
$\, {\bf B} \, $, which are actually invisible at the 
projection on the plane $\, \Pi \, $. This means actually
that the number of topologically nonequivalent cylinders
of long closed trajectories (separated by singular trajectories)
is in fact greater than 3. However, if the length of the 
closed trajectories is very big, we can still introduce three
sets of similar long closed trajectories corresponding to three
essentially different periods $\, T_{1}$, $\, T_{2}$, $\, T_{3}$
as it is shown at Fig. \ref{FullLongCyl}. We can note also that
when the long closed trajectories become shorter we should have
a ``splitting'' of the periods $\, T_{1}$, $\, T_{2}$, $\, T_{3}$
into a bigger set $\, \{ T_{i} \} \, $, corresponding to
the number of extremal trajectories on all topologically
different cylinders of closed trajectories described above.

 It can be shown also, that in the case when the number
of nonequivalent cylinders connecting former carriers of open 
trajectories is equal to 2, we can introduce five periods
$\, T_{1}$, $\, T_{2}$, $\, T_{3}$, $\, T_{4}$, $\, T_{5}$
(the number of the cylinders of long closed trajectories is 
actually greater), which can be numerated in such a way, that we 
can write the relations
\begin{equation}
\label{T5Rel}
T_{4} \,\, \simeq \,\, T_{2} \, + \, T_{3}  \quad , \quad \quad
T_{5} \,\, \simeq \,\, T_{1} \, + \, T_{2} \, + \, T_{3}
\end{equation}
or
\begin{equation}
\label{S5Rel}
S_{4} \,\, \simeq \,\, S_{2} \, + \, S_{3}  \quad , \quad \quad
S_{5} \,\, \simeq \,\, S_{1} \, + \, S_{2} \, + \, S_{3}
\end{equation}

 As we said already, this situation can arise in special regions
in the domain $\, \Omega_{\alpha}^{\prime} \, $
(regions II at Fig. \ref{Examples}, e, f, g), which are 
separated by special curves, observable in the study of
oscillation phenomena at high frequencies 
($\Omega \simeq \omega_{B}$). We should note also, that the 
relations (\ref{T5Rel}) and (\ref{S5Rel}) can be written only 
for long enough closed trajectories and we should also observe
here a splitting of the periods
$\, T_{1}$, $\, T_{2}$, $\, T_{3}$, $\, T_{4}$, $\, T_{5}$
into a bigger set when the trajectories become shorter. 
Theoretically we can consider also even more complicated cases,
where the number of typical periods is even greater, however,
they are very unlikely for real metals. In particular, we will
always have the situation with three periods
$\, T_{1}$, $\, T_{2}$, $\, T_{3}$ in the vicinity of 
``regular'' points of the first boundary of a Stability Zone 
(regions I at Fig. \ref{Examples}, e, f, g).

\vspace{1cm}

\section
{The geometry of the long closed trajectories and the 
behavior of conductivity in the domain
$\, \Omega^{\prime} \, $.}
\setcounter{equation}{0}

 What can we say about the behavior of the 
magneto-conductivity or, more generally, the effects
connected with the presence of a strong magnetic field
in the case when the direction of $\, {\bf B} \, $
belongs to the domain 
$\, \Omega^{\prime}_{\alpha} \, = \, 
\Sigma_{\alpha} \backslash \Omega_{\alpha} \, $?
Let us say first of all that the domain
$\, \Sigma_{\alpha} \backslash \Omega_{\alpha} \, $
always contains a narrow region near the boundary
of the Zone $\, \Omega_{\alpha} \, $ where the length
of the long closed trajectories is very large. As a corollary,
the condition $\, \tau / T \, \gg \, 1 \, $
in this region is obviously violated and the long closed
trajectories are indistinguishable here from the open
trajectories from the experimental point of view. 
The conductivity tensor (for available values of $\, B $)
has here the same geometric properties as in the Stability
Zone $\, \Omega_{\alpha} \, $ and the same is also valid 
for its analytic properties (see \cite{JETP2017}). 
As we have said already, the position of the boundary of
$\, \Omega_{\alpha} \, $ can be detected in general by the
study of special features of the cyclotron resonance phenomenon 
(at high external frequencies) or other (quantum) oscillation
phenomena (see \cite{OscPhen}), which are connected with the 
disappearance of a cylinder of (short) closed trajectories
on the Fermi surface. 

\begin{figure}[t]
\begin{center}
\vspace{5mm}
\includegraphics[width=0.9\linewidth]{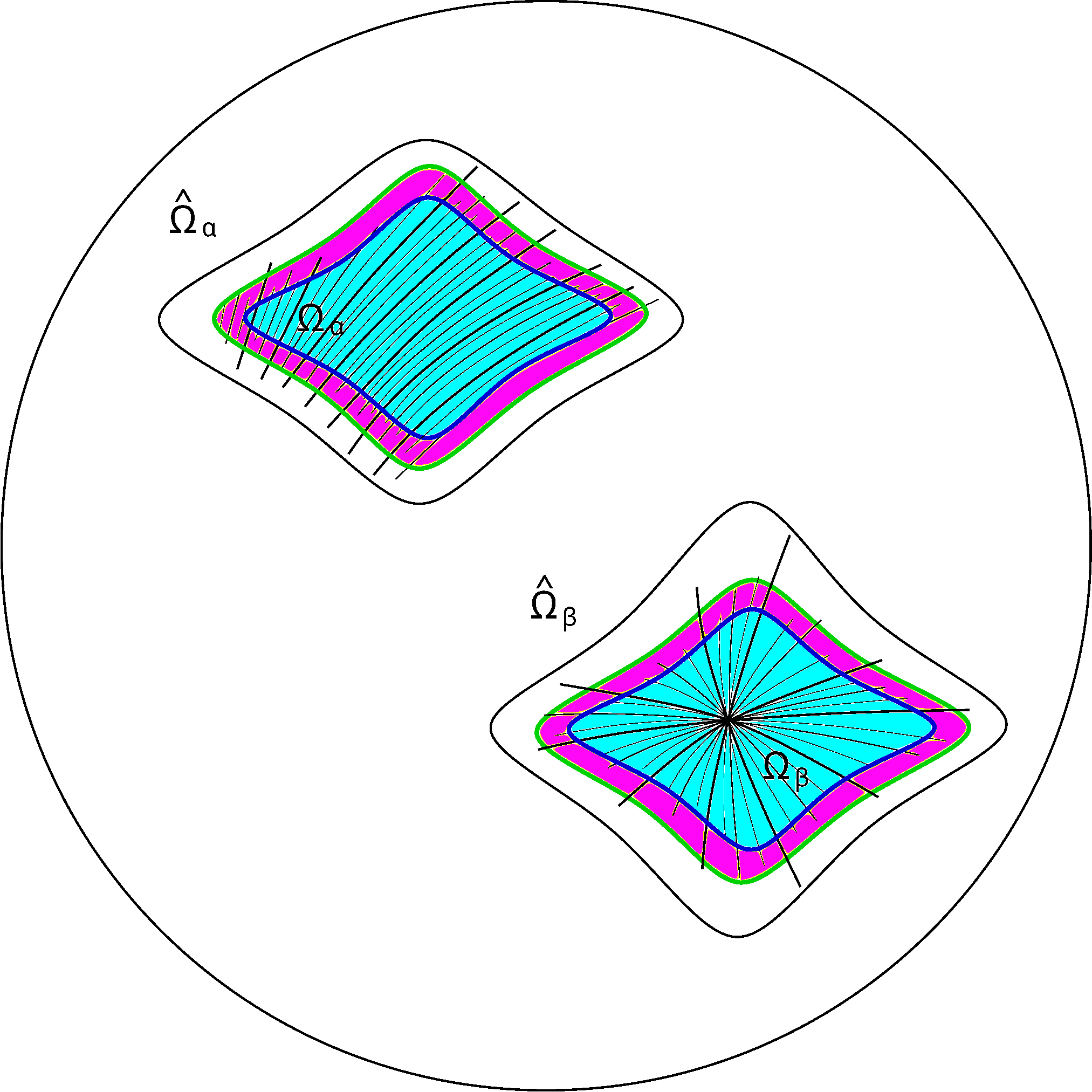}
\end{center}
\caption{Narrow domains 
$\, \Omega^{\prime}_{\alpha}$, $\, \Omega^{\prime}_{\beta}$
(pink) inside the ``experimentally observable'' Stability 
Zones $\, \hat{\Omega}_{\alpha}$, $\, \hat{\Omega}_{\beta}$
(schematically).}
\label{NarrowRegion}
\end{figure}

 As we have also said already, the position of the second
boundary of a Stability Zone is to some extent independent
of the position of the first boundary. In particular, we can
have the situation when the second boundary of a Stability
Zone is very close to its first boundary, such that the 
domain $\, \Omega^{\prime}_{\alpha} \, $ represents 
a narrow region in the vicinity of the boundary of 
$\, \Omega_{\alpha} \, $. In this case the period $\, T \, $
of the electron motion over the long closed trajectories can 
be very big and the condition 
$\, \tau / T \, \gg \, 1 \, $ will be in fact violated
everywhere in $\, \Omega^{\prime}_{\alpha} \, $
(Fig. \ref{NarrowRegion}). In this case we can just observe 
the first and the second boundaries of a Stability Zone with 
the aid of study of the oscillation phenomena inside the 
``experimentally observable'' Stability Zone
$\, \hat{\Omega}_{\alpha} \, $.

 Let us note here that in the standard setting of the 
experimental study of the cyclotron resonance phenomenon 
it is often convenient to choose the direction of the 
electric field $\, {\bf E}_{inc} \, $ in the incident wave 
orthogonal to the direction of the constant magnetic 
field $\, {\bf B} \, $. In this case the endpoints on 
the ``trivial'' cylinders of closed trajectories 
(Fig. \ref{TrOnComplFermiSurf}) give no contribution to 
the picture of oscillations, so, only the extremal 
trajectories on the topologically nontrivial cylinders 
of closed trajectories bring oscillating terms in the 
dependence of physical quantities on the value of
$\, B \, $ (or $\, \Omega$). If we assume also that every
topologically nontrivial cylinder of closed trajectories
contains just one extremal trajectory then the number
of the oscillating terms in a Stability Zone should 
be exactly $\, g - 1 \, $, where $\, g \, $ is the genus
of the Fermi surface. 

\vspace{1mm}

\begin{figure}[t]
\begin{center}
\vspace{5mm}
\includegraphics[width=\linewidth]{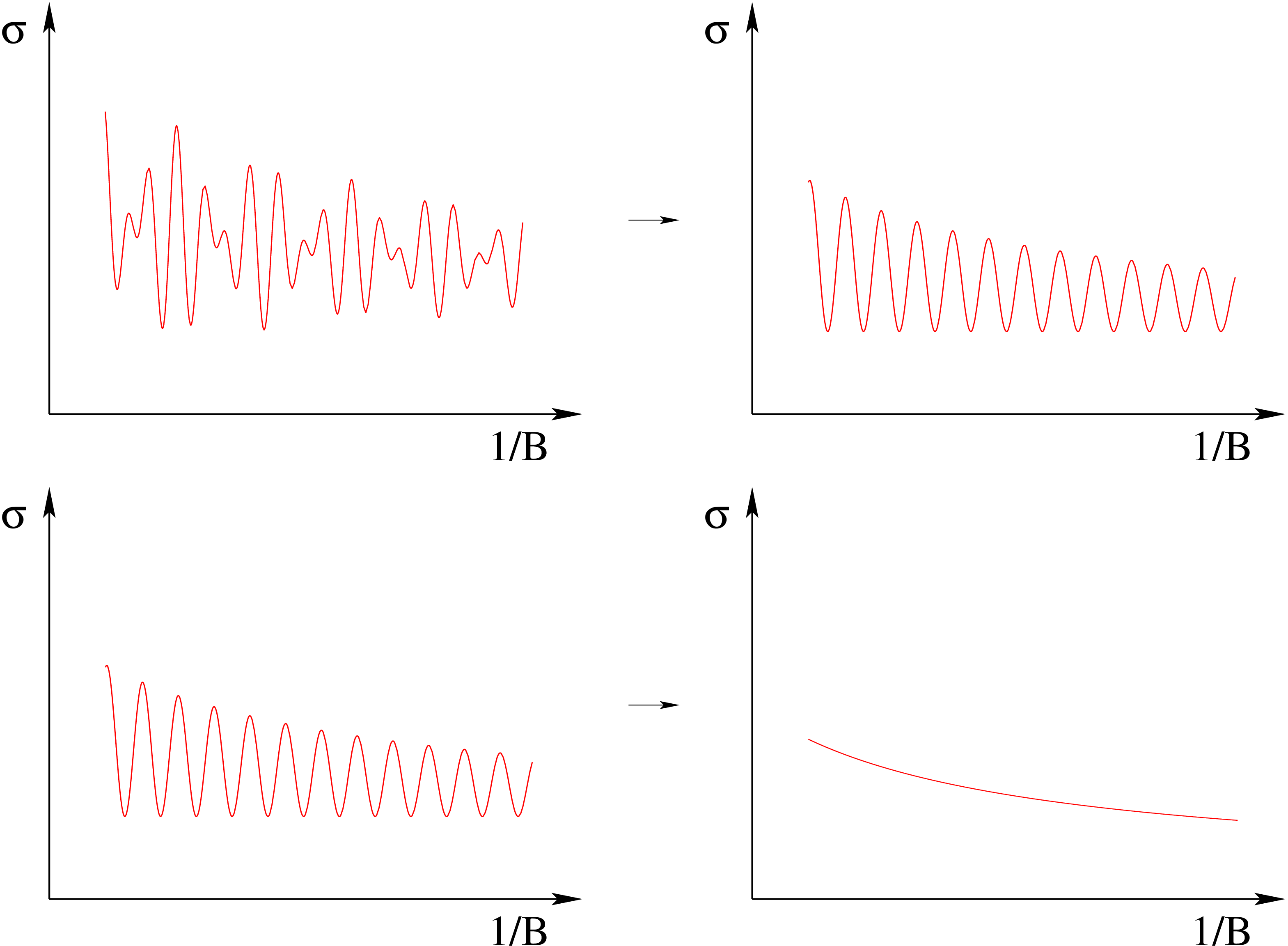}
\end{center}
\caption{The transformation of the picture of the cyclotron 
resonance oscillations at the internal and the  
external boundaries of a narrow domain
$\, \Omega^{\prime}_{\alpha}$ around a Stability Zone 
$\, \Omega_{\alpha} \, $ for a Fermi surface having
genus 3. (The direction of electric field in the 
incident wave is orthogonal to the direction of the 
constant magnetic field $\, {\bf B}$).}
\label{Gen3OscPic}
\end{figure}

\begin{figure}[t]
\begin{center}
\vspace{5mm}
\includegraphics[width=\linewidth]{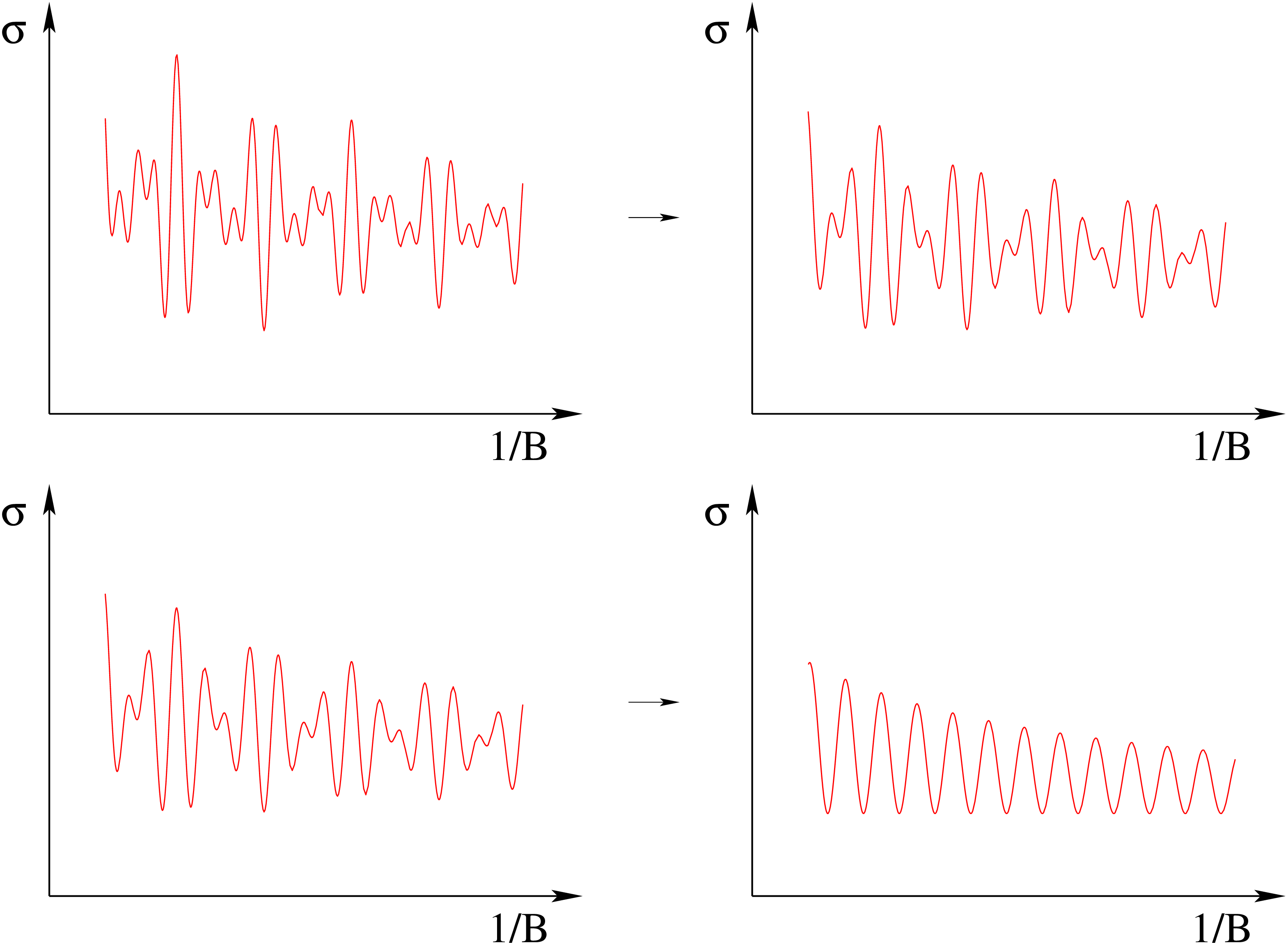}
\end{center}
\caption{The transformation of the picture of the cyclotron 
resonance oscillations at the internal and the  
external boundaries of a narrow domain
$\, \Omega^{\prime}_{\alpha}$ around a Stability Zone 
$\, \Omega_{\alpha} \, $ for a Fermi surface having
genus 4. (The direction of electric field in the 
incident wave is orthogonal to the direction of the 
constant magnetic field $\, {\bf B}$).}
\label{Gen4OscPic}
\end{figure}

 As was pointed out in \cite{OscPhen}, the picture of 
oscillations should undergo a rapid transformation
on the boundary of any Stability Zone due to the 
disappearance of a cylinder of closed trajectories after 
crossing the boundary of $\, \Omega_{\alpha} \, $ on the 
angular diagram. Thus, for a Fermi surface of genus 3
we should observe a rapid transformation of a picture,
containing 2 oscillating terms, to a picture, containing
just one oscillating term, at the first boundary of 
a Stability Zone. If we assume now that the second boundary
is very close to the first boundary of a Stability Zone
and the long closed trajectories do not give any 
contribution to the oscillation phenomena inside the 
domain $\, \Sigma_{\alpha} \backslash \Omega_{\alpha} \, $,
we expect then a rapid transformation of the picture,
containing just one oscillating term, to a picture, 
containing no oscillating terms, at the second boundary
of a Stability Zone $\, \Omega_{\alpha} \, $ 
(Fig. \ref{Gen3OscPic}). For Fermi surfaces having higher
genera the picture of oscillations should be more
complicated. For comparison, for a Fermi
surface of genus 4 we should observe a rapid transformation 
of a picture, containing 3 oscillating terms, to a picture, 
containing 2 oscillating terms, at the first boundary of 
a Stability Zone and a rapid transformation of the picture,
containing 2 oscillating terms, to a picture, containing 
just one oscillating term, at the second boundary
(Fig. \ref{Gen4OscPic}). Let us note also that the pictures
represented at Fig. \ref{Gen3OscPic} and Fig. \ref{Gen4OscPic}
can be somewhat schematic in nature since the form of the 
oscillation peaks in the oscillating terms can be different
in different situations.

\begin{figure}[t]
\begin{center}
\vspace{5mm}
\includegraphics[width=\linewidth]{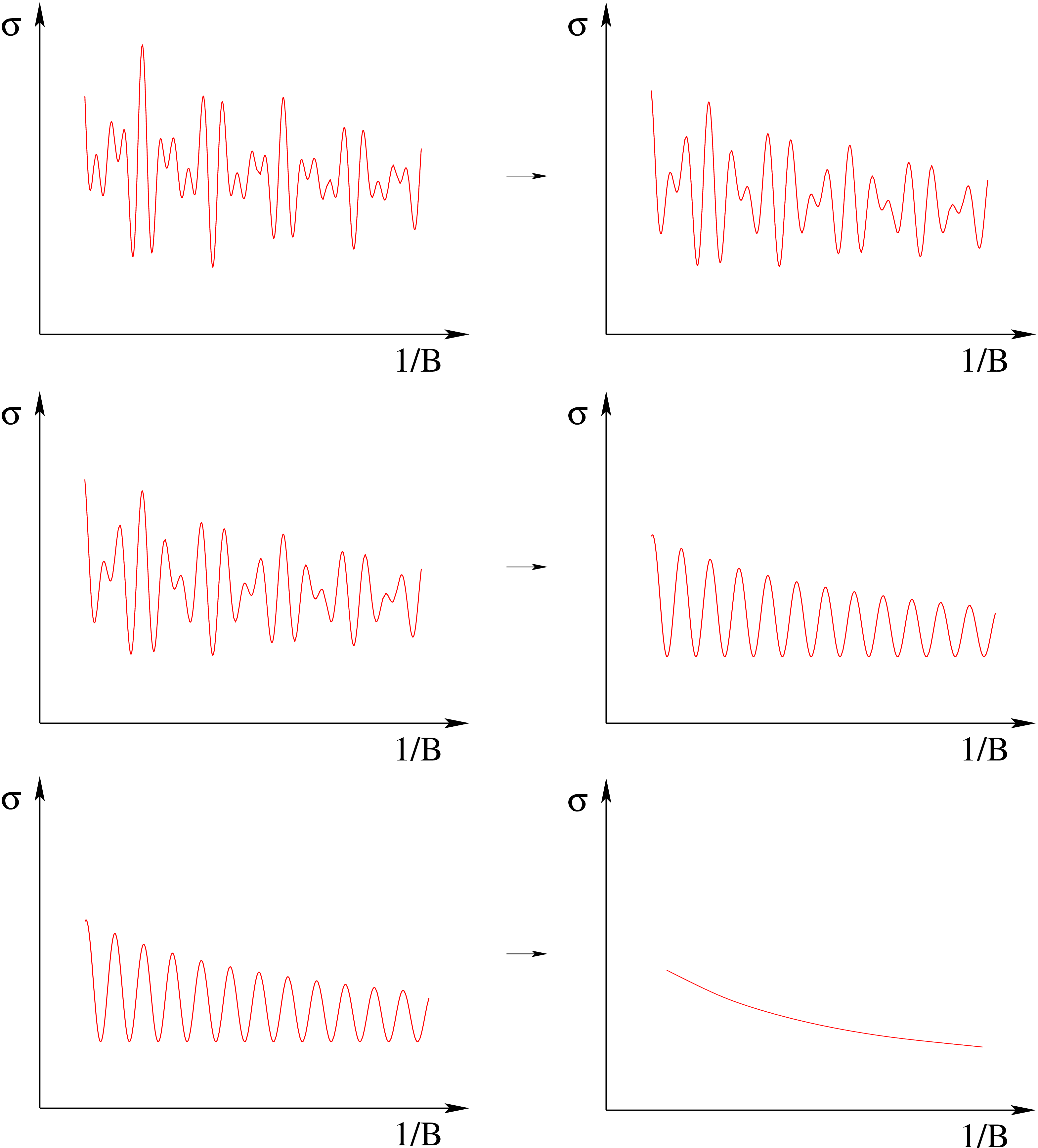}
\end{center}
\caption{The transformation of the picture of the cyclotron 
resonance oscillations at the internal boundary,
special curve inside the domain $\, \Omega^{\prime}_{\alpha}$,
and the external boundary of a narrow domain
$\, \Omega^{\prime}_{\alpha}$ around a Stability Zone 
$\, \Omega_{\alpha} \, $ for a Fermi surface having
genus 4. (The direction of electric field in the 
incident wave is orthogonal to the direction of the 
constant magnetic field $\, {\bf B}$).}
\label{ThreeTrans}
\end{figure}

 Another special feature, which can arise on complicated
Fermi surfaces is the disappearance of additional cylinders
of closed trajectories between the first and the second
boundaries of a Stability Zone (Fig. \ref{Examples}, e, f, g).
In this case the full picture of transformations of the 
cyclotron oscillations caused by the rotation of 
$\, {\bf B} \, $ can resemble the situation shown at 
Fig. \ref{ThreeTrans}.

  As we said already, in the case of a narrow domain
$\, \Omega^{\prime}_{\alpha} \, = \, 
\Sigma_{\alpha} \backslash \Omega_{\alpha} \, $,
lying entirely in the experimentally observable
Stability Zone $\, \hat{\Omega}_{\alpha} \, $, 
rapid transformations of the oscillation picture,
similar to those shown at Fig. \ref{Gen3OscPic}, 
\ref{Gen4OscPic}, \ref{ThreeTrans}, give the only evidence 
of the existence of the first and the second boundaries of 
a Stability Zone $\, \Omega_{\alpha} \, $.

\vspace{1mm}

\begin{figure}[t]
\begin{center}
\vspace{5mm}
\includegraphics[width=0.9\linewidth]{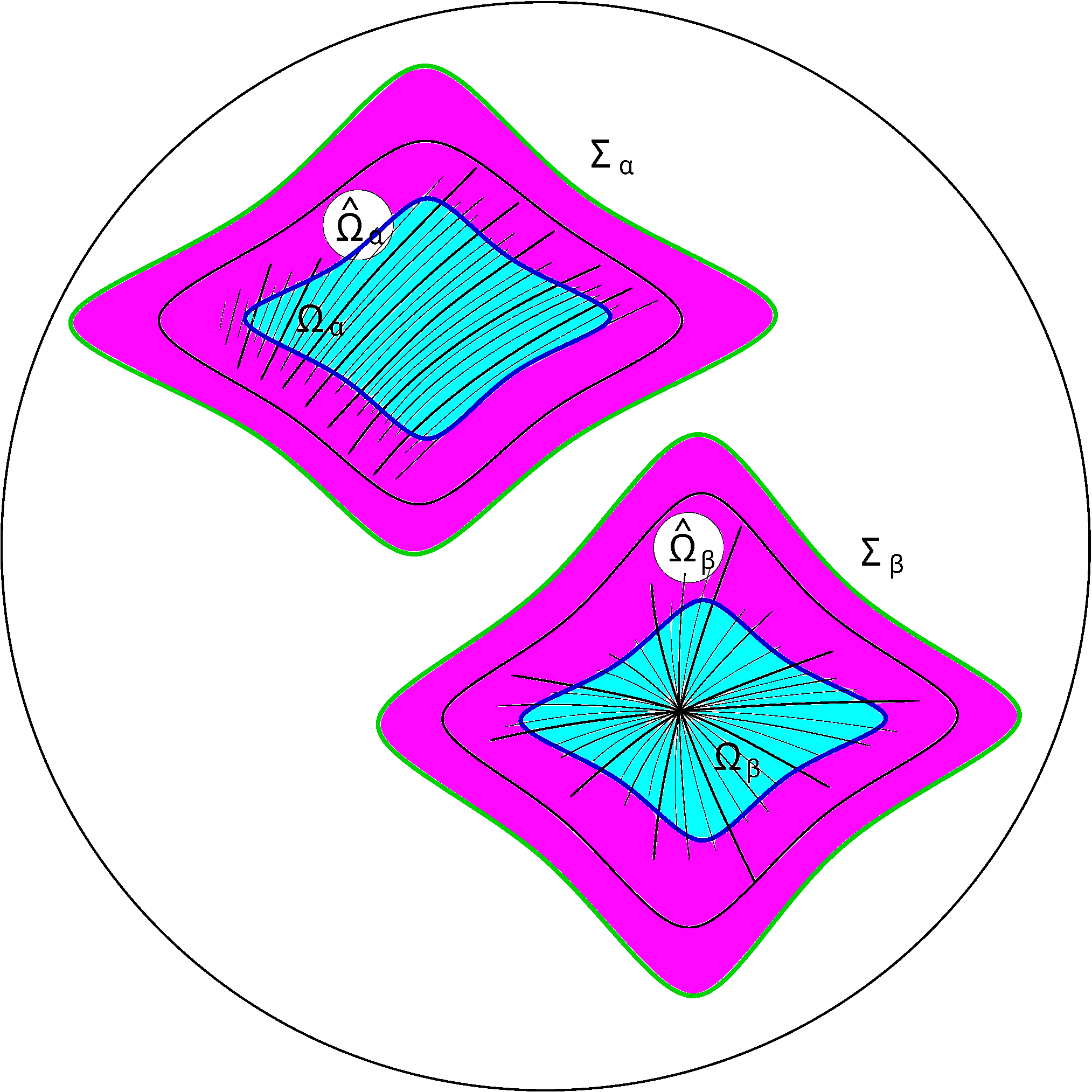}
\end{center}
\caption{The domains 
$\, \Omega^{\prime}_{\alpha} \,\,\, = \,\,\, 
\Sigma_{\alpha} \backslash \Omega_{\alpha} \,$, 
$\, \Omega^{\prime}_{\beta} \,\,\, = \,\,\, 
\Sigma_{\beta} \backslash \Omega_{\beta} \,$ (pink),
having ``intermediate'' size on the angular diagram
(schematically).}
\label{InterRegion}
\end{figure}

 We can consider now the ``intermediate situation''
when the long closed trajectories in the domain 
$\, \Omega^{\prime}_{\alpha} \, $ can have ``intermediate'' 
length, such that we have the relations
\begin{equation}
\label{InterCond}
T^{-1} \,\,\, \ll \,\,\, \omega_{B} 
\, = \, {e B \over m^{*} c}
\quad , \quad \quad 
\tau \, / \, T \, \gg \, 1 \,\,\, , 
\end{equation}
where $\, T \, $ is the period of the electron motion over
the long closed trajectory. This means that the domain 
$\, \Omega^{\prime}_{\alpha} \, $ has only a partial 
intersection with the ``experimentally observable''
Stability Zone $\, \hat{\Omega}_{\alpha} \, $ 
(Fig. \ref{InterRegion}) and we can
observe some special features in the conductivity behavior
in the domain $\, \Omega^{\prime}_{\alpha} \, $.

 It is easy to see that we should observe here a special
picture of oscillation phenomena (classical or quantum)
connected with the long closed trajectories arising in
$\, \Omega^{\prime}_{\alpha} \, $. Let us specially discuss 
here the picture of the cyclotron resonance oscillations
connected with the closed trajectories satisfying
condition (\ref{InterCond}). 

\begin{figure}[t]
\begin{center}
\vspace{1cm}
\includegraphics[width=0.95\linewidth]{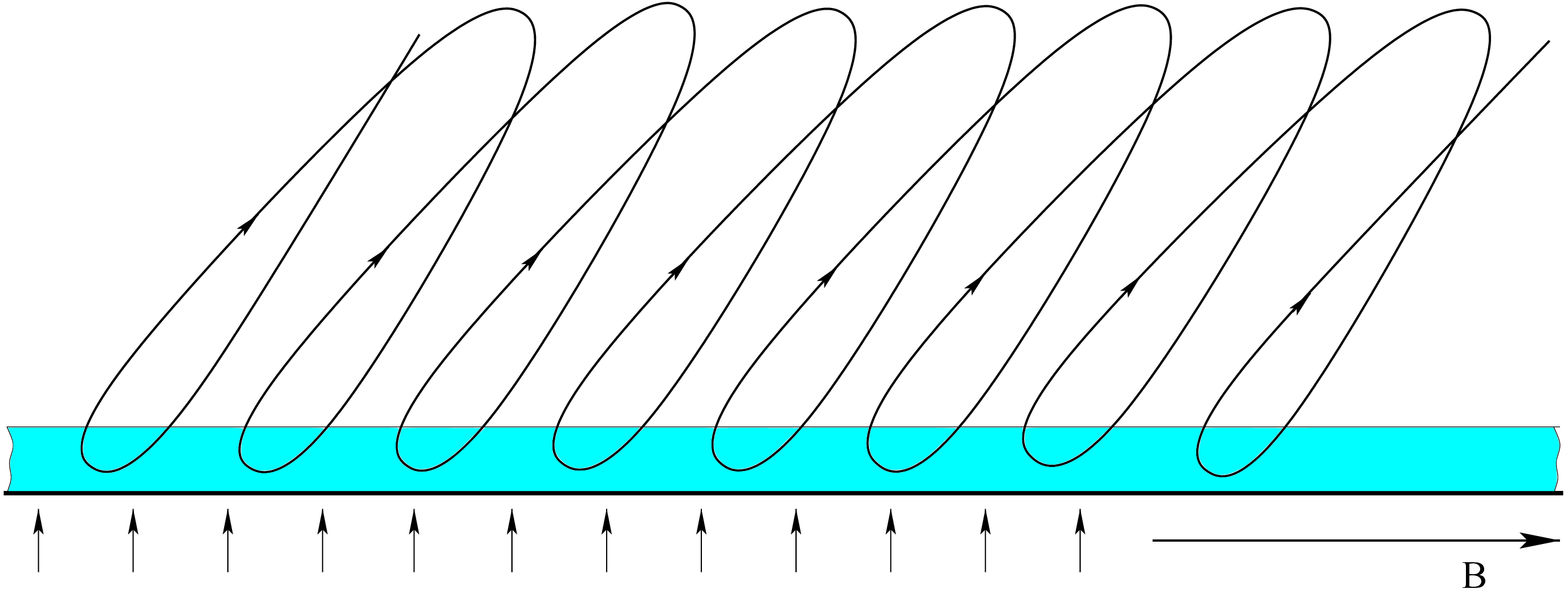}
\end{center}
\caption{The electron motion through the skin layer 
in the $\, {\bf x}$ - space corresponding to the motion
over the long closed trajectories in the domain
$\, \Omega^{\prime}_{\alpha}$.}
\label{SkinLayer}
\end{figure}

 In the standard setting of experimental investigation of 
the cyclotron resonance phenomenon we assume that the 
magnetic field $\, {\bf B} \, $ is parallel to the surface 
of a metal sample and we explore the behavior of the surface
current in the field of an incident electromagnetic wave. 
The interaction of electrons with the wave field occurs 
in a narrow skin layer which is usually much less than the 
size of the cyclotron orbit in the $\, {\bf x}$ - space. 
In generic case the long closed trajectories are inclined 
with respect to the surface of a sample in the 
$\, {\bf x}$ - space, so we have the picture
of electron motion through the skin layer schematically
represented at Fig. \ref{SkinLayer}. We can see, that in the 
$\, {\bf x}$ - space the electron trajectory has in general 
also a drift along the direction of $\, {\bf B} \, $. Let us
note here that for the long closed trajectories this drift is 
actually rather small since the average of the value
$\, v^{z}_{gr} \, $ is close to zero on the trajectories of 
this type. (Strictly speaking, this property can be in fact
violated for very complicated Fermi surfaces of very high
genera).

 According to standard theory of the cyclotron resonance
(see e.g. \cite{AzbelKaner,Kittel,etm,Ziman,Abrikosov}), 
every extremal closed trajectory gives an oscillating
term in the surface conductivity (in the situation of 
anomalous skin effect), which is caused by the coincidence 
of the frequency $\, \Omega \, $ of the incident wave with 
the values $\, 2 \pi n / T_{i} \, $, $\, n \in \mathbb{N} \, $,
where $\, T_{i} \, $ is the time of the electron motion
over the extremal closed trajectory. We can see then,
that every extremal closed trajectory gives an oscillating 
contribution $\, \Delta_{i} \, \sigma (\Omega) \, $ to the 
surface conductivity with the period
$$\Delta_{i} \Omega \,\,\, = \,\,\, 2 \pi \, T^{-1}_{i} $$ 

 Equivalently, these contributions can be considered as 
oscillating functions of $\, 1/B \, $ under fixed value of 
$\, \Omega \, $ having periods defined by the geometry of
closed extremal trajectories in the $\, {\bf p}$ - space.

 We can see, however, that in the case of the long 
closed trajectories, arising near the boundary of a
Stability Zone, all the trajectories belonging to the 
same layer (except trajectories which are very close 
to singular ones) correspond to almost the same
value of the period $\, T_{i} \, $. As a consequence,
the oscillating terms in this case are created actually
by the whole layers of the long closed trajectories
and represent the contributions of finite parts of the 
Fermi surface. In the simplest situation, described above, 
the layers of the long closed trajectories should
give three oscillating terms with three different periods,
satisfying the condition
$$\left( \Delta_{1} \Omega \right)^{-1} \,\,\, = \,\,\,
\left( \Delta_{2} \Omega \right)^{-1} \,\,\, + \,\,\,
\left( \Delta_{3} \Omega \right)^{-1} $$
or
$$\left( \Delta_{1} \left({1 \over B}\right) \right)^{-1}
\,\, = \,\,\,
\left( \Delta_{2} \left({1 \over B}\right) \right)^{-1}
\,\, + \,\,\,
\left( \Delta_{3} \left({1 \over B}\right) \right)^{-1} $$

 The oscillating terms given by the long closed 
trajectories should be actually added with the oscillating
terms coming from short extremal closed trajectories arising
on the Fermi surface. It is not difficult to see, however,
that the oscillations given by the long closed trajectories
should actually be observed at lower frequencies in comparison 
with the oscillations given by short extremal trajectories. 
Indeed, the first peaks of oscillations given by the long 
closed trajectories arise at the frequencies
$$\Omega_{1,2,3} \,\,\, = \,\,\, 2 \pi / T_{1,2,3} 
\,\,\, , $$
which are supposed to be much smaller than 
$\, \omega_{B} \, $ according to (\ref{InterCond}).

 Let us point out here also one more special feature
of the contribution of the long closed trajectories to 
the surface conductivity in strong magnetic fields. 
As we said already, the oscillations brought by the 
long closed trajectories are created by finite parts
of the Fermi surface. At the same time, the oscillations
created by short trajectories are given by the contributions 
of the extremal trajectories on the Fermi surface, satisfying 
the condition
$$d T_{i} / d p_{z} \,\,\, = \,\,\, 0 $$

 We can see then, that the oscillating 
terms, coming from the long closed trajectories, can
have actually bigger amplitude in comparison with 
standard situation due to the effect pointed above.
At the same time, we can see that the part of time
spent by electrons in the skin layer 
(Fig. \ref{SkinLayer}) should additionally be multiplied 
in this case by a small parameter $\, r_{B} / L \, $,
where $\, L \, $ is the length of the long closed
trajectory in the $\, {\bf x}$ - space. So, the two 
effects considered above actually play against each 
other in the situation represented at 
(Fig. \ref{SkinLayer}). Let us say also that in the 
region of the high frequencies 
$\, \Omega \, \sim \, \omega_{B} \, $ the first effect
disappears since the typical period variation on each
cylinder of closed trajectories has actually the order
$\, \Delta T_{i} \, \sim \, 1 / \omega_{B} \, $.
It it easy to see, that this variation has a small 
relative value in the range 
$\, 1 / \omega_{B} \, \ll \, 1 / \Omega $,
however, it becomes important if
$\, \Omega \, \sim \, \omega_{B} \, $.
As a result, we can expect that the oscillating terms 
given by long closed trajectories have rather small
amplitude (or almost invisible) in the region
$\, \Omega \, \sim \, \omega_{B} \, $.

 Another distinctive feature of the oscillations
brought by the long closed trajectories is that the change
of the oscillations picture does not have here a sharp form
with changes of the set of the corresponding periods
(say, $\, (T_{1}, T_{2}, T_{3})$) caused by reconstructions
of the layers of such trajectories under rotations of
the direction of $\, {\bf B}$. Indeed, it is not difficult
to see, that the contribution of a vanishing layer of
long closed trajectories tends to zero together with the 
area covered by the corresponding trajectories on the Fermi 
surface.  The situation is completely different for 
oscillating terms brought by extremal closed trajectories
since the trajectories of this kind remain almost unchanged
up to the disappearance of the corresponding cylinder of closed 
trajectories.

 Let us discuss also the situation when we still assume 
that in the region $\, \Omega \, \gtrsim \, \omega_{B} \, $
the oscillations generated by the long and the short 
closed trajectories can coexist, so we have here the 
oscillation picture given by the sum of the oscillations
of two kinds. It is easy to see, that the oscillations,
generated by the long closed trajectories, have much smaller 
period than the oscillations of the second type, so we
can actually easily separate these oscillating terms
from the terms generated by short closed trajectories.
Thus, any procedure of local averaging of oscillations
on the graph of $\, \sigma (\Omega) \, $ 
(or $\, \sigma (1/B)$) on an appropriate length 
will suppress the oscillations of the first origin 
and leave a clean picture of the oscillations 
of the second type. As a result, the same
scheme of determination of the second boundary of a
Stability Zone $\, \Omega_{\alpha} \, $, based on the 
pictures, analogous to Fig. \ref{Gen3OscPic}, 
\ref{Gen4OscPic}, can be applied also in the 
``intermediate situation'' (\ref{InterCond})
without significant changes. Let us note that the 
same is valid also in the situation of the presence
of ``internal boundaries'' within the domain
$\, \Omega^{\prime}_{\alpha} \, $ and appearance
of a bigger number of large periods
$\, (T_{1}, T_{2}, T_{3}, T_{4}, T_{5}) \, $
near the boundary of a Stability Zone.

\vspace{1mm}

\begin{figure}[t]
\begin{center}
\vspace{5mm}
\includegraphics[width=\linewidth]{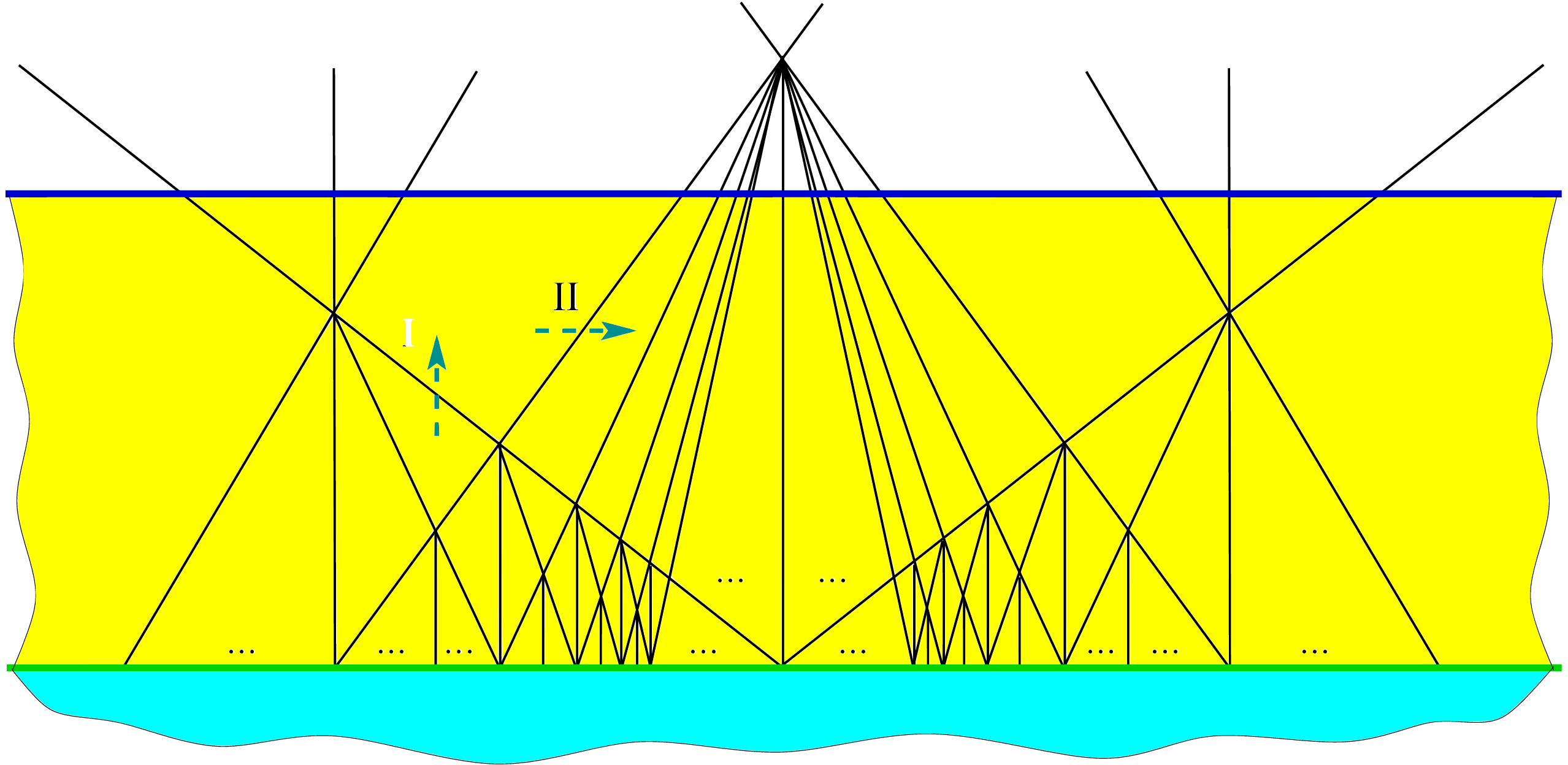}
\end{center}
\caption{The division of the domain 
$\, \Omega^{\prime}_{\alpha} \, $
into parts, corresponding to different values
of $\, T_{1}$, $\, T_{2}$, $\, T_{3}$,
between the first and the second boundaries
of a Stability Zone (very schematically).}
\label{ZonesNearBound}
\end{figure}

 The division of the domain 
$\, \Omega^{\prime}_{\alpha} \, $ into ``islands'',
corresponding to different values of 
$\, T_{1}$, $\, T_{2}$, $\, T_{3}$, is actually
rather complicated since the density of the 
``islands'' becomes infinite near the first boundary
of a Stability Zone (Fig. \ref{ZonesNearBound}). 
At the same time, we can see that the corresponding 
part of $\, \Omega^{\prime}_{\alpha} \, $ usually 
belongs to the ``experimentally observable''
Stability Zone $\, \hat{\Omega}_{\alpha} \, $, so
the corresponding long closed trajectories are
actually indistinguishable here from the open 
trajectories in the experiment.

 In that part of the domain 
$\, \Omega^{\prime}_{\alpha} \, $, which does not 
belong to the Zone $\, \hat{\Omega}_{\alpha} \, $,
the boundaries of the islands, corresponding 
to different values of $\, T_{1}$, $\, T_{2}$, 
$\, T_{3}$, can be detected approximately
by observing changes of the cyclotron resonance
picture in the frequency range, corresponding to
the contribution of the long closed trajectories.
As we have said already, the changes of the 
oscillations picture do not have here a sharp form
on the corresponding boundaries. In general, the 
values of $\, T_{1}$, $\, T_{2}$, $\, T_{3} \, $ 
are {\it apriori} unpredictable, still, we can say 
something about the behavior of these values when 
crossing the boundary between two different islands 
in $\, \Omega^{\prime}_{\alpha} \, $.

 Let us introduce now the ``vertical'' and the 
``horizontal'' deviations of the direction of
$\, {\bf B} \, $ on the angular diagram near the 
boundary of a Stability Zone. Namely, we will
call a deviation of a direction of $\, {\bf B} \, $
in the domain $\, \Omega^{\prime}_{\alpha} \, $
``vertical'' if it does not change the direction 
of the intersection of the plane, orthogonal
to $\, {\bf B} \, $, and the integral plane
$\, \Gamma_{\alpha} \, $ (arrow I at Fig.
\ref{ZonesNearBound}). Easy to see, that the 
``vertical'' deviations of the directions of
$\, {\bf B} \, $ do not change the mean directions
of the straight strips and can change just their 
widths (and the diameters of the circles) at the 
diagram \ref{PerNet}. It is easy to see also
that in the vicinity of the first boundary of
a Stability Zone the diameters of the circles
at Fig. \ref{PerNet} grow if we rotate the direction
of $\, {\bf B} \, $ away from the boundary 
of $\, \Omega_{\alpha} \, $ and decrease if we 
rotate $\, {\bf B} \, $ towards the boundary. 

 In the same way, we introduce the ``horizontal'' 
deviations of the direction of $\, {\bf B} \, $ in 
$\, \Omega^{\prime}_{\alpha} \, $, which do not 
change the diameters of the circles and can 
change the mean directions of the straight strips 
at the diagram \ref{PerNet} (arrow II at Fig.
\ref{ZonesNearBound}).

 Let us assume now that
$\, T_{1} > T_{2} > T_{3} \, $
(and $\, T_{1} \simeq T_{2} + T_{3} $)
in the vicinity of the first boundary of a 
Stability Zone. Then we can claim that every time 
when we cross a boundary between two islands in 
$\, \Omega^{\prime}_{\alpha} \, $
in a vertical deviation of $\, {\bf B} \, $
away from the boundary of the Zone
$\, \Omega_{\alpha} \, $ (arrow I at Fig. 
\ref{ZonesNearBound}) we will have one of the 
following simple transformations of the triple
$\, (T_{1}, T_{2}, T_{3})$:
$$(T_{1}, \, T_{2}, \, T_{3}) 
\,\,\, \rightarrow \,\,\,
(T_{2}, \, T_{2} - T_{3}, \, T_{3}) $$
(if $\, T_{2} - T_{3} \, > \, T_{3} $), or
$$(T_{1}, \, T_{2}, \, T_{3}) 
\,\,\, \rightarrow \,\,\,
(T_{2}, \, T_{3}, \, T_{2} - T_{3}) $$
(if $\, T_{3} \, > \, T_{2} - T_{3} $).

 In the same way, 
every time when we cross 
a boundary between two islands 
in a vertical deviation of $\, {\bf B} \, $
towards the boundary of the Zone
$\, \Omega_{\alpha} \, $ 
we should observe one of the following inverse 
transformations of the triple 
$\, (T_{1}, T_{2}, T_{3})$:
$$(T_{1}, \, T_{2}, \, T_{3}) 
\,\,\, \rightarrow \,\,\,
(T_{1} + T_{3}, \, T_{1}, \, T_{3}) $$
or
$$(T_{1}, \, T_{2}, \, T_{3}) 
\,\,\, \rightarrow \,\,\,
(T_{1} + T_{2}, \, T_{1}, \, T_{2}) $$

 It is easy to see that the values of 
$\, T_{i} \, $ decrease if we shift the direction 
of $\, {\bf B} \, $ away from the boundary of 
$\, \Omega_{\alpha} \, $ and increase if we 
shift it towards the boundary of a Stability
Zone. Let us say that this situation takes place
just in the vicinity of the first boundary
of a Stability Zone and can change at some distance
from $\, \Omega_{\alpha} \, $. In the last case
we can actually observe all the above 
transformations in any ``vertical'' deviation
of the direction of $\, {\bf B} \, $. It is not 
difficult to show also, that when 
we cross a boundary between two islands in 
$\, \Omega^{\prime}_{\alpha} \, $ in a horizontal
deviation of $\, {\bf B} $ (arrow II at Fig.
\ref{ZonesNearBound}) all the transformations
of the triple $\, (T_{1}, T_{2}, T_{3}) \, $
listed above become possible. 

 We can say again, that the picture represented above
keeps its key features also in the situation of 
the presence of ``internal boundaries'' within the 
domain $\, \Omega^{\prime}_{\alpha} \, $ and 
appearance of a bigger number of large periods
$\, (T_{1}, T_{2}, T_{3}, T_{4}, T_{5}) \, $ near 
the boundary of a Stability Zone. It can be also seen, 
that all the main features of classical oscillations
described above also take place for quantum oscillations 
(De Haas - Van Alphen or Shubnikov - De Haas
oscillations) in the described situation.

\vspace{1mm}

\begin{figure}[t]
\begin{center}
\vspace{5mm}
\includegraphics[width=0.95\linewidth]{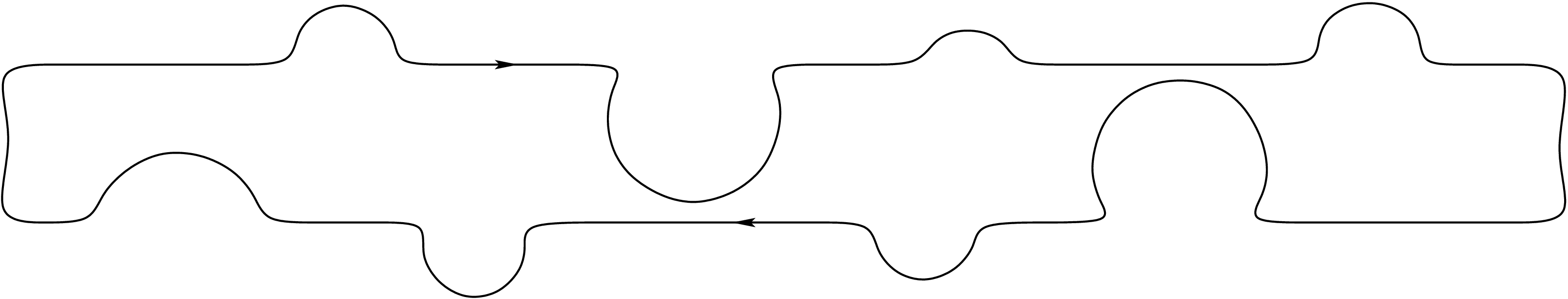}
\end{center}
\caption{More precise typical form of the long closed
trajectories arising near the boundary of a Stability
Zone.}
\label{PresForm}
\end{figure}

\begin{figure}[t]
\begin{center}
\vspace{1cm}
\includegraphics[width=0.95\linewidth]{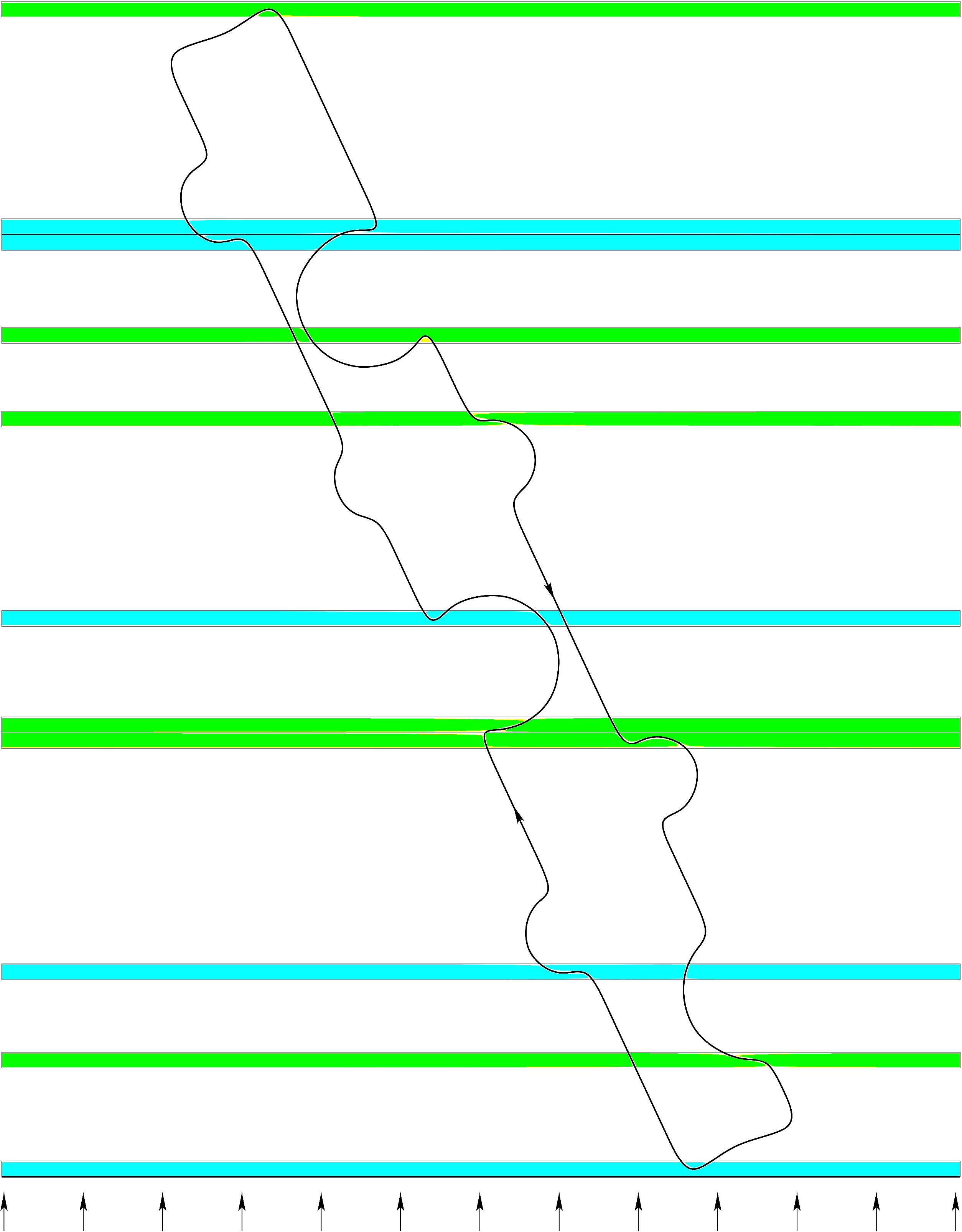}
\end{center}
\caption{The current layers arising at different distances
from the sample boundary in the presence of the long
closed trajectories.}
\label{PresFormSkinLay}
\end{figure}

 Among other features of the cyclotron resonance
picture specific to the above situation we can indicate
deep penetration of electric current and electromagnetic
field inside the metal in the form of ``splashes''
at rather big distances from the surface. Indeed,
due to the special form of the trajectories shown
at Fig. \ref{SkinLayer} we should observe the formation
of the second (and the next) current layers at the distances
of order of $\, L \, $ which is supposed to be
much bigger than the value of $\, r_{B} \, $. Let us say
also that the formation of the current layers in the presence
of the long closed trajectories has in fact many other
special features. Indeed, the precise form of the long
closed trajectories is complicated enough and can be
schematically represented by Fig. \ref{PresForm}.
According to standard kinetic approach, every point
of a trajectory where the velocity of electron is parallel
to the boundary of the metal sample creates its own
current layer at the correspondent distance from the
sample boundary (see e.g. \cite{etm,Abrikosov}).
We can see then, that in the case of presence of the
trajectories shown at Fig. \ref{PresForm} we should have
rather complicated structure of many current layers inside
the metal at rather big distances from its boundary
(Fig. \ref{PresFormSkinLay}). This phenomenon
can be detected in experimental study of the
``size effects'', such as the cutoff of the cyclotron
resonance orbits in thin metal films in strong magnetic
fields (see e.g. \cite{etm,Abrikosov}).

\begin{figure}[t]
\begin{center}
\vspace{5mm}
\includegraphics[width=\linewidth]{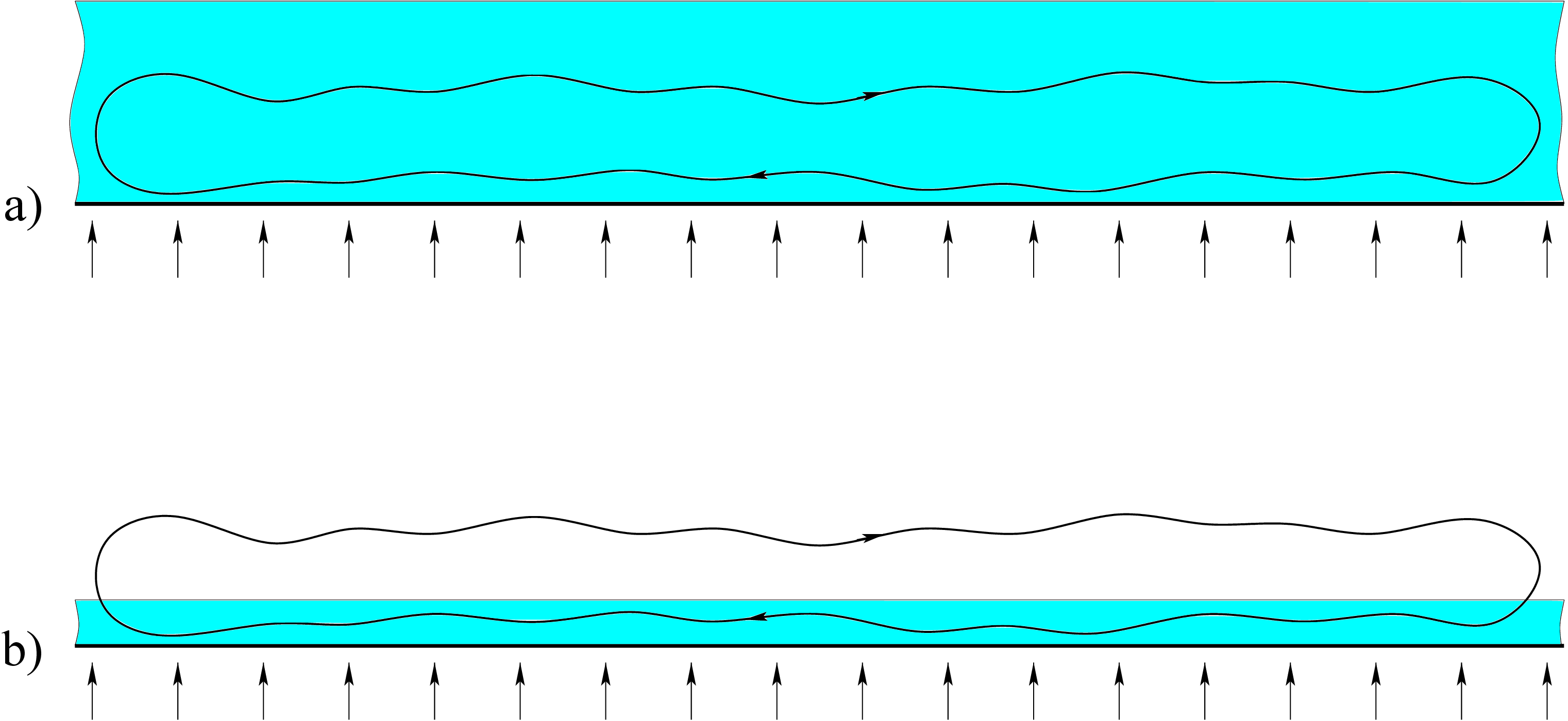}
\end{center}
\caption{A long closed trajectory (projection on the 
plane orthogonal to $\, {\bf B}$) oriented along
the boundary of a metal sample in the 
$\, {\bf x}$ - space. The situations 
$\, \delta > r_{B} \, $ (a) and 
$\, \delta < r_{B} \, $ (b).}
\label{SkinLayerSpecial}
\end{figure}

 Let us now make one more remark about the cyclotron
resonance on long closed trajectories satisfying the 
condition (\ref{InterCond}). Namely, we can consider
also the situation when the direction of the open 
orbits in $\, \Omega_{\alpha} \, $ is parallel
to the surface of the metal sample in the 
$\, {\bf x}$ - space. This will mean automatically
that the long closed trajectories in the domain
$\, \Omega^{\prime}_{\alpha} \, $ will be also 
oriented along the boundary of the metal sample
in the $\, {\bf x}$ - space 
(Fig. \ref{SkinLayerSpecial}), which will give some
additional features in the picture of oscillations.

\begin{figure}[t]
\begin{center}
\vspace{1cm}
\includegraphics[width=\linewidth]{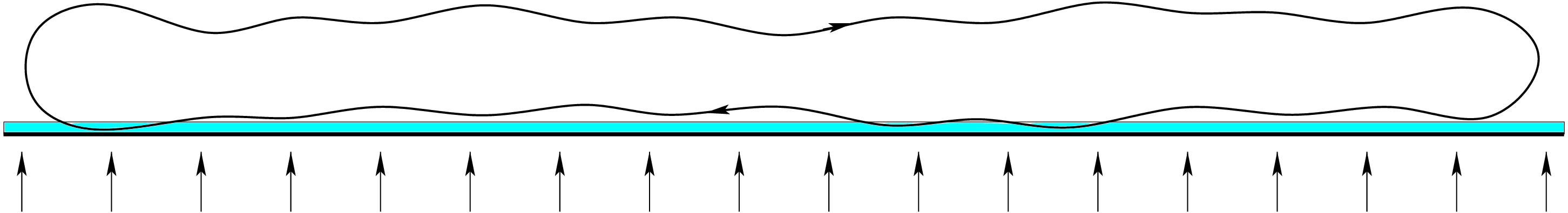}
\end{center}
\caption{A long closed trajectory oriented along
the boundary of a metal sample in the case  
$\, \delta \ll r_{B} \, $.}
\label{SkinLayerNarrow}
\end{figure}

 Let us note first of all, that the width of the 
skin layer depends on the frequency of the incident
wave and we consider here the frequencies corresponding
to the periods $\, T_{i} \, $, which are much lower
than the cyclotron frequency $\, \omega_{B} \, $. 
As a theoretical possibility, we could consider here 
a skin layer of a width $\, \delta \, $, which is 
compatible with the standard cyclotron radius 
$\, r_{B} \, = \, m^{*} v_{gr} / eB \, $ in a metal 
(Fig. \ref{SkinLayerSpecial}). From the standard
kinetic theory of the cyclotron resonance we get then
that in the case, shown at Fig. \ref{SkinLayerSpecial} a,
the first resonance peak ($\Omega = 2\pi / T_{i}$) in 
every oscillating term should have much bigger amplitude
compared with the other peaks. In the situation,
represented at Fig. \ref{SkinLayerSpecial} b,
we can claim that in every (of the three) oscillating 
term the even ($\Omega = 4\pi n / T_{i}$) resonance
peaks will be strongly suppressed in compare with the
odd ($\Omega = 2\pi (2n-1) / T_{i}$) resonance peaks.
Let us say, however, that the situation represented
at Fig. \ref{SkinLayerSpecial} is actually rather
unlikely in metals and we should expect much more 
stronger relation $\, \delta \ll r_{B} \, $ in the 
experiment. In this case we should assume in fact that 
only some ``pieces'' of a trajectory fall into the skin 
layer, while its most part lies outside the skin 
layer (Fig. \ref{SkinLayerNarrow}).

\begin{figure}[t]
\begin{center}
\vspace{5mm}
\includegraphics[width=\linewidth]{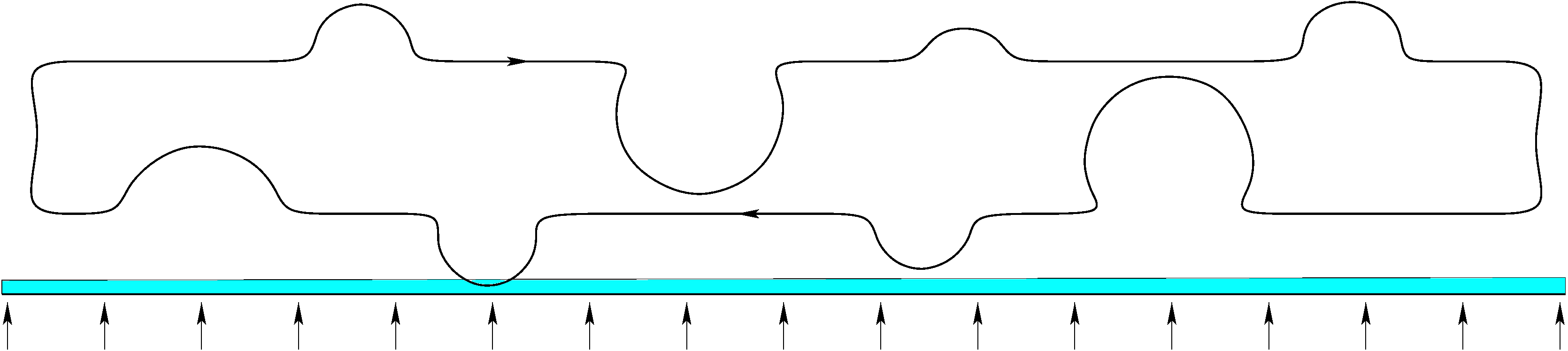}
\end{center}
\caption{A short piece of a long closed trajectory
falling into the skin layer near the boundary of 
a metal sample.}
\label{OneShortPiece}
\end{figure}

 Moreover, coming back to the precise form of the 
long closed trajectories (Fig. \ref{PresForm}),
we can assume in fact that the skin layer contains
just one very short piece of a trajectory for most
of the long closed trajectories 
(Fig. \ref{OneShortPiece}). As a result, the frequency
analysis of the dependence $\, \sigma (\Omega) \, $
in the range $\, \Omega \, \sim \, 2\pi / T_{i} \, $
should actually show here the same picture as in the 
case of ``inclined'' long closed trajectories 
(Fig. \ref{SkinLayer}). At the same time, the picture
of the current layers near the boundary of a sample
should be different here from generic case, since all 
the zeros of the value $\, v^{\perp}_{gr} \, $ are 
located now at a distance of the order of $\, r_{B} \, $
from the boundary. This difference, in particular,
can be detected in the study of the ``size effects''
in thin metal films, mentioned above. 

\vspace{1mm}

\begin{figure}[t]
\begin{center}
\vspace{1cm}
\includegraphics[width=0.9\linewidth]{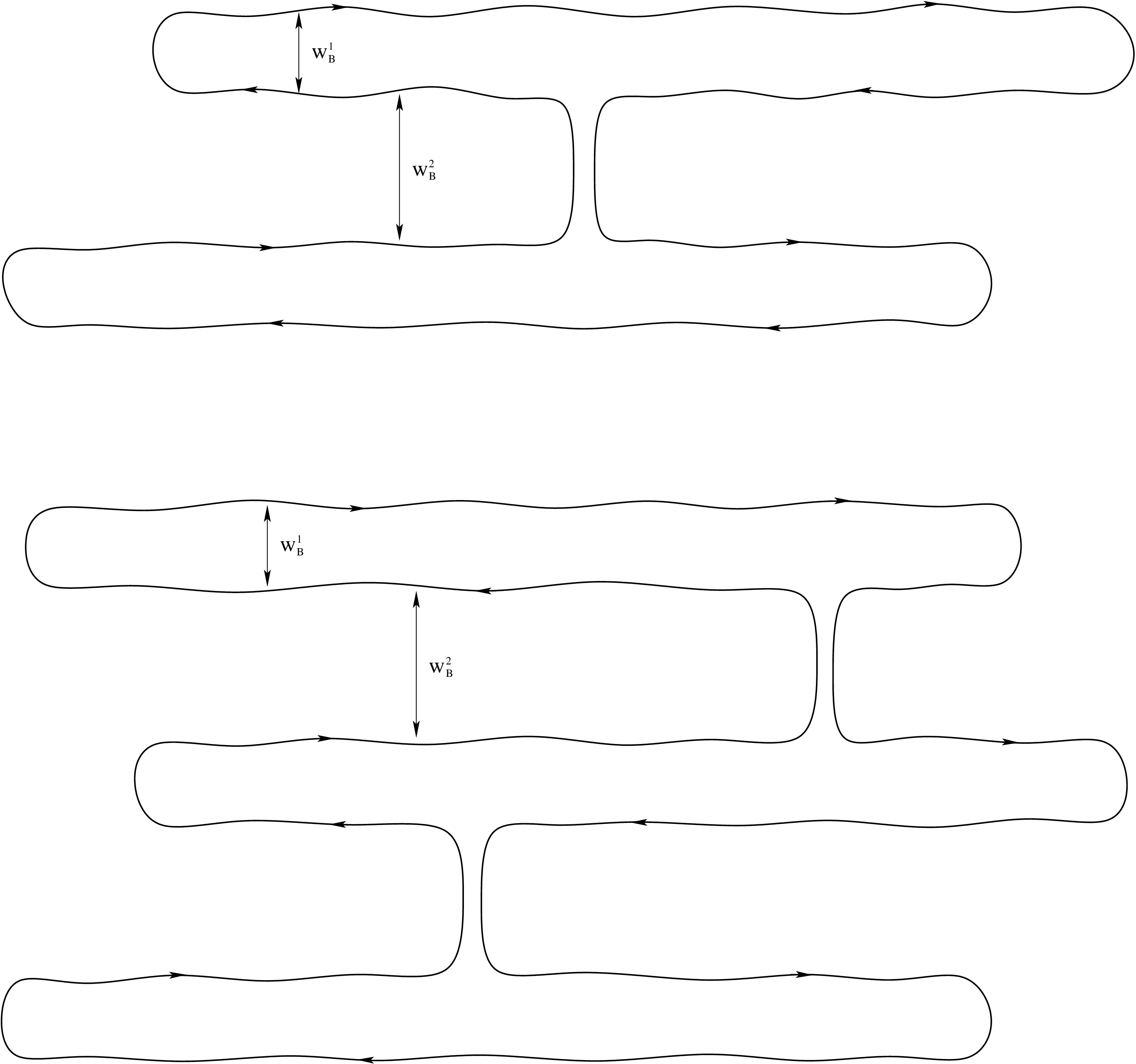}
\end{center}
\caption{Complex long closed trajectories arising after
crossing the second boundary of a Stability Zone.}
\label{CombLongTraject}
\end{figure}

 As we told already, one of the properties of
the domain $\, \Omega^{\prime}_{\alpha} \, $ 
is that we can not have stable open trajectories
on the Fermi surface for 
$\, {\bf B}/B \in  \Omega^{\prime}_{\alpha} \, $.
Let us say here, that stable open trajectories 
do not appear also just after crossing the 
second boundary of a Stability Zone in generic
points. Thus, for generic directions of
$\, {\bf B} \, $ we should first observe the 
appearance of the ``complex long closed trajectories''
(Fig. \ref{CombLongTraject}) after crossing the 
boundary of $\, \Sigma_{\alpha} \, $, which can
at the end be transformed into open trajectories
(stable or unstable).

 In general case we can state that the 
crossing of the second boundary of a Stability Zone leads 
to (abrupt) disappearance of one oscillating term in the 
frequency range $\, \Omega \sim \omega_{B} \, $
(Fig. \ref{Gen3OscPic}, \ref{Gen4OscPic}) and appearance
of an additional oscillating term, corresponding to appearance 
of the new period $\, T \, \simeq \, T_{i_{1}} + T_{i_{2}}\, $, 
in the range of ``low'' frequencies. 

\vspace{1mm}

\begin{figure}[t]
\begin{center}
\vspace{5mm}
\includegraphics[width=\linewidth]{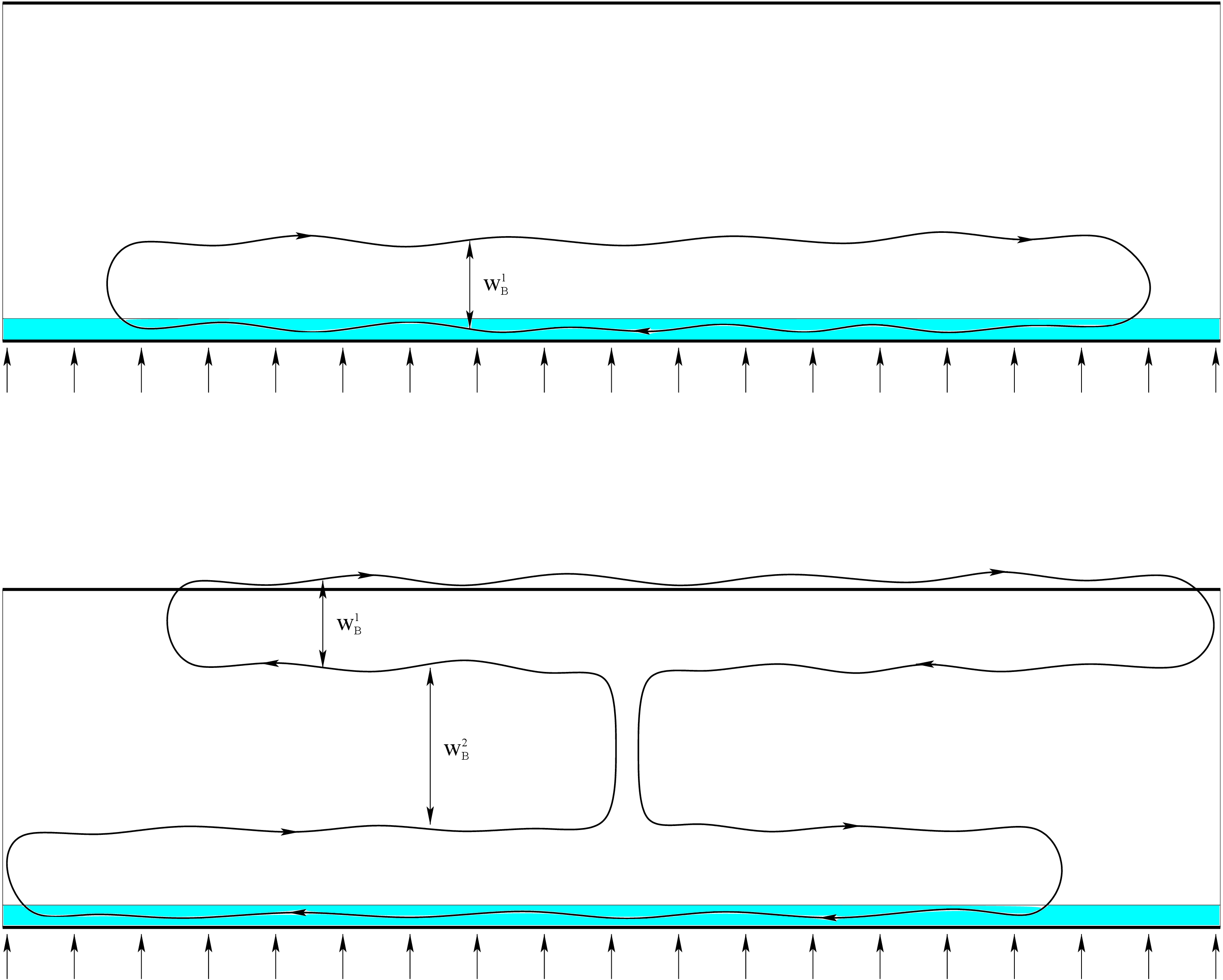}
\end{center}
\caption{The cutoff of a cyclotron resonance
orbit after crossing the second boundary of a Stability 
Zone in the case when the direction of 
the open trajectories in $\, \Omega_{\alpha} \, $ is 
parallel to the surface of the metal sample.}
\label{CutOff}
\end{figure}

 One of the special features arising in the presence of
the complex closed trajectories when the direction of 
the open trajectories in $\, \Omega_{\alpha} \, $ is 
parallel to the surface of a metal sample is the 
formation of rather complicated picture of current
splashes inside the metal sample. Another obvious feature 
observable in the study of size effects is a rapid change 
of the sample thickness corresponding to the cutoff of the 
cyclotron resonance orbits after crossing of the second 
boundary of a Stability Zone (Fig. \ref{CutOff}). In this
situation for appropriate thickness of the sample
we should observe the disappearance of
an oscillating term in the range of the high 
frequencies $\, \Omega \sim \omega_{B} \, $
without appearance of any oscillating 
term in the range of the ``low'' frequencies
just after crossing of the second boundary of
$\, \Omega_{\alpha} \, $.

\vspace{1mm}

\begin{figure}[t]
\begin{center}
\includegraphics[width=0.9\linewidth]{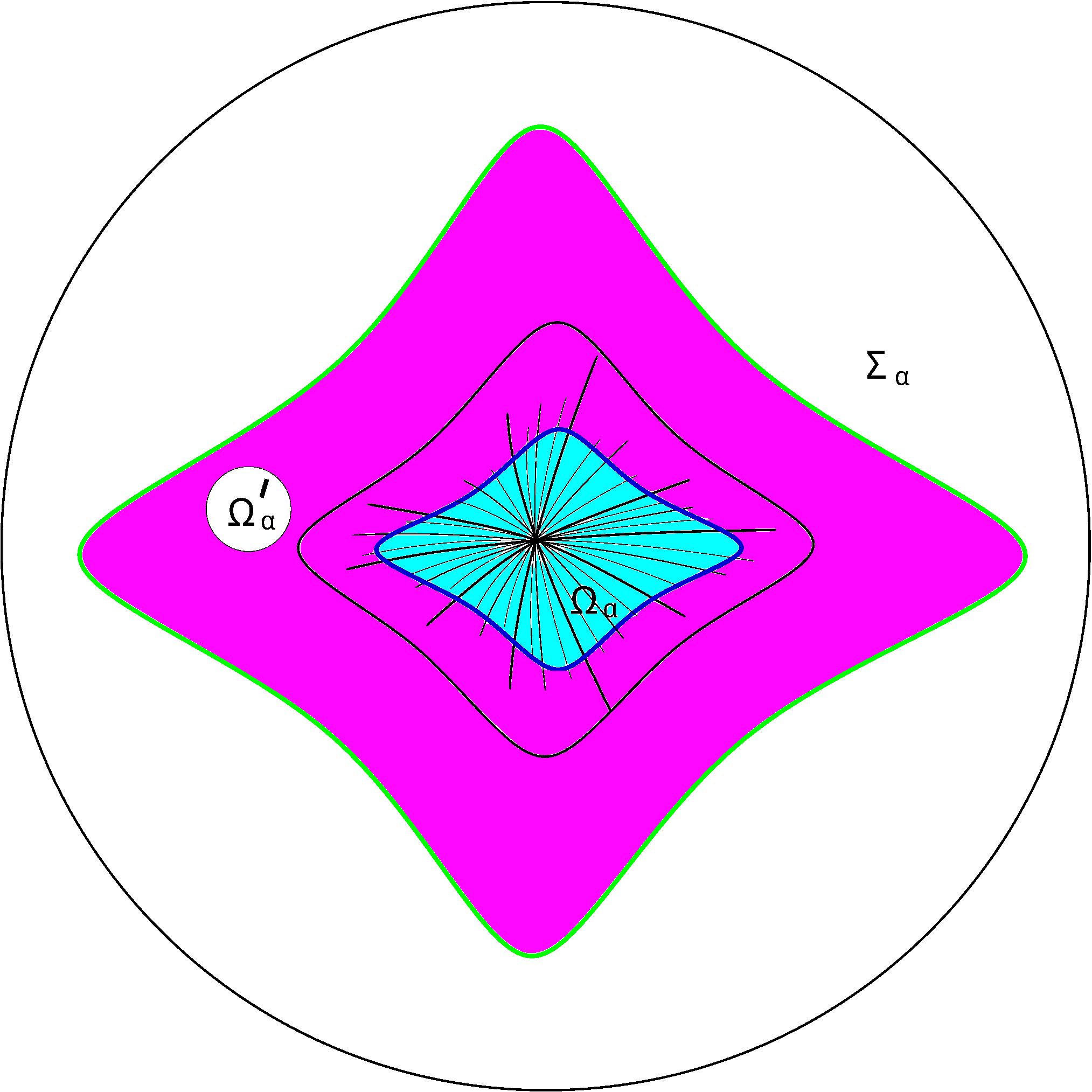}
\end{center}
\caption{A large domain
$\, \Omega^{\prime}_{\alpha} \,\,\, = \,\,\, 
\Sigma_{\alpha} \backslash \Omega_{\alpha} \,$
(pink) on the angular diagram (schematically).}
\label{LargeRegion}
\end{figure}

 Finally, let us consider the situation when the 
second boundary of a Stability Zone is far from 
its first boundary (Fig. \ref{LargeRegion}). It is easy 
to see, that in this case we can have the situation when 
the ``long closed trajectories'' get the ``normal size''
($\sim p_{F}$) in the $\, {\bf p}$ - space.
The oscillating terms coming from all closed 
trajectories appear now in the same frequency range,
so we have in general a more complicated oscillations
picture than those shown at Fig. \ref{Gen3OscPic},
\ref{Gen4OscPic} near the second boundary of a Stability 
Zone. All the oscillating terms are brought now by extremal 
closed trajectories, so the oscillations picture undergoes 
here a rapid transformation at every boundary between two 
``islands'' in the domain  
$\, \Omega^{\prime}_{\alpha} \backslash \hat{\Omega}_{\alpha} \, $
as well as at the second boundary of a Stability Zone
(and also behind the second boundary). Most probably,
in this situation the most convenient way to distinguish 
the second boundary of a Stability Zone from a boundary 
between two islands is the study of the cutoff of the 
cyclotron resonance orbits in thin metal samples.
Thus, for appropriate thickness of the sample 
(Fig. \ref{CutOff}) we should observe a disappearance
of one oscillating term after crossing the second
boundary of a Stability Zone if the mean direction
of the open trajectories in $\, \Omega_{\alpha} \, $ 
is parallel to the surface of the metal sample.
At the same time, after crossing a boundary 
between two islands inside the domain
$\, \Omega^{\prime}_{\alpha} \, $ we should observe
a rapid change of the periods $\, T_{i} \, $
of the oscillating terms without change of their number.

\vspace{1mm}

 We should certainly point out one more general feature
specific to the conductivity behavior in the domain
$\, \Omega^{\prime}_{\alpha} \, $. Let us come back now
to the formula (\ref{Closed}) giving the asymptotic
behavior of the conductivity tensor $\, \sigma^{kl} (B) \, $
inside a metal sample in constant electric field
($\Omega = 0$). Formula (\ref{Closed}) gives the 
contribution of closed trajectories to the conductivity
for $\, \omega_{B} \tau \, \gg \, 1 \, $ and can be used
also in that part of the domain
$\, \Omega^{\prime}_{\alpha} \, $
where the conditions  \linebreak
$\, \tau \, \gg \, T_{i} \, $
are satisfied. The conductivity tensor in the plane
orthogonal to $\, {\bf B} \, $ can be written here in the 
form
\begin{multline*}
\sigma^{\alpha\beta} \,\,\,\, = \,\,\,\,
{n e^{2} \tau \over m^{*}} \, \left(
\begin{array}{cc}
c^{11} \, ( \omega_{B} \tau )^{-2} \, &  
\, c^{12} \, ( \omega_{B} \tau )^{-1}  \cr
- \, c^{12} \, ( \omega_{B} \tau )^{-1} \, &  
\, c^{22} \, ( \omega_{B} \tau )^{-2} 
\end{array}  \right) \,\, ,   \\
\omega_{B} \tau \,\, \rightarrow \,\, \infty  \,\,\, , 
\end{multline*}
where $\, c^{\alpha\beta} \, $ represent some dimensionless
constants. The constants $\, c^{\alpha\beta} \, $ have the 
order of 1 for closed trajectories of normal size, however,
they can have very special values for complicated extended
trajectories. In particular, for the long closed trajectories
shown at Fig. \ref{ThreeTraject}, \ref{PresForm} the values
of $\, c^{11} \, $ and $\, c^{22} \, $ differ significantly
if the $\, x $ - axis is chosen along the intersection
of the plane $\, \Gamma_{\alpha} \, $ and the plane, 
orthogonal to $\, {\bf B} $. As can be shown by a simple
analysis, the value of $\, c^{22} \, $ has in this case
the order of $\, (\omega_{B} T)^{2} \, \gg \, 1 \, $,
while the constant $\, c^{11} \, $ has the order of 1.
We can see then, that the tensor $\, \sigma^{kl} (B) \, $
keeps in general anisotropic structure in the plane
orthogonal to $\, {\bf B} \, $ in the domain
$\, \Omega^{\prime}_{\alpha} \, $ with the same geometric
properties as in the Zone $\, \Omega_{\alpha} \, $.
As can be also seen, the values of $\, c^{22} \, $
can vary widely within the domain 
$\, \Omega^{\prime}_{\alpha} \, $ while the constant
$\, c^{11} \, $ keeps almost the same value in this domain.
Thus, the second boundary of a Stability Zone
$\, \Omega_{\alpha} \, $ can be roughly defined also as
the boundary beyond which the conductivity along the 
$\, x $ - direction in our special coordinate system
can start to change substantially.

\vspace{1mm}

\begin{figure}[t]
\begin{center}
\includegraphics[width=\linewidth]{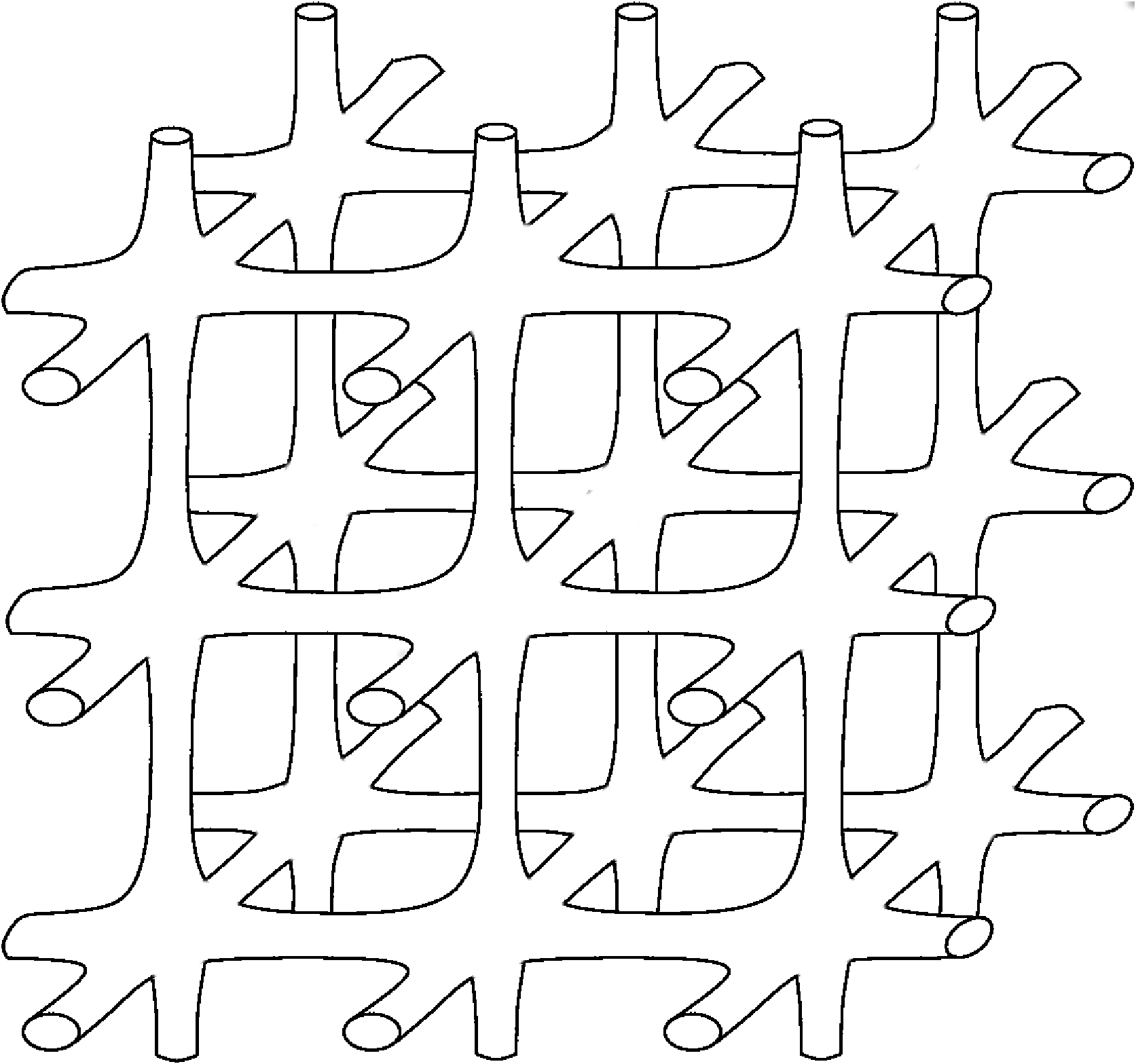}
\end{center}
\begin{center}
\vspace{5mm}
\includegraphics[width=0.9\linewidth]{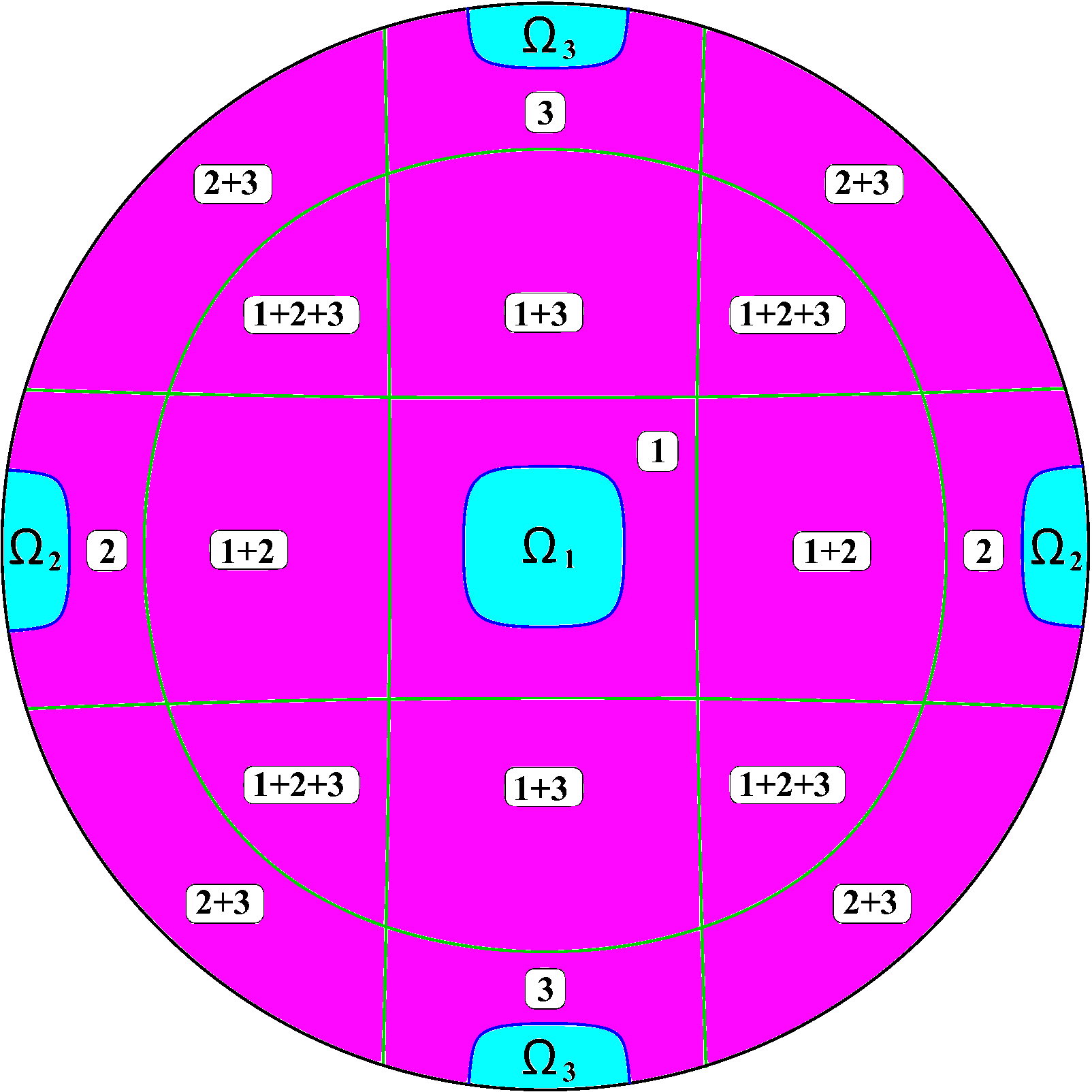}
\end{center}
\caption{The Fermi surface in the form of ``thin spatial net''
and the corresponding angular diagram.}
\label{ThinNet}
\end{figure}

 Let us say now that, in contrast to Stability Zones
$\, \Omega_{\alpha} \, $, different domains 
$\, \Omega^{\prime}_{\alpha} \, $ can overlap on the angular 
diagram. The corresponding example can be easily constructed
if we consider the Fermi surface in the form of the
``thin spatial net'' and the corresponding angular diagram
for the magneto-conductivity (Fig. \ref{ThinNet}). It is
not difficult to see that the angular diagram contains in this
case six rather small regions that represent three
Stability Zones $\, \Omega_{1,2,3} \, $
and six rather large connected regions representing the 
domains $\, \Omega^{\prime}_{1,2,3} \, $. 
The domains $\, \Omega_{\alpha} \, $ and
$\, \Omega^{\prime}_{\alpha} \, $ cover the whole
unit sphere and we can easily show the regions on 
$\, \mathbb{S}^{2} \, $, which are covered by two or even 
three different domains $\, \Omega^{\prime}_{\alpha} \, $ 
simultaneously. The closed trajectories of system 
(\ref{MFSyst}) can be described in this case in different 
ways according to different pictures, corresponding to 
different representations of the Fermi surface. 
In general, the long closed trajectories in the region, 
covered by two different domains 
$\, \Omega^{\prime}_{\alpha} \, $,
$\, \Omega^{\prime}_{\beta} \, $ can be extended just 
in one integer direction (in $\, {\bf p}$ - space),
given by the intersection of the planes
$\, \Gamma_{\alpha} \, $ and $\, \Gamma_{\beta} \, $.
In the same way, in the region covered by three
(or more) domains $\, \Omega^{\prime}_{\alpha} \, $,
we should not expect closed trajectories extended in 
one special direction if the corresponding planes
$\, \Gamma_{\alpha} \, $ do not have a common integer
vector in $\, {\bf p}$ - space. It is not difficult 
to see that the latter situation can arise only at a 
considerable distance from the boundaries of the 
Stability Zones, so that the long closed trajectories 
have disappeared as a result of the reconstructions 
described above.

Let us note here, that we can
also have in general the situation where the second
boundary is actually absent on the angular diagram.
This will mean actually, that the Zone 
$\, \Sigma_{\alpha} \, $ covers the whole sphere
$\, \mathbb{S}^{2} \, $ and the Stability Zone
$\, \Omega_{\alpha} \, $ represents the only Stability 
Zone on the angular diagram. It is not difficult 
to point also a situation where the domain
$\, \Omega^{\prime}_{\alpha} \, $ connects the opposite 
parts of the Zone $\, \Omega_{\alpha} \, $ but the 
corresponding Zone $\, \Sigma_{\alpha} \, $ does not 
cover the whole sphere $\, \mathbb{S}^{2} \, $.

\section
{The boundaries of Stability Zones and the global
structure of the angular diagram.}
\setcounter{equation}{0}

 In this section we will consider some features of
the global structure of the angular diagram and its connection 
with the types of the boundaries of Stability Zones.
Our consideration will be of a somewhat topological nature and 
will be related to a large extent with the behavior of the Hall 
conductivity on the angular diagram. 
 
  Let us say that the Hall conductivity in metals is traditionally 
associated with the concentration and the type of current carriers 
in a metal, in particular, it can have positive or negative sign 
depending on the type of the Fermi surface. In most cases this 
classification of carriers is actually used for metals in the 
presence of an external magnetic field in the case when only 
closed trajectories are present on the Fermi surface. We will 
need, however, somewhat more precise formulations about the 
behavior of the Hall conductivity in this situation, so let us 
consider this question in more detail below.

 Let us assume here that the Fermi surface is represented
by a smooth (connect or disconnect) 3-periodic surface 
in the $\, {\bf p}$ - space, given by the equation
\begin{equation}
\label{FermiSurface}
\epsilon ({\bf p}) \,\,\, = \,\,\, \epsilon_{F}
\end{equation}
for some periodic function 
$\, \epsilon ({\bf p}) \, $. 

\begin{figure}[t]
\begin{center}
\vspace{5mm}
\includegraphics[width=\linewidth]{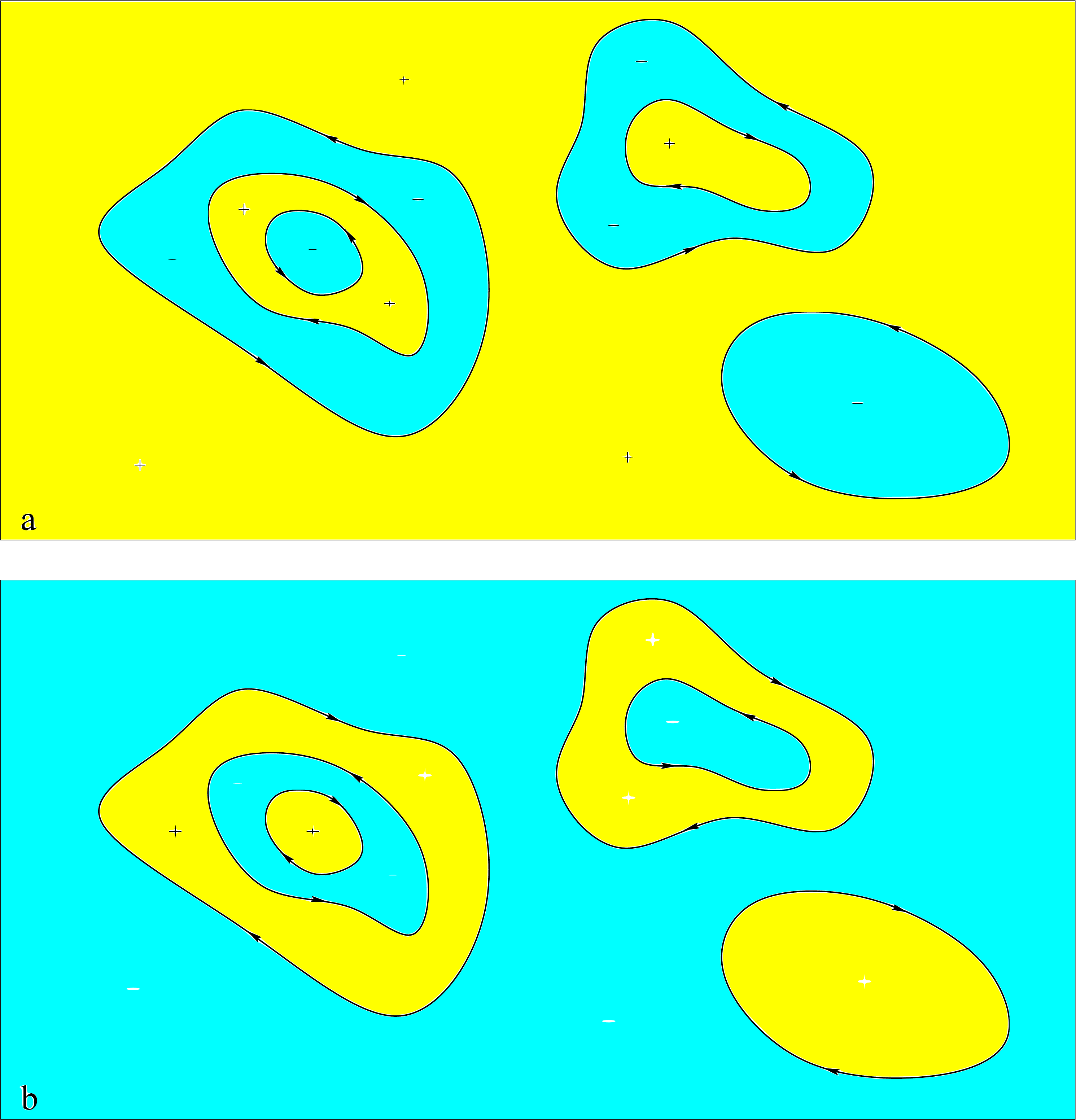}
\end{center}
\caption{Two possible pictures of electron trajectories
in the planes orthogonal to $\, {\bf B} \, $ if
the Fermi surface contains only closed trajectories
in the $\, {\bf p}$ - space.}
\label{PlaneSea}
\end{figure}

 Suppose now that the Fermi surface (\ref{FermiSurface})
contains only closed (in $\, {\bf p}$ - space)
trajectories of system (\ref{MFSyst})
for some fixed direction of $\, {\bf B} \, $.
Let us also assume that the direction of $\, {\bf B} \, $
is not purely rational, i.e. the plane orthogonal
to $\, {\bf B} \, $ is not generated by two reciprocal
lattice vectors in $\, {\bf p}$ - space.
In this case in all the planes orthogonal to 
$\, {\bf B} \, $ we will have one of the following 
situations:

\vspace{1mm}

I) The region of the higher values of energy
$\, \epsilon ({\bf p}) \, > \, \epsilon_{F} \, $
represents a ``sea'' in the plane, orthogonal to
$\, {\bf B} \, $, while the regions of the lower
energy $\, \epsilon ({\bf p}) \, < \, \epsilon_{F} \, $
represent finite ``islands'' in this sea.
(The ``islands'' can contain ``lakes'' 
of the higher values of energy etc. 
Fig. \ref{PlaneSea}, a).

\vspace{1mm}

II) The region of the lower values of energy
$\, \epsilon ({\bf p}) \, < \, \epsilon_{F} \, $
represents a ``sea'' in the plane, orthogonal to
$\, {\bf B} \, $, while the regions of the higher
energy $\, \epsilon ({\bf p}) \, > \, \epsilon_{F} \, $
represent finite ``islands'' in this sea.
(The ``islands'' can contain ``lakes'' 
of the lower values of energy etc. 
Fig. \ref{PlaneSea}, b).

\vspace{1mm}

 The type of the picture (I or II) does not depend on
the plane orthogonal to $\, {\bf B} \, $ and is locally 
stable with respect to small rotations of 
$\, {\bf B} \, $ under the requirements formulated
above. At the same time, the picture can change
after a big deviation of the direction of $\, {\bf B} \, $
for complicated Fermi surfaces. We can see then that in 
general case we can select areas on the angular diagram,
corresponding to the presence of only closed trajectories 
on the Fermi surface and described by one of the two 
pictures above. It is not difficult to see also that
any two regions corresponding to different types 
of the pictures described above should be separated
by Zones corresponding to the presence of open
trajectories on the Fermi surface.

 If all the trajectories of system (\ref{MFSyst})
are closed, the conductivity tensor 
$\, \sigma^{kl} (B) \, $ can be represented in the 
form of regular series in powers of 
$\, (\omega_{B} \tau)^{-1} \, $ in the limit
$\, \omega_{B} \tau \, \rightarrow \, \infty \, $
(see e.g. \cite{etm, Ziman, Abrikosov}),
and we can write for its main terms 

\begin{multline}
\label{ClosedTraj}
\sigma^{\alpha\beta} \,\,\, = \,\,\, \left(
\begin{array}{cc}
a^{11} \, (\omega_{B} \tau )^{-2}  \, &  
\, a^{12} \, ( \omega_{B} \tau )^{-1}  \cr
- a^{12} \, ( \omega_{B} \tau )^{-1}  \, &  
\, a^{22} \, ( \omega_{B} \tau )^{-2}
\end{array}  \right) \,\, ,  \\
\omega_{B} \tau \, \rightarrow \, \infty
\end{multline}
in the plane orthogonal to $\, {\bf B} \, $.

\vspace{1mm}

 For the value 
$\, \sigma^{12} \, = \, 
a^{12} \, ( \omega_{B} \tau )^{-1} \, $ 
we can use the formula
\begin{equation}
\label{SigmaNeNh}
\sigma^{12} \,\,\, = \,\,\, {ec \over B} \, 
\left( n_{e} - n_{h} \right) 
\end{equation}
where $\, n_{e} \, $ and $\, n_{h} \, $
represent the ``electron concentration''
and the ``hole concentration'' respectively.
Taking into account the spin variables, the values
$\, n_{e} \, $ and $\, n_{h} \, $ can be defined as
$$n_{e} \,\,\, = \,\,\, 2 V_{e} / (2\pi \hbar)^{3} 
\quad , \quad \quad
n_{h} \,\,\, = \,\,\, 2 V_{h} / (2\pi \hbar)^{3} $$
where $\, V_{e} \, $ and $\, V_{h} \, $ are the volumes
restricted by all nonequivalent cylinders of the 
electron-type and the hole-type closed trajectories
in the $\, {\bf p}$ - space.

\vspace{1mm}

 After a simple analysis, we can see actually
that in the situation described above we can
write the formula for $\, \sigma^{12} \, $ 
in one of the following forms:
\begin{equation}
\label{Sigma12Elekt}
\sigma^{12} \,\,\, = \,\,\, 
{2 ec \over (2\pi \hbar)^{3} B} \,\,\,  V_{-}
\quad \quad {\rm (situation \,\,\, I)}
\end{equation}
\begin{equation}
\label{Sigma12Hole}
\sigma^{12} \,\,\, = \,\,\, 
- \, {2 ec \over (2\pi \hbar)^{3} B} \,\,\,  V_{+}
\quad \quad {\rm (situation \,\,\, II)}
\end{equation}
where $\, V_{-} \, $ and $\, V_{+} \, $ are the 
volumes defined by the conditions 
$\, \epsilon ({\bf p}) \, < \, \epsilon_{F} \, $
and 
$\, \epsilon ({\bf p}) \, > \, \epsilon_{F} \, $
in the Brillouen zone.

\vspace{1mm}

 Thus, we can attribute even an extended Fermi surface 
to the electron or the hole type if it contains only 
closed trajectories of system (\ref{MFSyst}) and this 
property is stable under small rotations of 
$\, {\bf B} \, $. The value of $\, \sigma^{12} B \,  $ 
in strong magnetic fields is constant for a given type of 
the Fermi surface and is defined by the formula 
(\ref{Sigma12Elekt}) or (\ref{Sigma12Hole}). Let us note 
again that for extended Fermi surfaces their type can 
depend on the direction of $\, {\bf B} \, $ and is not 
defined if the Fermi surface contains open trajectories 
of system (\ref{MFSyst}).

\vspace{1mm}

\begin{figure}[t]
\begin{center}
\vspace{5mm}
\includegraphics[width=0.9\linewidth]{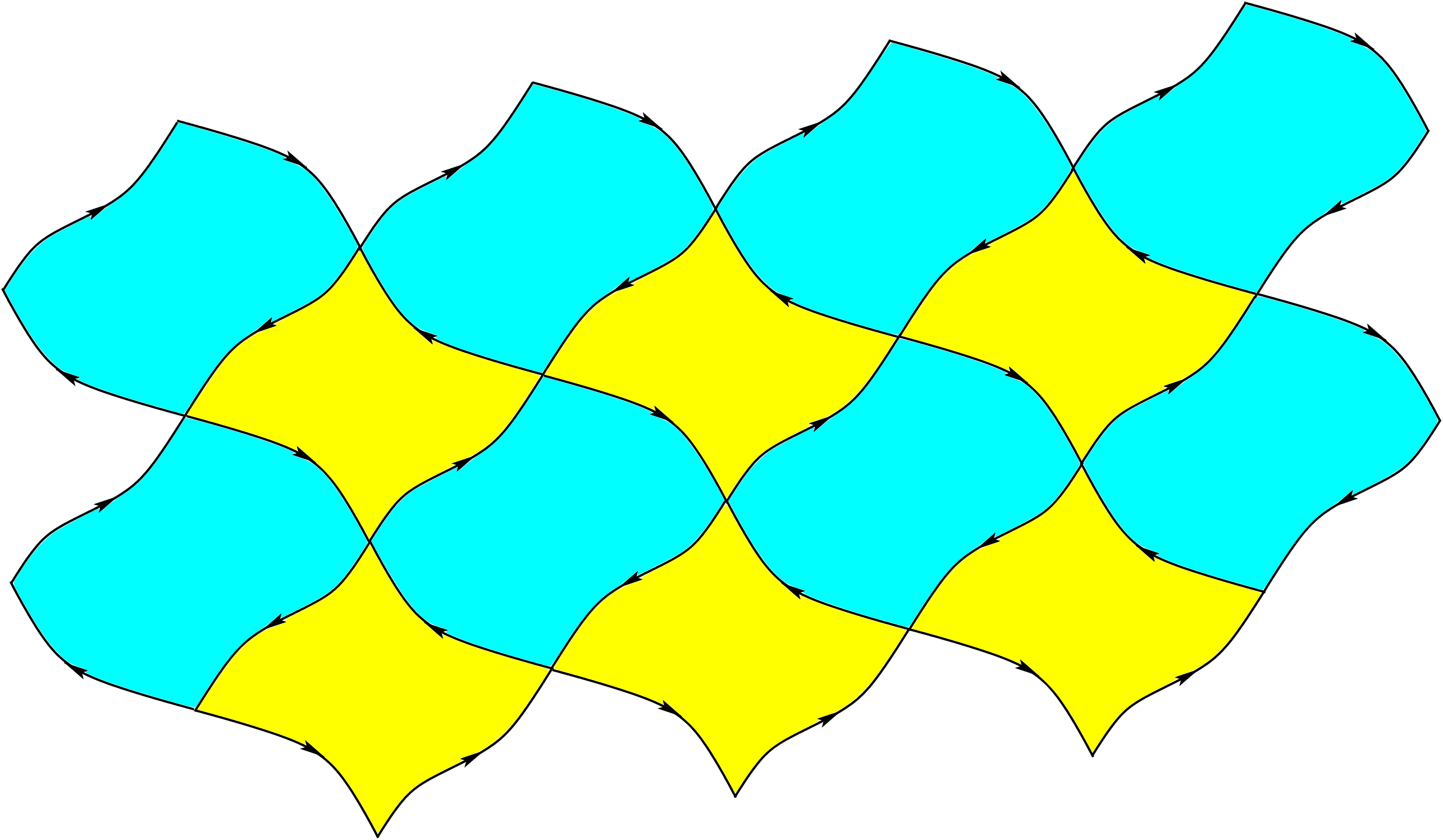}
\end{center}
\caption{A periodic net of singular trajectories 
which can arise for purely rational directions of 
$\, {\bf B} \, $.}
\label{PeriodicNet}
\end{figure}

 Let us say now that the picture described above can be
incorrect if the direction $\, {\bf B}_{0} \, $ is
purely rational. Namely, in this case we can have the 
situation when the Fermi surface contains only closed 
trajectories of system (\ref{MFSyst}), however, the types 
of the pictures arising in the planes orthogonal to 
$\, {\bf B}_{0} \, $ can be different in different planes. 
The ``layers'' of the planes, corresponding to different 
types (I or II), are separated by special planes in the
$\, {\bf p}$ - space, containing ``periodic nets''
of singular trajectories of system (\ref{MFSyst})
(Fig. \ref{PeriodicNet}). In this case neither
formula (\ref{Sigma12Elekt}) nor formula 
(\ref{Sigma12Hole}) can be used for the value of
$\, \sigma^{12} \, $ although formula (\ref{SigmaNeNh})
is still valid in this situation. It is not difficult
to see actually, that the corresponding direction 
$\, {\bf B}_{0} \, $ belongs in this case to a Stability 
Zone with the topological numbers defined by the 
plane $\, \Gamma_{0} \, $, orthogonal to 
$\, {\bf B}_{0} \, $. As a result, any small deviation
of the direction of $\, {\bf B} \, $ causes the appearance 
of open trajectories on the Fermi surface, which prevents 
the determination of its type. 

 It can be seen that the situation described above
can be considered in a sense as the situation, opposite
to those where the direction of $\, {\bf B} \, $ lies 
outside any of the Stability Zones. As we have said 
already, in the last case we can define uniquely the type 
of the Fermi surface (electron or hole type) even for 
extended Fermi surfaces in the $\, {\bf p}$ - space.

\begin{figure}[t]
\begin{center}
\vspace{5mm}
\includegraphics[width=\linewidth]{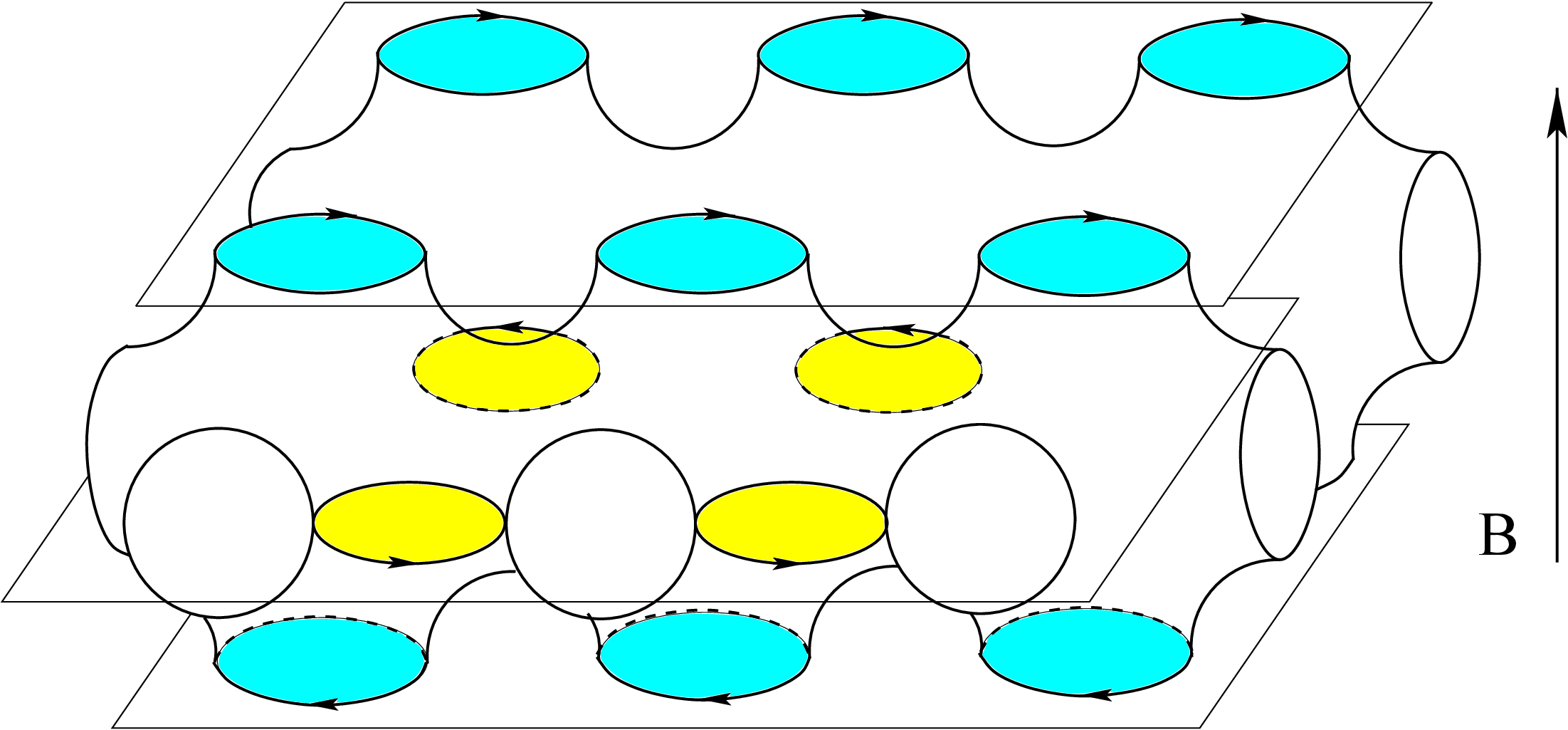}
\end{center}
\caption{Pictures of different types (I or II) arising
in different planes orthogonal to $\, {\bf B} \, $
for special rational direction of $\, {\bf B} \, $.}
\label{FermiSurfSingNet}
\end{figure}

 The example of the surface mentioned above
can be easily constructed if we consider rather simple
Fermi surface (of rank 3) and the magnetic field with the 
direction $\, (0, 0, 1) \, $ in the reciprocal lattice 
basis (Fig. \ref{FermiSurfSingNet}). It's not difficult
to see that the corresponding direction 
$\, {\bf B}_{0} \, $ belongs in this case to a Stability 
Zone with the topological numbers corresponding to the 
plane $\, \Gamma_{0} \, $, orthogonal to 
$\, {\hat z} \, $. At the same time, for the direction
$\, {\bf B}_{0} \, = \, (0, 0, 1) \, $ open trajectories
are absent and only closed or singular trajectories
exist on the Fermi surface. 

 We can see also that the special situation described 
above is in a sense exceptional and corresponds just
to special isolated directions of $\, {\bf B} \, $.
At the same time, this situation can be quite frequent
in experiments since it corresponds to the central
points of the most symmetric Stability Zones on
the angular diagram. As a result, it can appear quite 
often due to a special setting of an experiment. 
Due to this circumstance, we must, therefore, consider 
this case on a par with the generic cases I and II.

\vspace{1mm}

\begin{figure}[t]
\begin{center}
\vspace{5mm}
\includegraphics[width=0.9\linewidth]{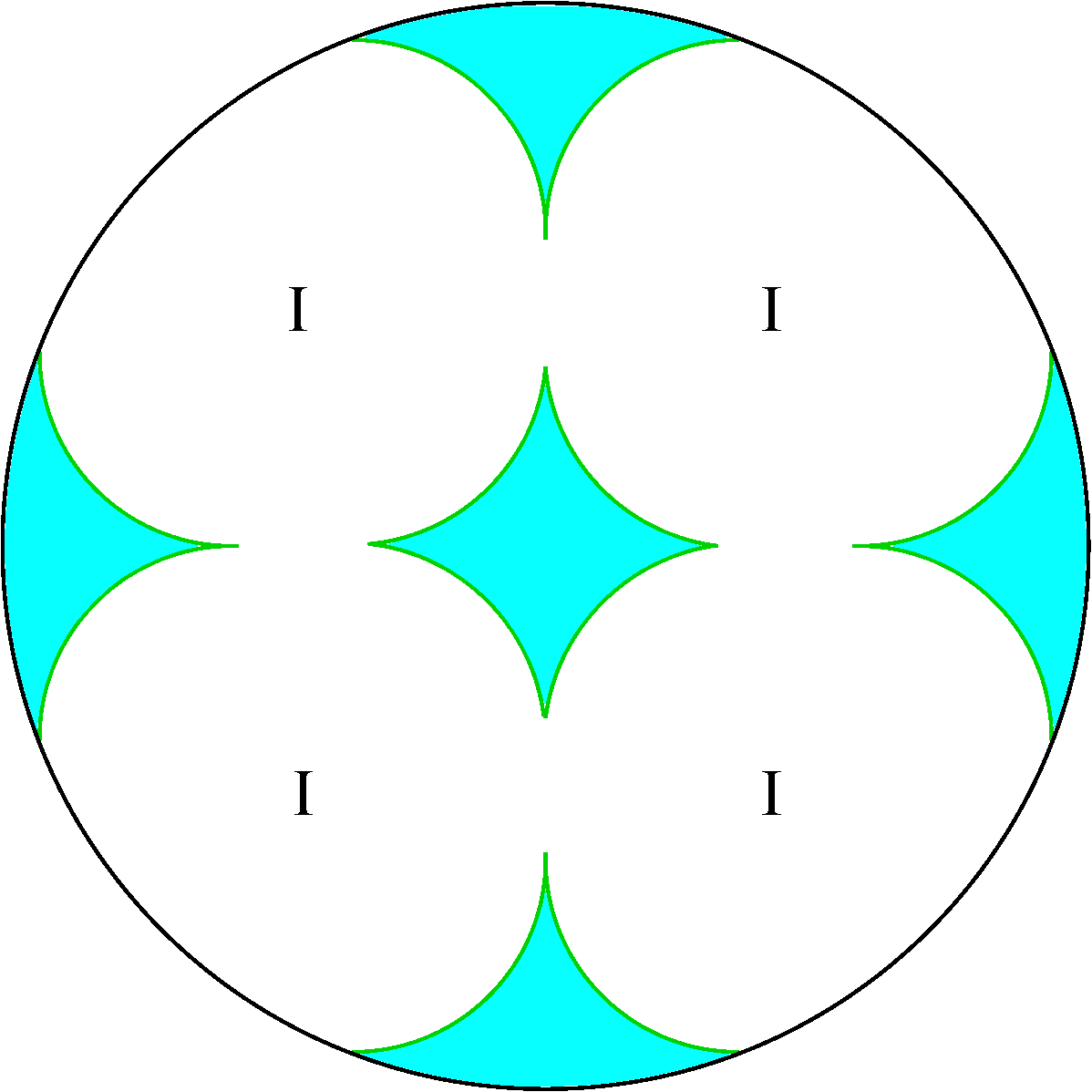}
\end{center}
\caption{A ``simple'' angular diagram containing just
a finite number of Stability Zones and corresponding 
to the picture of the same type (I) on the Fermi surface 
outside the Stability Zones (schematically, only 
mathematical boundaries of Stability Zones are shown).}
\label{TypeA}
\end{figure}

 Let us say now, that for many Fermi surfaces we can have
the situation when we observe just the case I or the case
II everywhere outside the Stability Zones on the angular
diagram. This situation arises when all the Stability
Zones have the first boundaries (simple or compound),
corresponding to disappearance of a cylinder of closed
trajectories of the same (electron or hole) type. 
Indeed, it is not difficult to see, that every time
when the domain 
$\, \Omega^{\prime}_{\alpha} \,\,\, = \,\,\, 
\Sigma_{\alpha} \backslash \Omega_{\alpha} \,$
is separated from the Zone $\, \Omega_{\alpha} \, $ 
by a line, corresponding to disappearance of a cylinder 
of closed trajectories of a certain type, we will have 
the situation I or II of the opposite type on the Fermi 
surface for any generic direction 
$\, {\bf B}/B \, \in \, \Omega^{\prime}_{\alpha} \, $.
In particular, the Fermi surface does not change its 
type under the reconstructions of the closed trajectories
inside the domain $\, \Omega^{\prime}_{\alpha} \, $
described in the previous chapter. It can also be seen, 
that the same situation persists after crossing the 
second boundary of a Stability Zone, as it was described 
in the previous section. We can see, in particular, that 
the situation described above should take place for angular 
diagrams containing just a finite number of Stability Zones, 
which do not divide the sphere $\, \mathbb{S}^{2} \, $ into 
unrelated domains (Fig. \ref{TypeA}).

\vspace{1mm}

 Let us call here angular diagrams corresponding to
the above situation the angular diagrams of the Type A.
In the same way, we can call an extended Fermi
surface the Fermi surface of the Type A, if it has
the angular diagram of the Type A.

\vspace{1mm}

\begin{figure}[t]
\begin{center}
\includegraphics[width=0.9\linewidth]{CompoundPair}
\end{center}
\begin{center}
\vspace{1cm}
\includegraphics[width=0.8\linewidth]{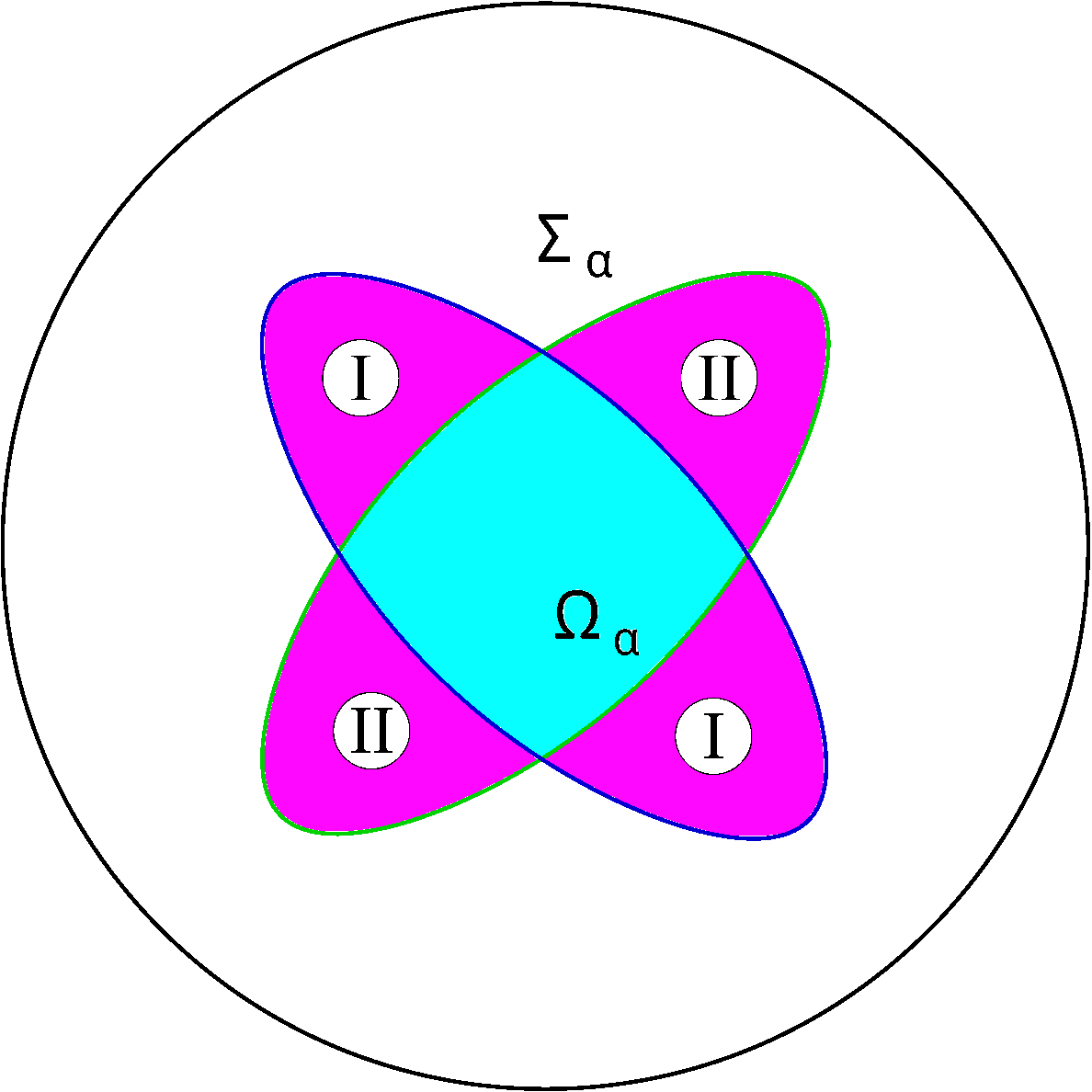}
\end{center}
\caption{An example of the Fermi surface and a
Stability Zone with a compound boundary defined
by disappearance of cylinders of closed trajectories
of different types on its different parts. We have
here different situations (I or II) arising on
the Fermi surface in different connected components
of the domain 
$\, \Sigma_{\alpha} \backslash \Omega_{\alpha} \,$.}
\label{SitIandII}
\end{figure}

 At the same time, the situation when the angular
diagram has different regions where we can observe 
both the situation I and the situation II on the 
Fermi surface also corresponds to the general case.
In particular, this situation should arise whenever 
at least one Stability Zone has a compound boundary 
such that its different parts correspond to disappearance 
of cylinders of closed trajectories of different 
(electron or hole) types. Indeed, it is not difficult
to see that the connected components of the domain
$\, \Omega^{\prime}_{\alpha} \, $, adjacent to the 
corresponding parts of the first boundary of
$\, \Omega_{\alpha} \, $, represent domains,
corresponding to different pictures (I or II)
on the Fermi surface. An example of such a Fermi 
surface can be represented by 
Fig. \ref{CompoundFirstBound}. Thus, if we represent
both the Zones $\, \Omega_{\alpha} \, $ and
$\, \Sigma_{\alpha} \, $ for the representation
of the Fermi surface, corresponding to
Fig. \ref{CompoundFirstBound}, we will see that
we have different pictures (I or II) in different
connected parts of the domain
$\, \Omega^{\prime}_{\alpha} \,\,\, = \,\,\, 
\Sigma_{\alpha} \backslash \Omega_{\alpha} \,$
(Fig. \ref{SitIandII}).

\begin{figure}[t]
\begin{center}
\vspace{5mm}
\includegraphics[width=0.9\linewidth]{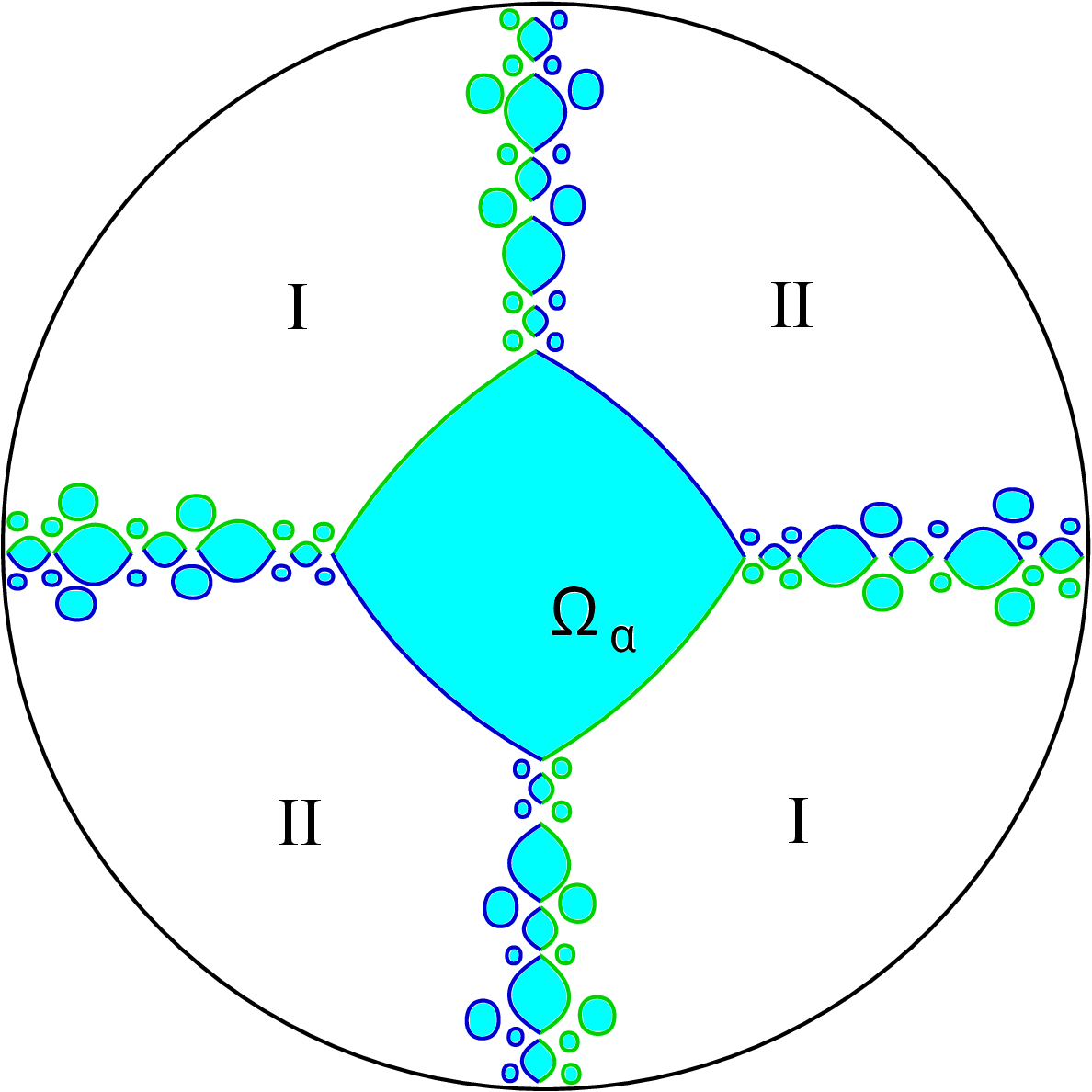}
\end{center}
\caption{A schematic sketch of the angular diagram
containing regions corresponding to different
regimes (I and II) on the Fermi surface (very schematically,
only boundaries of a finite number of Stability Zones 
are shown).}
\label{TypeB}
\end{figure}

 As we said already, the type of the picture of
closed trajectories on the Fermi surface does not 
change after immediate crossing of the second boundary
of a Stability Zone (in generic points), so, the domains 
I and II at Fig. \ref{SitIandII} should be actually extended
outside the Zone $\, \Sigma_{\alpha} \, $ and 
represent in fact larger areas on the angular diagram. 
In general, the situations I and II can coexist on the 
angular diagram if the corresponding domains are separated 
by a set of Stability Zones, which can have rather 
complicated structure in general case. So, we should
expect here that the complete regions corresponding to 
the regimes I and II should be separated on the angular
diagram by chains of Stability Zones, which can be quite 
non-trivial upon careful consideration. In particular, 
these chains must start at the corner points of a Stability 
Zone and we can have a situation where the next Stability Zone 
simply adjoins the corner point and the situation where an 
infinite number of tiny Stability Zones are concentrated 
near the corner point. The first situation corresponds
to non-generic case (when the direction of
$\, {\bf B} \, $ at the corner point corresponds
to appearance of periodic trajectories on the Fermi
surface) while the second situation is generic
(and corresponds to the case when the direction of
$\, {\bf B} \, $ at the corner point corresponds
to irrational mean direction of the open trajectories).
In general, the regions, corresponding to the regimes I 
and II, are separated by an area containing an infinite
number of diminishing Stability Zones, which are 
concentrated along some curves on the angular diagram
(Fig. \ref{TypeB}).

\begin{figure}[t]
\begin{center}
\vspace{5mm}
\includegraphics[width=0.9\linewidth]{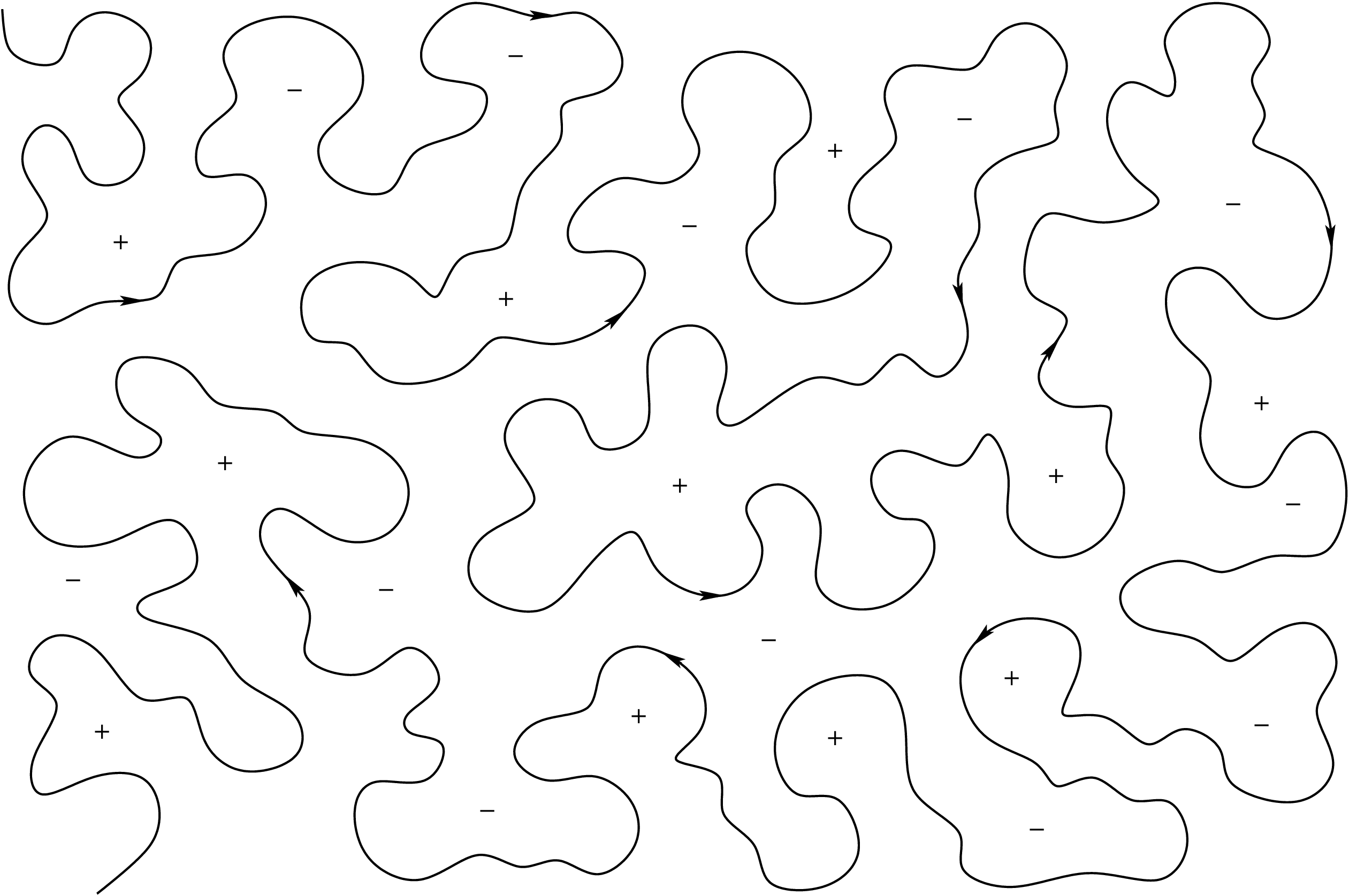}
\end{center}
\caption{A schematic representation of a chaotic 
trajectory of system (\ref{MFSyst}) in the plane,
orthogonal to $\, {\bf B} \, $.}
\label{ChTr}
\end{figure}

\begin{figure}[t]
\begin{center}
\vspace{5mm}
\includegraphics[width=0.9\linewidth]{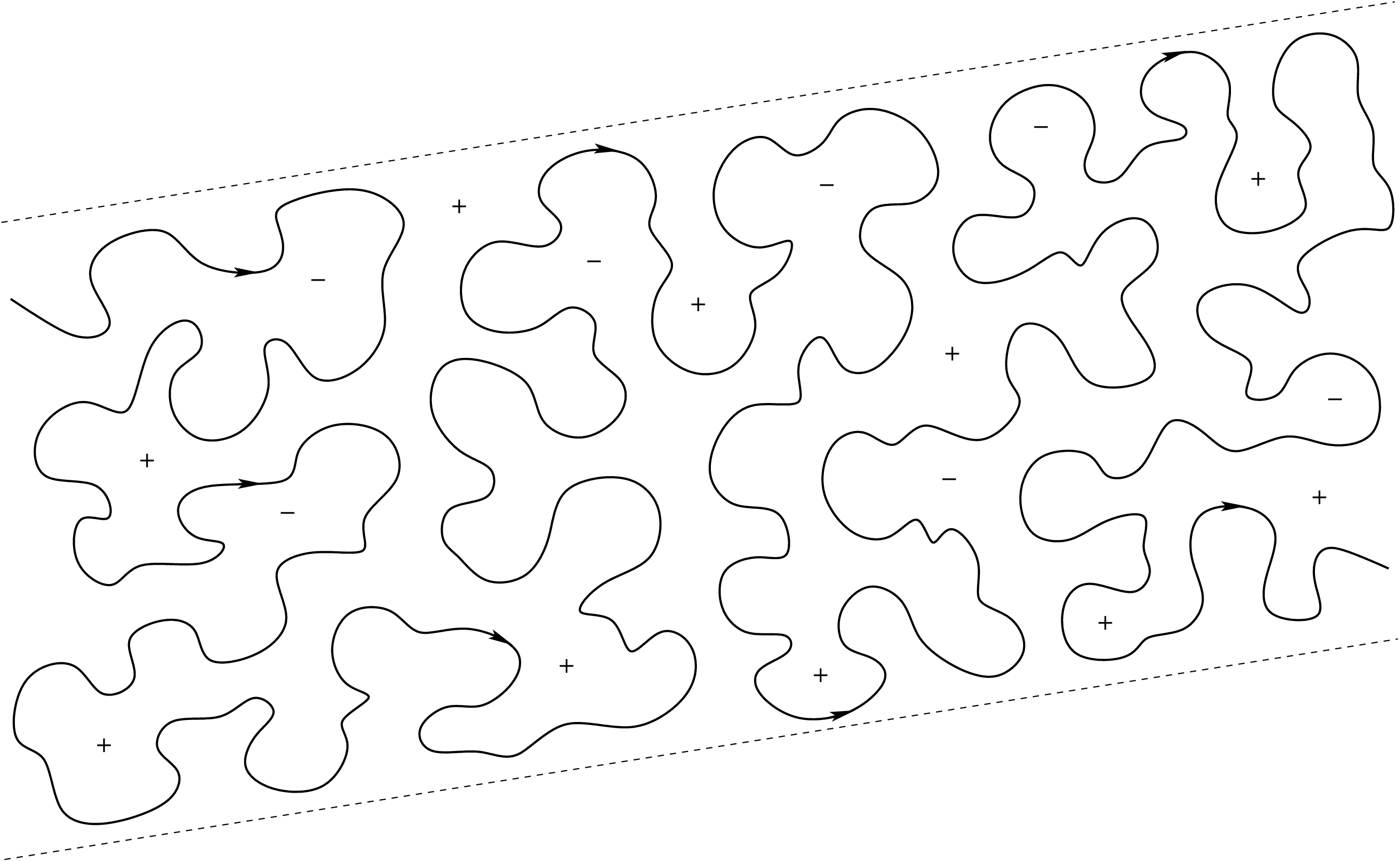}
\end{center}
\caption{A schematic representation of a stable open
trajectory of system (\ref{MFSyst}) in the plane,
orthogonal to $\, {\bf B} \, $, for a small Stability 
Zone.}
\label{WideStr}
\end{figure}

 Let us note here that the regions of concentration of 
tiny Stability Zones can also contain special directions
of $\, {\bf B} \, $ which correspond to appearance of 
more complicated chaotic trajectories on the Fermi 
surface. The behavior of such trajectories in the plane 
orthogonal to $\, {\bf B} \, $ can be schematically 
represented by Fig. \ref{ChTr} and resembles a random walk 
in the plane.  As a result, the conductivity in strong 
magnetic fields also has very special behavior in the 
limit $\, \omega_{B} \tau \, \rightarrow \, \infty \, $
for the corresponding directions of $\, {\bf B} \, $
(\cite{ZhETF1997}). Besides that, the behavior
of stable open trajectories in small Stability Zones
has in fact both the features of ``regular'' and
chaotic behavior, depending on the time scale 
(Fig. \ref{WideStr}). As a result, the conductivity
in strong magnetic fields can also reveal here both the
features of ``regular'' and ``special'' behavior
depending on the value of $\, \omega_{B} \tau \, $.
Let us call angular diagrams containing both the regions 
corresponding to situation I and the regions corresponding 
to situation II the angular diagrams of the Type B.

\vspace{1mm}

 Let us note now that we have considered here the situation
when the Fermi surface is represented by the relation
$\, \epsilon_{s} ({\bf p}) \, = \, \epsilon_{F} \, $
for some dispersion law $\, \epsilon_{s} ({\bf p}) \, $.
In general case we can expect actually that the full
Fermi surface is given by the union of several components,
corresponding to several dispersion relations. As we said
already, we will always assume in this case that different
components of the Fermi surface do not intersect each other.

 Quite often we can actually have the situation when the 
``extended'' part of the Fermi surface is given by just
one component
$\, \epsilon_{s_{0}} ({\bf p}) \, = \, \epsilon_{F} \, $,
while the other components are compact in the 
$\, {\bf p}$ - space. Easy to see that in this case
the angular diagram for conductivity is defined completely
by the dispersion relation $\, \epsilon_{s_{0}} ({\bf p}) \, $.
The only difference arising here is that the formulae
(\ref{Sigma12Elekt}) - (\ref{Sigma12Hole}) for the Hall
conductivity outside Stability Zones should be replaced
now by their sum over all the significant dispersion 
relations. We can see that we can say here that the angular
diagram has Type A if the Hall conductivity can take just
one constant value (at a fixed value of $\, B \, $,
$\,\, \omega_{B} \tau \gg 1 $) everywhere outside the 
Stability Zones and the angular diagram has Type B if
it can take two different values outside the Stability Zones.

 In the most general case we can suppose that the extended
part of the Fermi surface is represented by several extended
components 
$\, \epsilon_{s} ({\bf p}) \, = \, \epsilon_{F} \, $,
corresponding to different dispersion relations
$\, \epsilon_{s} ({\bf p}) \, $. Easy to see that the full
angular diagram for conductivity is given here by the 
superposition of the angular diagrams for the corresponding
dispersion relations. It is important here that under our
requirements above Stability Zones for different dispersion 
relations can overlap only if they correspond to the same
Topological numbers, defining the mean directions of the 
open trajectories. As a result, we can describe in fact
the full angular diagram in the same terms as an angular
diagram for a particular dispersion relation if we admit 
that some Stability Zones can have ``compound'' structure 
but the same main features of conductivity behavior as
the ``simple'' Stability Zones. It is natural to say here
that the full angular diagram has Type B if at least one 
extended component of the Fermi surface has the angular
diagram of Type B. It can be seen also, that we can say 
here that an angular diagram has type B if the Hall
conductivity can take at least two different values
(at a fixed value of $ \, B \, $, 
$\,\, \omega_{B} \tau \gg 1 $)
outside the full Stability Zones on the angular diagram.

\vspace{1cm}

\section{The full angular diagram of a dispersion law and 
the second boundaries of Stability Zones of normal metals.}
\setcounter{equation}{0}

 In this section we will just make some remarks on 
the connection between the full angular diagram of 
a dispersion law and the objects considered in the previous 
sections. The full angular diagram for a dispersion law 
(dispersion relation) was introduced by I.A. Dynnikov 
in \cite{dynn3} and can be described in the following way:

\vspace{1mm}

 Let us consider an arbitrary dispersion relation given by
a three-periodic function $\, \epsilon ({\bf p}) \, $
in the $\, {\bf p}$ - space, satisfying the condition
$\, \epsilon_{min} \, \leq \, \epsilon ({\bf p}) \, \leq \,
\epsilon_{max} \, $. Let us fix an arbitrary direction
of $\, {\bf B} \, $ and consider energy levels
$\, \epsilon ({\bf p}) \, = \, const \, $ containing
non-closed trajectories of system (\ref{MFSyst}).

 Then:
 
\vspace{1mm}
 
 The energy levels containing non-closed trajectories
of system (\ref{MFSyst}) represent either a connected 
interval
$$\epsilon_{1} ({\bf B}/B) \,\,\, \leq \,\,\, \epsilon
\,\,\, \leq \,\,\, \epsilon_{2} ({\bf B}/B) $$
or just one isolated point 
$\, \epsilon \, = \, \epsilon_{0} ({\bf B}/B) $.

\vspace{1mm}

 Everytime, when non-closed trajectories appear in a finite
energy interval 
$\, [ \epsilon_{1} ({\bf B}/B) \, , \,
\epsilon_{2} ({\bf B}/B) ] \, $, all the non-singular
open trajectories of system (\ref{MFSyst}) have the 
regular form shown at Fig. \ref{StableTr} with the same
mean direction, given by the intersection of the plane,
orthogonal to $\, {\bf B} \, $, and some integral plane
$\, \Gamma \, $ in $\, {\bf p}$ - space.

\begin{figure}[t]
\begin{center}
\vspace{5mm}
\includegraphics[width=0.9\linewidth]{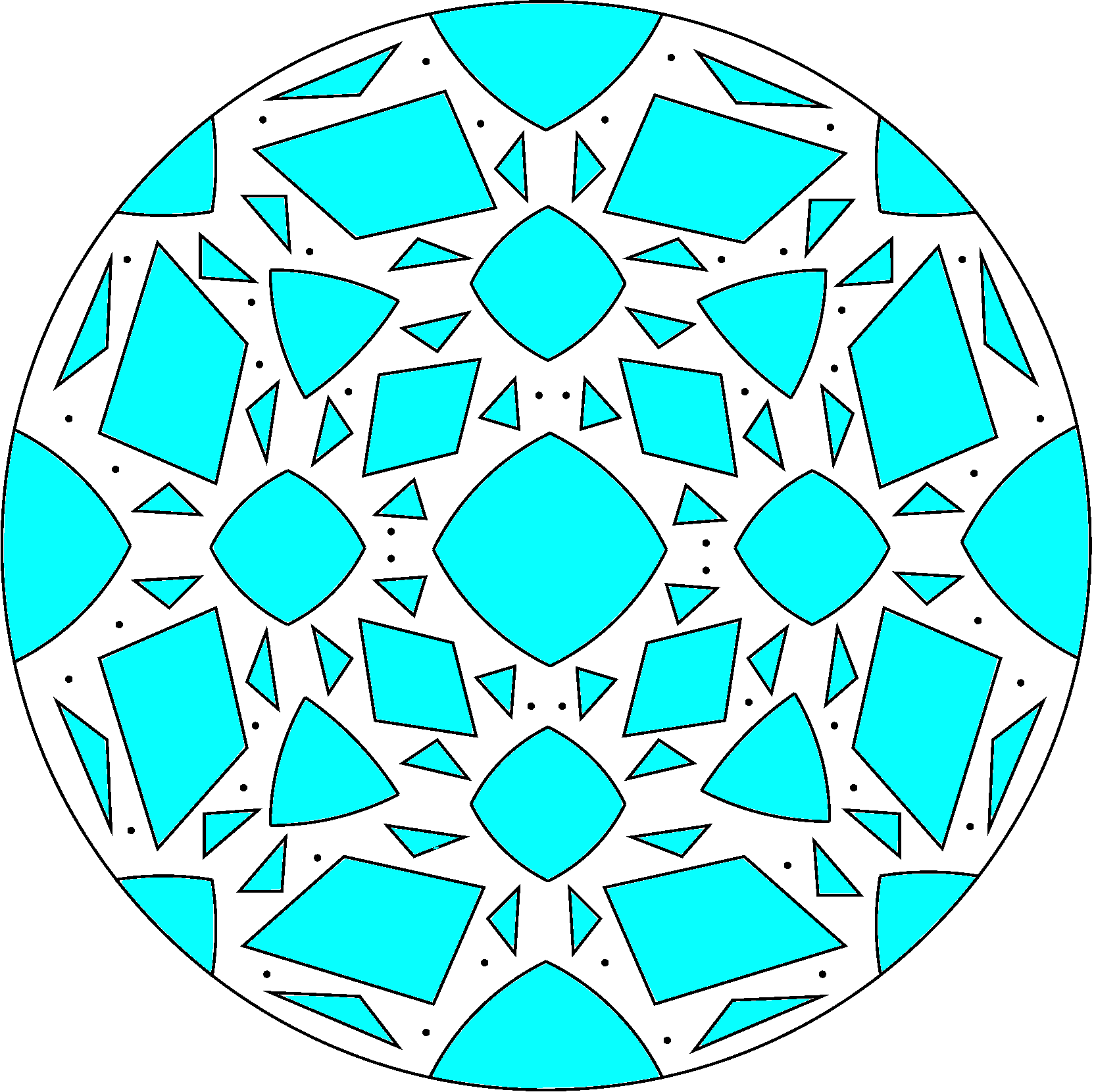}
\end{center}
\caption{An angular diagram for a whole dispersion
relation (very schematically, only a finite number
of Stability Zones $\, \Omega^{*}_{\alpha} \, $
and ``chaotic directions'' is shown).}
\label{DispRel}
\end{figure}

 The open trajectories and the integral plane $\, \Gamma \, $
are stable with respect to small rotations of $\, {\bf B} \, $,
such that the total set of directions of $\, {\bf B} \, $,
corresponding to the same plane $\, \Gamma \, $, represents
a finite region with a piecewise smooth boundary on the 
angular diagram (the unit sphere $\, \mathbb{S}^{2}$).
The region $\, \Omega^{*}_{\alpha} \, $, corresponding to the 
presence of the stable open trajectories with the same integral 
plane $\, \Gamma_{\alpha} \, $, can be called a Stability Zone 
for the dispersion relation $\, \epsilon ({\bf p}) \, $.
Let us also note that the boundaries of Stability Zones
have in this case both the ``electron'' and the ``hole'' 
type since they correspond here to the disappearance of 
simultaneously two cylinders of closed trajectories of 
opposite types. All the Stability Zones 
$\, \Omega^{*}_{\alpha} \, $ form
an everywhere dense set on the angular diagram
(Fig. \ref{DispRel}), which can contain either just one 
or infinitely many different Stability Zones.

\vspace{1mm}

 For generic directions of $\, {\bf B} \, $ the values
$\, \epsilon_{1} ({\bf B}/B) \, $ and
$\, \epsilon_{2} ({\bf B}/B) \, $ coincide with the 
values of some globally defined continuous functions
$\, \tilde{\epsilon}_{1} ({\bf B}/B) \, $ and
$\, \tilde{\epsilon}_{2} ({\bf B}/B) \, $,
such that
$\, \tilde{\epsilon}_{1} ({\bf B}/B) \, = \,
\tilde{\epsilon}_{2} ({\bf B}/B) \, $ everywhere
on the boundaries of Stability Zones. At the same time,
for those directions of $\, {\bf B} \, $ which correspond
to purely rational mean directions of the open trajectories,
we always have the relations
\begin{equation}
\label{SpecialRel}
\epsilon_{1} ({\bf B}/B) \,\,\, \leq \,\,\, 
\tilde{\epsilon}_{1} ({\bf B}/B) \,\,\, , \quad 
\epsilon_{2} ({\bf B}/B) \,\,\, \geq \,\,\, 
\tilde{\epsilon}_{2} ({\bf B}/B)
\end{equation}

\vspace{1mm}

 The complement to the set $\, \{ \Omega^{*}_{\alpha} \} \, $
on the unit sphere represents a very complicated set 
(of Cantor type) and consists of the directions of 
$\, {\bf B} \, $ corresponding to appearance of chaotic open
trajectories on just one energy level
$$\epsilon_{0} ({\bf B}/B) \,\,\, = \,\,\, 
\tilde{\epsilon}_{1} ({\bf B}/B) \,\,\, = \,\,\,
\tilde{\epsilon}_{2} ({\bf B}/B) $$

\vspace{1mm}

 Let us just say here that the structure of the set of these
special directions and the properties of the chaotic trajectories
are being intensively investigated at present and represent
very interesting branch of the dynamical systems theory
(see e.g. \cite{dynn2,dynn3,ZhETF1997,zorich2,Zorich1996,
ZorichAMS1997,zorich3,DeLeo1,DeLeo2,DeLeo3,ZorichLesHouches,
DeLeoDynnikov1,DeLeoDynnikov2,Skripchenko1,Skripchenko2,
DynnSkrip1,DynnSkrip2,AvilaHubSkrip1,AvilaHubSkrip2,DeLeo2017}).

\vspace{1mm}

 Coming back to the theory of normal metals, we can 
see that we get non-closed trajectories on the Fermi surface
if one of the following relations
$$\epsilon_{1} ({\bf B}/B) \,\, < \,\, \epsilon_{F}
\,\, < \,\, \epsilon_{2} ({\bf B}/B) \,\, \quad 
{\rm or} \quad \,\,
\epsilon_{F} \,\, = \,\, \epsilon_{0} ({\bf B}/B) $$
is satisfied.

 As a result, in real metals we can observe just some 
parts $\, \Omega_{\alpha} \, $ of the Zones
$\, \Omega^{*}_{\alpha} \, $, which are 
not everywhere dense now on the unit sphere. It can be
seen as well, that the relation (\ref{SpecialRel}) gives
also a possibility of existence of periodic open 
trajectories on the Fermi surface outside the 
Stability Zones $\, \Omega_{\alpha} \, $ on the 
corresponding angular diagram.

 It is not difficult to see, that for any Stability Zone
$\, \Omega_{\alpha} \, $ we can always write the inclusion
$\, \Omega_{\alpha} \, \subset \, \Omega^{*}_{\alpha} \, $
on the angular diagram. It can be shown also that we always
have in this case the inclusion
$\, \Omega^{*}_{\alpha} \, \subset \, \Sigma_{\alpha} \, $
for the regions $\, \Sigma_{\alpha} \, $ introduced in the 
previous sections. Thus, we can claim in fact that the 
derivative
$\, \Omega^{\prime}_{\alpha} \, = \,
\Sigma_{\alpha} \backslash \Omega_{\alpha} \, $
of any Stability Zone $\, \Omega_{\alpha} \, $ always
contains the entire boundary of the corresponding Zone
$\, \Omega^{*}_{\alpha} \, $.

 Another connection between the full angular diagram for 
a dispersion law and the angular diagram for a fixed Fermi
surface can be detected in the Type of a diagram (A or B) 
introduced in the previous section. Indeed, it can be seen
that the Type of the angular diagram of a real
metal is closely connected with the position of the 
Fermi level $\, \epsilon_{F} \, $ with respect to the 
values of $\, \tilde{\epsilon}_{1} ({\bf B}/B) \, $ and
$\, \tilde{\epsilon}_{2} ({\bf B}/B) \, $ at different 
directions of $\, {\bf B}$. Thus, we can see that the 
angular diagram of a metal has Type A if one of the 
following relations is satisfied everywhere on 
$\, \mathbb{S}^{2} \, $:
$$\epsilon_{F} \,\,\, \leq \,\,\, 
\tilde{\epsilon}_{2} ({\bf B}/B) \quad \quad
{\rm (electron \,\,\, type)} $$
or
$$\epsilon_{F} \,\,\, \geq \,\,\,
\tilde{\epsilon}_{1} ({\bf B}/B) \quad \quad
{\rm (hole \,\,\, type)} $$

 At the same time, the angular diagram of a metal has Type B
if in different parts of $\, \mathbb{S}^{2} \, $ we can find
both the relations
$$\epsilon_{F} \,\,\, > \,\,\, 
\tilde{\epsilon}_{2} ({\bf B}/B) $$
and
$$\epsilon_{F} \,\,\, < \,\,\,
\tilde{\epsilon}_{1} ({\bf B}/B)$$

\vspace{1cm}

\section
{Conclusions.}
\setcounter{equation}{0}

 We have considered a number of general questions concerning 
the behavior of the electrical conductivity of normal metals 
in the presence of a strong magnetic field. The main discussions 
were devoted to special features of the angular diagram for 
conductivity in the case of the appearance of stable quasiclassical 
open electron trajectories on the Fermi surface under certain 
directions of $\, {\bf B} $. In particular, it was shown that every 
Stability Zone $\, \Omega_{\alpha} \, $ on the angular diagram 
corresponding to the presence of stable open trajectories on 
the Fermi surface has an additional second boundary restricting 
a special domain $\, \Omega^{\prime}_{\alpha} \, $ around the 
Zone $\, \Omega_{\alpha} \, $. For directions of $\, {\bf B} \, $, 
belonging to the domain $\, \Omega^{\prime}_{\alpha} \, $, 
quasiclassical electron trajectories on the Fermi surface admit 
an effective description closely related to the description of 
stable open trajectories for the Zone $\, \Omega_{\alpha} \, $.
The effective description of the quasiclassical electron trajectories 
makes it possible to indicate a number of special features in the 
behavior of conductivity and other quantities in the domain
$\, \Omega^{\prime}_{\alpha} \, $, which allows to select this 
domain from other areas on the angular diagram. From the experimental 
point of view, the most convenient method for determining the 
boundaries (the first and the second) of the Zone 
$\, \Omega_{\alpha} \, $ and investigation the internal structure 
of the domain $\, \Omega^{\prime}_{\alpha} \, $ is to study oscillation 
phenomena (classical and quantum) for the corresponding directions 
of the magnetic field.

 Another issue discussed in the paper is the complexity of the 
distribution of the Zones $\, \Omega_{\alpha} \, $ and 
$\, \Omega^{\prime}_{\alpha} \, $ on the angular diagram for a metal 
with an arbitrary Fermi surface. In particular, we indicate here a 
theoretical possibility of separating of all angular diagrams into two 
types, which presuppose a simpler and more complex structure for the 
arrangement of these Zones in the space of directions of $\, {\bf B} $.
As can be also shown, the complexity of an angular diagram is closely
connected with the behavior of the Hall conductivity outside the 
Stability Zones in the presence of a strong magnetic field. 

 In conclusion, we compare the described properties of the angular 
diagrams for conductivity in metals with the properties of angular 
diagrams for the full dispersion relations that can be introduced 
at the theoretical level. As can be seen in this case, it is possible 
to indicate a number of significant interrelations between the angular 
diagrams of both types.

\vspace{2cm}

The study was carried out at the expense of a grant from the 
Russian Science Foundation (project \textnumero $\, $ 18-11-00316).

\end{document}